%% file: newModel.tex
\documentclass[preprint,3p,times,onecolumn]{elsarticle}
\usepackage{amsmath,amssymb,pennames,fixltx2e,lineno}

\def\d{\mathrm{d}}
\def\p{\partial}
\def\erf{\operatorname{erf}}
\renewcommand{\vec}[1]{{\boldsymbol{#1}}}

\newcommand*\patchAmsMathEnvironmentForLineno[1]{%
  \expandafter\let\csname old#1\expandafter\endcsname\csname #1\endcsname
  \expandafter\let\csname oldend#1\expandafter\endcsname\csname end#1\endcsname
  \renewenvironment{#1}%
     {\linenomath\csname old#1\endcsname}%
     {\csname oldend#1\endcsname\endlinenomath}}%
\newcommand*\patchBothAmsMathEnvironmentsForLineno[1]{%
  \patchAmsMathEnvironmentForLineno{#1}%
  \patchAmsMathEnvironmentForLineno{#1*}}%
\AtBeginDocument{%
\patchBothAmsMathEnvironmentsForLineno{equation}%
\patchBothAmsMathEnvironmentsForLineno{align}%
\patchBothAmsMathEnvironmentsForLineno{flalign}%
\patchBothAmsMathEnvironmentsForLineno{alignat}%
\patchBothAmsMathEnvironmentsForLineno{gather}%
\patchBothAmsMathEnvironmentsForLineno{multline}%
}

\begin{document}

\begin{frontmatter}

\title{A simple energy loss model and its applications for silicon detectors}

\author{Ferenc Sikl\'er}
\ead{sikler@rmki.kfki.hu}
\address{KFKI Research Institute for Particle and
Nuclear Physics, Budapest, Hungary \\ CERN, Geneva, Switzerland}

\begin{abstract}

The energy loss of charged particles in silicon can be approximated by a simple
analytical model. With help of measured charge deposits in individual channels
of hit clusters their position and energy can be estimated. Deposits below
threshold and saturated values are treated properly, resulting in a wider
dynamic range. The proposed method gives improvements on both hit position and
energy residuals. The model is successfully applied to track differential
energy loss estimation and to detector gain calibration tasks.

\end{abstract} 

\begin{keyword}
Energy loss \sep Silicon
\PACS 29.40.Gx \sep 29.85.-c \sep 34.50.Bw
\end{keyword}

\end{frontmatter}

\section{Introduction}

The identification of charged particles is essential in several fields of
particle and nuclear physics: particle spectra, correlations, selection of
daughters of resonance decays and for reducing the background of rare physics
processes \cite{Ullaland:2003cm,Yamamoto:1999uc}. Silicon detectors can be
employed for identification by proper use of energy deposit measurements along
the trajectory of the particle.
The aim of this study is to provide a simple method to evaluate energy deposits
and other derived quantities for a wide range of particle momenta ($\beta\gamma
= p/m = 0.56 - 10.0$) and detector thickness, based on the precise knowledge of
the known underlying physics processes.

This article is organized as follows: the model is motivated and discussed in
detail in Sec.~\ref{sec:model}. Among the various applications we will deal
with the estimation of hit position (Sec.~\ref{sec:position}), hit energy
deposit (Sec.~\ref{sec:energy}), differential energy loss for tracks
(Sec.~\ref{sec:differential}), as well as detector gain calibration for
tracks (Sec.~\ref{sec:gain}). This work ends with conclusions.

\begin{figure}[!t]
 
 \begin{center}
  \input{density_1}
 \end{center}

 \caption{Probability density functions for particles with $\beta\gamma =
1.00$, at given path lengths of 20, 50, 100 and 200~$\mu$m.}

 \label{fig:densityElossBichsel}

\end{figure}
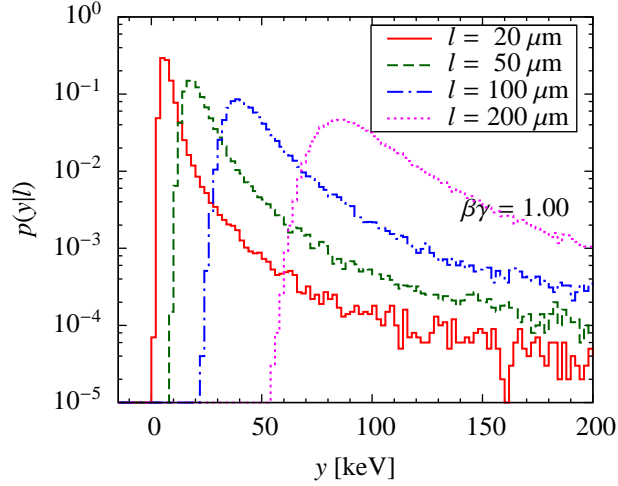

\section{Simple energy loss model}
\label{sec:model}

When a charged particle traverses material it loses energy in several discrete
steps, dominantly by resonance excitations ($\delta$-function) and Coulomb
excitations (truncated $1/E^2$ term). This latter is the reason for the long
tail observed in energy deposit distributions.
It is generally difficult to describe the process with a simple function.
The most probable energy loss $\Delta$, and the
full width of the energy loss distribution at half maximum $\Gamma_\Delta$
\cite{Bichsel:1988if} can be approximated by
\begin{align}
 \label{eq:Delta}
 \Delta &= \xi \left[\log\frac{2 mc^2 \beta^2 \gamma^2 \xi}{I^2}
            + 0.2000 - \beta^2 - \delta \right] \\
 \Gamma_\Delta &= 4.018 \xi \nonumber
\end{align}

\begin{figure}

 \begin{center}
  \input{bg_3.16_all}
 \end{center}

 \caption{Fits of the conditional probability density $f_y(l)$ as function of
the path length $l$ for particles with $\beta\gamma = 3.16$ (see
Eq.\eqref{eq:pyl1}). The simulated values (points) and the fitted curves
(lines) corresponding to $y = 50, 100, 200$ and $300$~keV are shown, from left
to right.}

 \label{fig:fitBichsel1}

\end{figure}
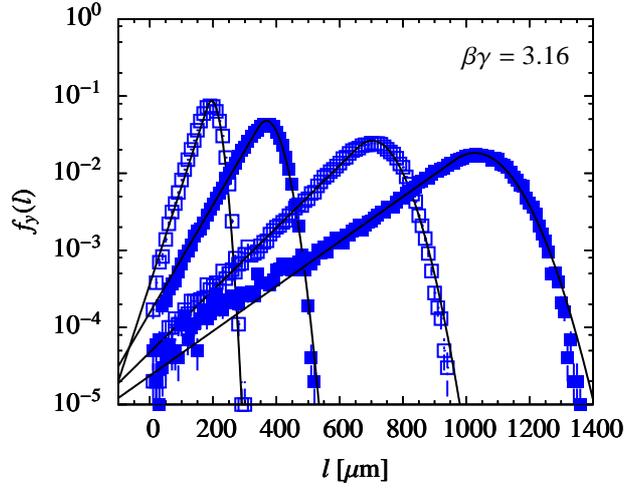

\noindent where
\begin{equation*}
 \xi = \frac{K}{2} z^2 \frac{Z}{A} \rho \frac{l}{\beta^2}
\end{equation*}
 
\noindent is the Landau parameter; $K = 4\pi N_A r_e^2 m_e c^2 =
0.307~075~\mathrm{MeV~cm^2/mol}$; $Z$, $A$, $\rho$ and $l$ are the mass number,
atomic number, density and thickness of the material, respectively; and
$\delta$ is the density correction \cite{Nakamura:2010zzi}.

The probability of an excitation, energy deposit, along the path of the
incoming particle is a function of $\beta\gamma = p/m$ of the particle and
depends on properties of the traversed material. The conditional probability
density $p(y|l)$, deposit $y$ along a given path length $l$, can be built using
the above mentioned elementary excitations combined with an exponential
occurrence model. The details of the microscopical simulation for silicon can
be found in Refs.~\cite{Bichsel:1988if}, \cite{Bichsel:1990} and
\cite{Bichsel:2006cs}. The result of these recursive convolutions is a smooth
function (Fig.~\ref{fig:densityElossBichsel}).

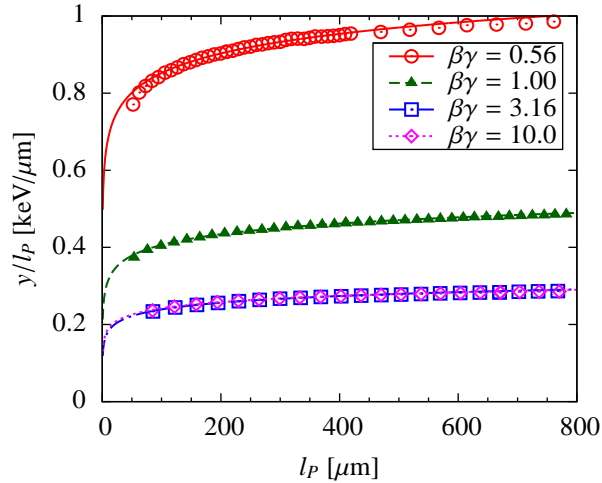
\begin{figure}[!h]

 \begin{center}
  \input{ylp_fit}
 \end{center}

 \caption{Relationship between the measured deposited energy $y$ and the
position of the path length peak $l_p$, shown together with fits using
Eq.~\ref{eq:ylp}.} 

 \label{fig:ylp_fit}

\end{figure}

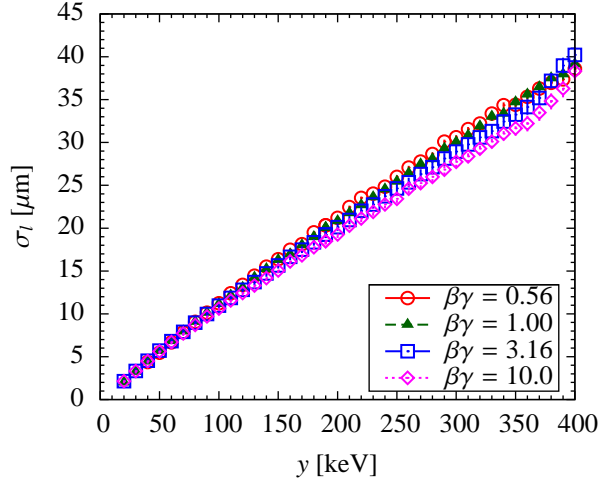
\begin{figure}[!h]

 \begin{center}
  \input{sllp_fit}
 \end{center}

 \caption{Relationship between the measured deposited energy $y$ and the
standard deviation $\sigma_l$ of $l$ distribution.}

 \label{fig:l_fit}

\end{figure}

In the experiment the deposit values $y$ are measured. For a fixed $y$ the
conditional probability can be regarded as function of $l$, thus $p(y|l) \equiv
f_y(l)$.
Values of $f_y(l)$ for particles with $\beta\gamma = 3.16$ are plotted in
Fig.~\ref{fig:fitBichsel1}, where the values corresponding to $y = 50, 100,
200$ and $300$~keV are shown. Other $\beta\gamma$ settings show very similar
behavior ($\beta\gamma = 0.56, 1.00$ and $10.0$, not plotted).
Apart from a multiplicative factor, they are well approximated by a combination
of simple functions, exponential and Gaussian:
\begin{equation}
 f_y(l) \propto
  \begin{cases}
    \exp\left(  \frac{\nu(l-l_P)}{\sigma_l} +
                 \frac{\nu^2}{2}\right),
    & \text{if $l <   l^*$} \\
     \exp\left(- \frac{(l - l_P)^2}{2\sigma_l^2} \right),
    & \text{if $l \ge l^*$}
  \end{cases}
  \label{eq:pyl1}
\end{equation}

\noindent where the limit is
\begin{equation*}
 l^* = l_P - \nu \sigma_l.
\end{equation*}

\noindent 
The peak position $l_P$ and standard
deviation $\sigma_l$ are both functions of $y$ and $\beta\gamma$, $\nu \approx
0.65$ is constant.
Note that $f_y$ is constructed such that the value and the derivative
are continuous at the limit $l^*$.
The peak position $l_P$ is roughly proportional to the deposit $y$,
their relationship can be approximated as
\begin{equation}
 y \approx \varepsilon l_p \left[1 + 0.08 \log(l_P/l_0)\right]
 \label{eq:ylp}
\end{equation}

\noindent where $l_0$ is some {\it reference path length}
(Fig.~\ref{fig:ylp_fit}). In this work $l_0 =$ 300~$\mu$m was chosen. At the
same time $\sigma_l$ has an approximate first order polynomial dependence on
$y$ (Fig.~\ref{fig:l_fit}).

As it was shown above, some interesting connections between the deposit and path
length can be observed, for wide range of $\beta\gamma$ values. Unfortunately
the dependence on $\beta\gamma$, and $l$, is still present. In the following
(Sec.~\ref{sec:mpde}) we will show how that can be suppressed or even
eliminated.

\subsection{Most probable differential energy loss}
\label{sec:mpde}

\begin{figure}
 
 \begin{center}
  \input{ylp_fit_2}
 \end{center}

 \caption{Relationship between the most probable energy deposit $\Delta$ and
the
path length $l$, shown together with fits using Eq.~\ref{eq:Delta_vs_l}.}

 \label{fig:l_fit_2}

\end{figure}
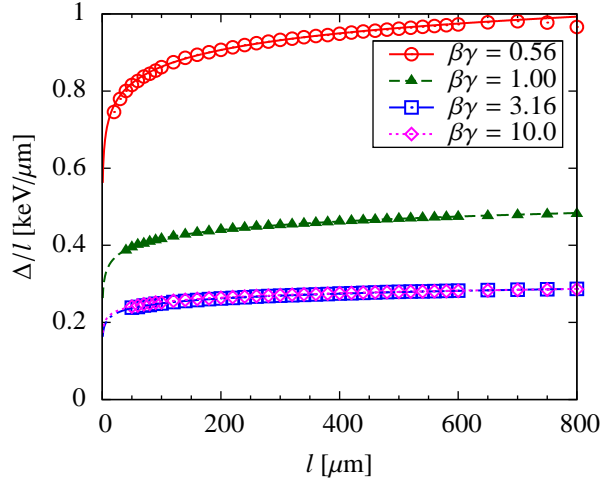

The path length dependence of the most probable energy loss $\Delta$
(Fig.~\ref{fig:l_fit_2}) has a form that is very similar to Eq.~\eqref{eq:ylp}.
The dependence can be approximated by
\begin{equation}
 \label{eq:Delta_vs_l}
 \Delta(l) \approx \varepsilon l \left[1 + a \log (l  /l_0)\right]
\end{equation}

\noindent where $\varepsilon$ is the {\it most probable differential energy
loss} along a reference length $l_0$, hence $\varepsilon = \Delta(l_0)/l_0$.
For the relevant $\beta\gamma$ region studied in this paper $a \approx 0.07$,
to a good approximation.\footnote{Based on Eq.~\ref{eq:Delta}, with some
approximations, the expected magnitude of $a$ would be $1/\log(2 m_e c^2 K z^2
\cdot Z/A \cdot \rho / I^2) \approx 0.064$.} For some applications,
such as hit position estimation, the assumption of linearity may be enough (see
Eq.~\eqref{eq:Delta_vs_l_linear} later).

\begin{figure*}[t]

 \begin{center}
  \input{bg_0.56_all_2}
  \input{bg_1.00_all_2}
 \end{center}
 
 \vspace{-0.3in}
 
 \begin{center}
  \input{bg_3.16_all_2}
  \input{bg_10.00_all_2}
 \end{center}

 \caption{Fits of the conditional probability density $f_y(\Delta)$ as function
of the most probable energy loss $\Delta$ for particles with $\beta\gamma =
0.56$, $1.00$, $3.16$, and $10.0$ (see Eq.~\eqref{eq:p_Delta}). The simulated
values (points) and the fitted curves (lines) corresponding to $y = 50, 100,
200$ and $300$~keV are shown, from left to right.}

 \label{fig:fitBichsel2}

\end{figure*}
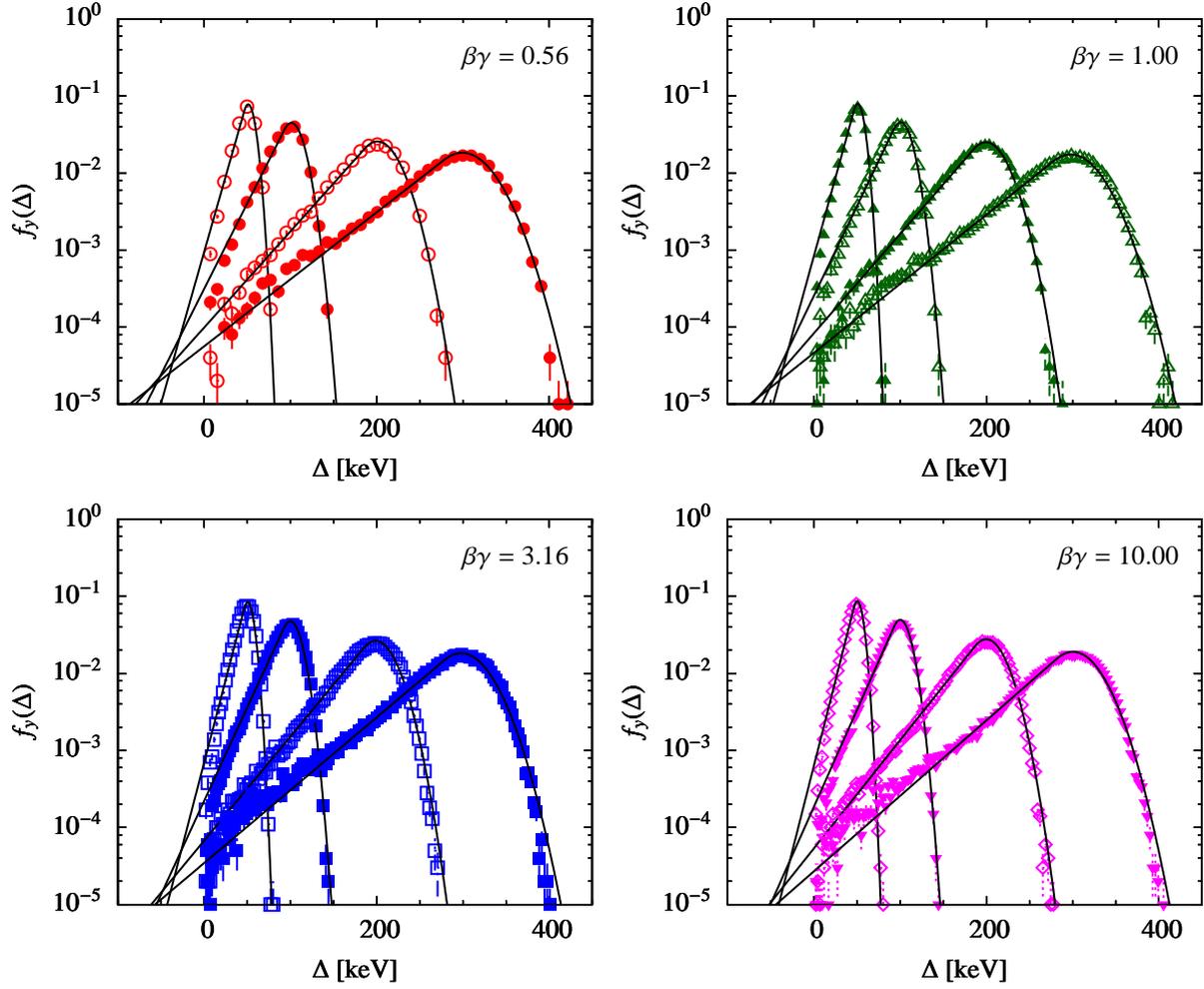

Hence it is worthwhile to check whether the dependence on $\beta\gamma$ and $l$
can be suppressed by taking the most probable deposit $\Delta$ as variable of
the conditional probability.
Values of $p(y|\Delta) \equiv f_y(\Delta)$ for particles with $\beta\gamma =
0.56, 1.00, 3.16$ and $10.0$ are plotted in Fig.~\ref{fig:fitBichsel2}, where
the values corresponding to $y = 50, 100, 200$, and $300$~keV are shown.
The distributions for a given $y$ value are remarkably similar for all
$\beta\gamma$ values.

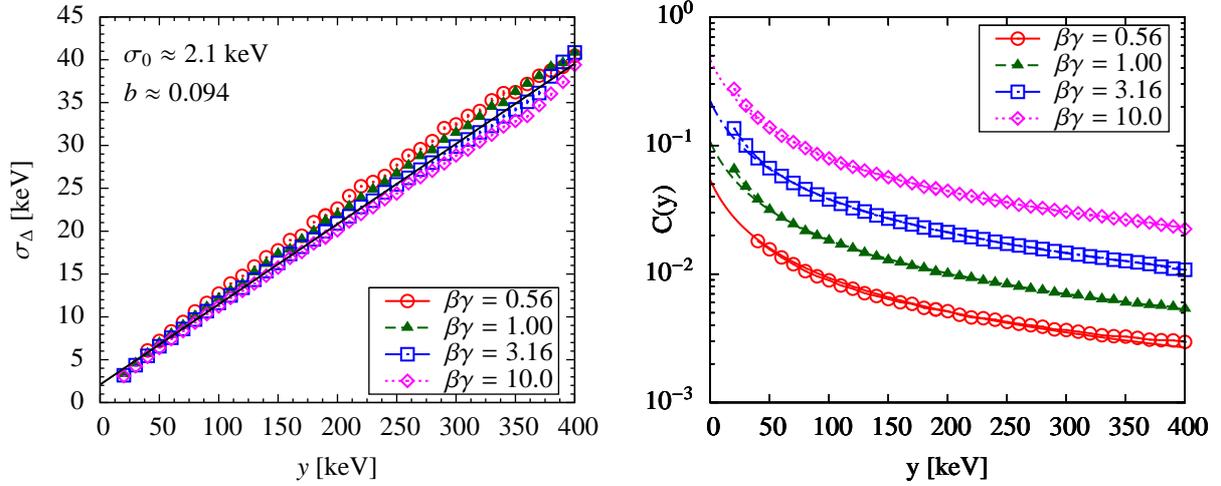
\begin{figure*}[!h]
 
 \begin{center}
  \input{sllp_fit_2}
  \input{exponent}
 \end{center}

 \caption{Left: Relationship between the measured deposited energy $y$ and the
standard deviation $\sigma_\Delta$ of $\Delta$ distribution. Right: The
measured value of $C(y)$ for particles with $\beta\gamma = 0.56$, $1.00$,
$3.16$, and $10.0$. The points and curves are successively multiplied by
factors of 2 for clarity. Note that the plotted function $1/\sigma_\Delta(y)$
is the same in all cases except for the multiplicative factor mentioned
before.}

 \label{fig:sllp_fit_2_exponent}

\end{figure*}

The conditional probability density for a given $y$, as a function of $\Delta$
can be described as
\begin{equation}
 \label{eq:p_Delta}
 f_y(\Delta) \approx C(y) \cdot
  \begin{cases}
    \exp\left(  \frac{\nu(\Delta - y)}{\sigma_\Delta} +
                 \frac{\nu^2}{2}\right),
    & \text{if $\Delta <   \Delta^*$} \\
      \exp\left(- \frac{(\Delta - y)^2}{2\sigma_\Delta^2} \right),
    & \text{if $\Delta \ge \Delta^*$}
  \end{cases}
\end{equation}

\noindent where $C(y)$ is a universal function of $y$. Note that $f_y$ is
constructed such that the value and the
derivative are continuous at the limit $\Delta^*$ that is given by
\begin{equation}
 \label{eq:Delta_star}
 \Delta^*(y) = y - \nu \sigma_\Delta(y)
\end{equation}

\noindent where $\nu \approx 0.65$ is constant.
Fig.~\ref{fig:sllp_fit_2_exponent}-left shows that $\sigma_\Delta$ is a
first order polynomial of $y$, practically independent of $\beta\gamma$, in the
form
\begin{equation}
 \label{eq:sigma_Delta}
 \sigma_\Delta(y) = \sigma_0 + b y
\end{equation}

\noindent where $\sigma_0 \approx$ 2~keV, $b \approx 0.095$.
According to Fig.~\ref{fig:sllp_fit_2_exponent}-right the $y$ dependent
coefficient of Eq.~\eqref{eq:p_Delta} is very well approximated as
\begin{align}
 \label{eq:C}
 C(y) &\propto 1/\sigma_\Delta(y).
\end{align}

The parameters of the above detailed energy loss model and their short
explanations are listed in Table~\ref{tab:model}. Note that these four numbers
are the only parameters of the description.
Usually the detector and readout noise can be neglected. If this is not the
case the term $\sigma_\Delta^2$ in Eq.~\eqref{eq:p_Delta} and \eqref{eq:C}
should be replaced by $\sigma_D^2 + \sigma_n^2$ where $\sigma_n$ is the
standard deviation of the Gaussian noise.

\begin{table}[!b]
 \begin{center}

 \caption{Parameters of the energy loss model.}

 \label{tab:model}

 \begin{tabular}{ccl}
  \hline
  $\nu$      & 0.65  & Gaussian vs exponential [$\sigma_\Delta(y)$] \\
  $a$        & 0.07  & coefficient of log term in $\Delta(l)$ \\
  $\sigma_0$ & 2~keV & constant term of $\sigma_\Delta(y)$ \\
  $b$        & 0.095 & linear coefficient of $\sigma_\Delta(y)$ \\
  \hline
 \end{tabular}

 \end{center}
\end{table}

\subsection{Left truncation and right censoring}

\label{sec:trunc_censor}

During readout the deposited energy is converted to measured ADC values through
several steps: primary and secondary electron-hole pairs, current signals,
front-end electronics, digitization \cite{Brigida:2004ff}. In this study we
assume that the response of the detector system is linear in the
threshold-to-censoring region. Signals below threshold are truncated, since
they produce no output. Signals above a certain level are censored, hence only
the fact that the deposit was above that level is known. Note that the
censoring level does not necessarily coincides with the saturation level of the
readout electronics: the former is simply chosen as the limit of linearity.

If the measured value $y$ is below or above a limit $t$, the corresponding
values can be calculated by integration
\begin{align*}
 f_{y<t}(\Delta) &= \int_{-\infty}^t f_y(\Delta) \d y, &
 f_{y>t}(\Delta) &= \int_t^\infty    f_y(\Delta) \d y.
\end{align*}

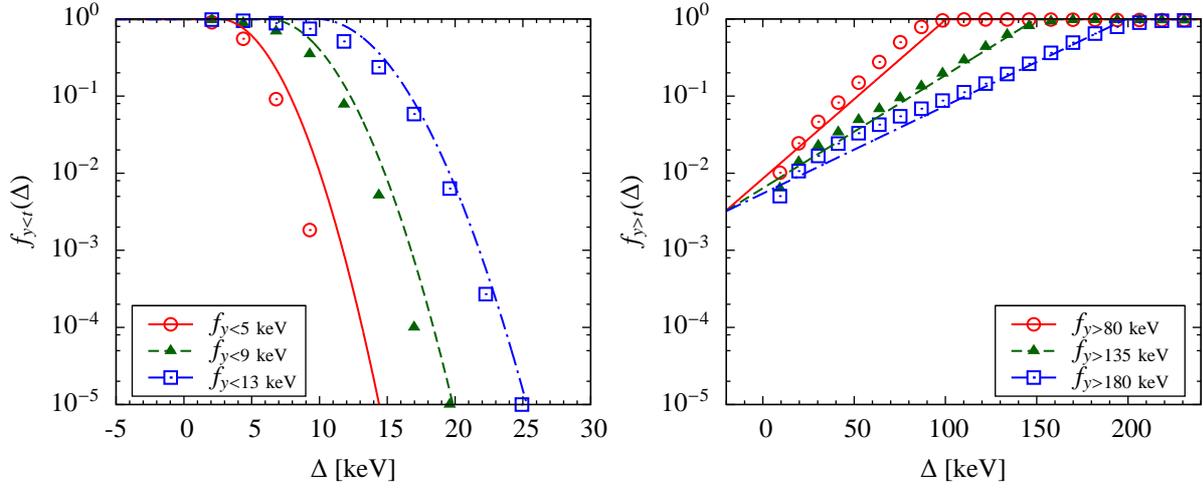
\begin{figure*}
 
 \begin{center}
  \input{threshold}
  \input{saturation}
 \end{center}

 \caption{Fits of the conditional probability $f_{y \lessgtr t}(\Delta)$ for
measurements below threshold (left) and above censoring level (right) as
function of $\Delta$, in case of several limiting values. Examples thresholds
are 5, 9 and 13~keV, while censoring levels are at 80, 135 and 180 keV. The
simulated values (points) are shown for particles with $\beta\gamma = 3.16$,
together with the fitted curves (lines, from Eqs.~\eqref{eq:p_truncation} and
\eqref{eq:p_censoring}).}

 \label{fig:limits}

\end{figure*}

\noindent Note that since $\sigma_\Delta$ is a function of $y$, the integrals
are difficult to perform. The simulated values with several truncation
thresholds and censoring levels are shown in Fig.~\ref{fig:limits}.
They are well approximated by the following functional forms:
\begin{align}
 \label{eq:p_truncation}
 f_{y<t}(\Delta) &=
  \begin{cases}
    1
    ,& \text{if $\Delta <   t - \sigma_\Delta$} \\
    \exp\left[-\frac{1}{2}
               \left(\frac{\Delta - t}{\sigma_\Delta} + 1\right)^2\right]
    ,& \text{if $\Delta \ge t - \sigma_\Delta$}
  \end{cases} \\
 \label{eq:p_censoring}
 f_{y>t}(\Delta) &=
  \begin{cases}
    \exp\left[\frac{1}{2} \left(\frac{\Delta-t}{\sigma_\Delta} - 1 \right)
       \right]
    ,& \;\;\;\; \text{if $\Delta <   t + \sigma_\Delta$} \\
    1
    ,& \;\;\;\; \text{if $\Delta \ge t + \sigma_\Delta$}.
  \end{cases}
\end{align}

\subsection{Log-likelihood minimization}

\label{sec:loglikelihood}

In summary, for a given measured deposit $y$ the probability density will be a
function of $\Delta$ only. The measure of goodness can be obtained with help of
the log-likelihood value as
\begin{equation*}
 \chi^2 = -2 \log f.
\end{equation*}

\noindent For a given channel, with measured ($y$) or limited deposit values
($t$), the corresponding $\chi^2$ can be derived from
Eqs.~\eqref{eq:p_Delta}-\eqref{eq:C} and
\eqref{eq:p_truncation}-\eqref{eq:p_censoring} and they are shown below:
\begin{align}
 \label{eq:chi2_a}
 \begin{split}
 \chi^2_y(\Delta) &= 2 \log \sigma_\Delta(y) + \\
  &+ \begin{cases}
   \frac{-2\nu(\Delta-y)}{\sigma_\Delta(y)} - \nu^2
    ,& \text{if $\Delta <   y - \nu\sigma_\Delta(y)$} \\
   \left(\frac{\Delta - y}{\sigma_\Delta(y)}\right)^2 
    ,& \text{if $\Delta \ge y - \nu\sigma_\Delta(y)$}
  \end{cases}
 \end{split} \\
 \label{eq:chi2_b}
 \chi^2_{y<t}(\Delta) &=
  \begin{cases}
    0
    ,& \;\, \text{if $\Delta <   t - \sigma_\Delta(t)$} \\
    \left(\frac{\Delta - t}{\sigma_\Delta(t)} + 1 \right)^2
    ,& \;\, \text{if $\Delta \ge t - \sigma_\Delta(t)$}
  \end{cases} \\
 \label{eq:chi2_c}
 \chi^2_{y>t}(\Delta) &=
  \begin{cases}
    -\frac{\Delta-t}{\sigma_\Delta(t)} + 1
    ,& \;\;\;\; \text{if $\Delta <   t + \sigma_\Delta(t)$} \\
    0
    ,& \;\;\;\; \text{if $\Delta \ge t + \sigma_\Delta(t)$}
  \end{cases}
\end{align}

\noindent where $\sigma_\Delta(y)$ is given by Eq.~\eqref{eq:sigma_Delta}.
The terms on the right side contain only linear or positive definite quadratic
functions of $\Delta$.
We can estimate $\varepsilon$ for single hits or for the whole particle
trajectory by minimizing the sum of corresponding chi-square values.

The proposed model has several applications, such as
 
\begin{itemize}
 \item for a given hit the estimation of position $\vec{P}$ and energy
deposit via optimizing $\varepsilon$ and the individual path lengths $l_i$ in
the various sensitive units, channels, of the detector (Sec.~\ref{sec:position}
and \ref{sec:energy}).
 \item for a given trajectory the estimation of most probable differential
energy loss via optimizing $\varepsilon$ (Sec.~\ref{sec:differential}).
 \item detector gain (cross-)calibration with tracks via the variation of the
chip by chip gain, hence by modifying the $y$ values
(Sec.~\ref{sec:gain}).
\end{itemize}

\subsection{Estimation of parameter errors}
\label{sec:fisher}

In case of normal approximations, the inverse of the observed, sample based,
Fisher information $\cal I$ can be used to estimate the
covariance of the fitted parameters $V$ \cite{Efron:1978}:
\begin{align*}
 {\cal I}_{jk} &= \frac{1}{2} \frac{\p^2 \sum_i \chi_i^2}{\p a_j \p a_k}, &
 V &\approx - {\cal I}^{-1}
\end{align*}

\noindent where $\sum_i \chi_i^2$ denotes the joint chi-square of hit channels
or trajectory hits.

\subsection{Generation of random deposits}
\label{sec:random}

If the expected shape of the distribution of the estimator has to be determined
using Monte Carlo simulation, the generation of deposits should be fast but precise.
The energy deposit distribution function
\begin{equation*}
 P(y|\Delta) \propto \frac{1}{\sigma_\Delta(y)} \cdot
  \begin{cases}
      \exp\left[- \frac{(\Delta - y)^2}{2\sigma_\Delta(y)^2} \right],
    & \text{if $y <   \Delta + \nu \sigma_\Delta(y)$} \\
    \exp\left[  \frac{\nu(\Delta - y)}{\sigma_\Delta(y)} +
                 \frac{\nu^2}{2}\right],
    & \text{if $y \ge \Delta + \nu \sigma_\Delta(y)$}
  \end{cases}
\end{equation*}

\noindent
does not have a closed antiderivative, hence a simple inverse
transform sampling is not possible.  Due to its long high energy tail the
acceptance-rejection method is also highly ineffective.

Let us suppose that we have generated a random deposit $y_0$ at a given
$\Delta_0$. The value $y_0$ can be simply transformed to get a
proper random deposit $y$ for $\Delta$ with a linear transformation as
\begin{equation*}
 \frac{y   - \Delta  }{\sigma_\Delta(\Delta)  } =
 \frac{y_0 - \Delta_0}{\sigma_\Delta(\Delta_0)}.
\end{equation*}

\noindent
The distribution of $y$ can be deduced from $P(y_0|\Delta_0)$ and it can be
shown that it equals with $P(y|\Delta)$. The key of the proof is the identity

\begin{equation*}
 \frac{\sigma_\Delta(\Delta_0)}{\sigma_\Delta(\Delta)}
 \frac{\sigma_\Delta(y)}       {\sigma_\Delta(y_0)} = 1.
\end{equation*}
 
\section{Estimation of hit position}
\label{sec:position}

During the passage of a charged particle through sensitive silicon volumes one
or more channels (pixels or strips) are hit. The charges created in the
adjacent channels are recorded (electrons or holes), they provide input for hit
cluster recognition. Here the task is to estimate the position $\vec{P}$ of the
hit, location of the trajectory in the central plane, using the measured
deposits.

Throughout this paper it is required that the local direction of the
trajectory, hence the vector of the projected passage $\vec{\lambda}$
(projected onto the surface layer by drift), is known (Fig.~\ref{fig:example}).
In other words only the re-estimation of cluster parameters is attempted.

\subsection{Standard estimation}

\label{sec:pos_standard}

The simplest way of hit position estimation (labeled as Weighted) is the weighted mean of
channel positions $\vec{p_i}$, where weights are the corresponding energy
deposits $y_i$:
\begin{align*}
 \vec{P}_\mathrm{Weighted} = \frac{\sum_i y_i \vec{p_i}}{\sum_i y_i}.
\end{align*}

Another widely used standard reconstruction technique (labeled as First-last) deals only with one
dimensional clusters \cite{Allkofer:2007ek,Swartz:2007zz}. This
treatment is natural for strip detectors. For pixels the cluster is projected
onto both directions ($x$ and $y$) and the two projections are analyzed
separately. Only the first and last, projected and summed, channels are used
for position determination, because that choice reduces the sensitivity to
fluctuations in energy deposition. If the energy would be lost steadily, the
estimated projected hit position would be
\begin{equation*}
 \vec{P}_\mathrm{First-last} =
  \frac{\vec{p_F} + \vec{p_L}}{2} +
  \frac{y_L - y_F}{2(y_L + y_F)} \vec{\lambda_{eff}}
\end{equation*}

\noindent where the $x$ and $y$ components of $\vec{\lambda_{eff}}$ are the sum
of the path lengths in the two corresponding edge channels. The above formula
can be further corrected for the drift direction if the silicon was placed in
magnetic field.

Due to the use of projections, relevant informations corresponding to the shape
of hit cluster and its deposits are lost. In this work a new method is
presented which takes each individual channel of the cluster into account.

\label{sec:pathLength}

\subsection{Estimation using the model}

The passage of a charged particle through the silicon layer is given by the
entry and exit points of its trajectory.  Even in magnetic field or for very
slow particles the passage is very well approximated by a straight line, the
entry and exit points being on different sides of the layer. The deposited
charge along the passage is projected onto the surface layer by drift, as
result of electric and magnetic fields. The local direction of the particle at
a silicon unit can be calculated using the parameters of the trajectory and the
vector of the projected passage $\vec{\lambda}$ is known.

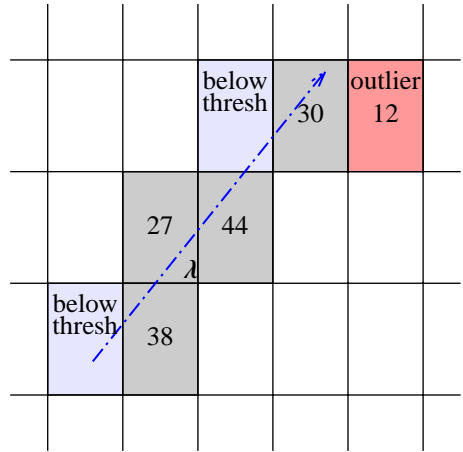
\begin{figure}[!h]

 \begin{center}
  \input{drawing2}
 \end{center}

 \caption{Example cluster with measured deposits (rounded numbers in keV) and
the the projected passage $\vec{\lambda}$ (arrow) at some stage of the
minimization. The pixels below threshold are indicated with a lighter fill
(below thresh.), while the outlier pixel has a darker fill than the others.}

 \label{fig:example}

\end{figure}

The following discussion deals with pixel silicon detectors, but it can be
easily applied to strips by taking one of the pixel dimensions to very large
values.
During clustering neighboring pixels (those with common edge or vertex) are
grouped to form a cluster, a reconstructed hit (Fig.~\ref{fig:example}).  For
the analysis of the corresponding cluster two types of pixels are of
importance:

\begin{itemize}

\item A {\it touched pixel} is touched by the projected passage, regardless of
its measured deposit. It can belong to the cluster (non-zero deposit) but it
may be an empty pixel as well, if the deposited charge was below the threshold.
The intersections of the projected passage and the edges of a pixel naturally
give the projected path length inside that pixel.

\item An {\it outlier pixel} is not crossed by the projected passage but it is
a member of the cluster (non-zero deposit). In other words this pixel is an
outlier, mostly left by a secondary particle, often $\delta$-electrons, created
during the passage of the charged particle in the silicon.
The projected passage does not have any section inside the pixel, but it is
possible to construct a measure of the pixel-passage distance, a sort of
negative path length (see Sec.~\ref{sec:ppl}). That notion is important in
order to properly take into account the effect of charge diffusion\footnote{In
case of 300 $\mu$m thickness its contribution is about 5~$\mu$m.} and to be
able to move the projected path towards the outlier. This extension also
ensures a smooth convergence of the minimization.

\end{itemize}

\subsection{Projected path length}

\label{sec:ppl}

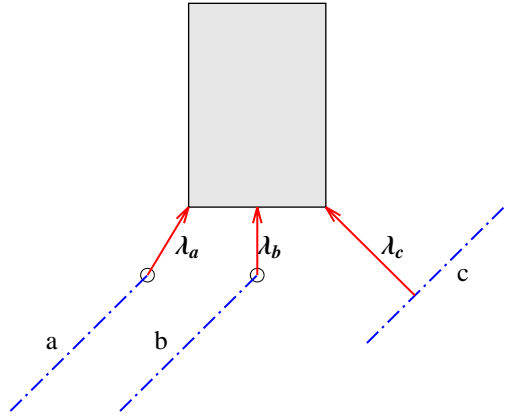
\begin{figure}

 \begin{center}
  \input{pathLength_drawing}
 \end{center}

 \caption{The three possible relative placements of projected paths (line
segments) and an outlier pixel (filled rectangle). The distance of the line
segment and the rectangle is shown by the corresponding arrows
($\vec{\lambda_{a,b,c}}$).}

 \label{fig:pathLength_drawing}

\end{figure}

The task is to determine the projected path lengths $\lambda_i$ for each pixel
$i$, for a given hit position $\vec{P}$. Using the coordinates of the two
endpoints, find points where the track crosses the horizontal (along $x$
direction) or vertical (along $y$ direction) pixel division lines. Determine
the dominant direction of the cluster by choosing longer projection of the
passage to the axes. Sort the endpoints and crossing points along their
coordinate in the dominant direction. The lengths of the resulting line
segments will give the projected path lengths $\lambda_i$ for touched pixels.
In case of outlier pixels determine the distance of the projected line segment
to the pixel (Fig.~\ref{fig:pathLength_drawing}) by evaluating the combinations

\begin{itemize}

 \item two endpoints of the projected path line segment with respect to the
four vertices and four edges of the pixel (cases a and b),

 \item projected path line segment with respect to the four vertices of the
pixel (case c),

\end{itemize}

\noindent and choose the {\it negative} of the smallest distance. The outlined
procedure ensures that by varying the hit position $\vec{P}$, the projected
path lengths of touched and outlier pixels will change continuously, without
jumps.

\begin{figure}
 
 \begin{center}
  \input{drawing}
 \end{center}

 \caption{Line segments (dash dotted lines) on the surface of a pixel (filled
rectangle). They can be bound by endpoints (open circles) or crossing points.
a: The length is insensitive to small change of the hit position. b: The length
changes when the hit is moved. c: Normal vectors $\vec{n_1}$, $\vec{n_2}$ and
the vector of the projected passage $\vec{\lambda}$.}

 \label{fig:pathDer}

\end{figure}
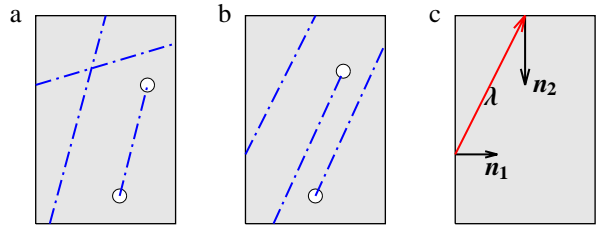

\subsection{Derivatives}

How does the projected path length $\lambda$ change if the hit position
$\vec{P}$ is varied? The corresponding line segment can be bound by endpoints
or crossing points. The length is insensitive to small changes of the hit
position if both points are endpoints or if both points are crossing points on
opposite sides of the pixel (Fig.~\ref{fig:pathDer} a). The length will change
only if one of the points is an endpoint or if both points are crossing points
on neighboring sides of the pixel (Fig.~\ref{fig:pathDer} b).  The derivative
$\p \lambda/\p \vec{P}$ can be computed with help of the inward directed normal
vectors $\vec{n_j}$ at crossing points and the vector of the projected passage
$\vec{\lambda}$ (Fig.~\ref{fig:pathDer} c):
\begin{equation}
 \label{eq:pderTouched}
 \frac{\p \lambda}{\p \vec{P}} =
  \sum\limits_j \frac{\vec{n_j} \lambda}{|\vec{n_j} \vec{\lambda}|}
\end{equation}

\noindent where the index $j$ runs for all (0, 1 or 2) crossing points
belonging to the line segment.
For outliers the derivative is the unit vector
\begin{equation}
 \label{eq:pderOutlier}
 \frac{\p \lambda}{\p \vec{P}} = \frac{\vec{\lambda}}{\lambda}.
\end{equation}

\subsection{Minimization}

The joint chi-square for all channels in a hit cluster can be written as
\begin{equation}
 \label{eq:joint}
 \chi^2(\varepsilon,\vec{P}) =
  \sum_i \chi^2_{y_i}\left(\Delta(\varepsilon,l_i(\vec{P}))\right)
\end{equation}

\noindent where the index $i$ runs for all the touched and untouched channels
of the hit. Note that while the deposits $y_i$ are given, the $\Delta$s depend
on $\varepsilon$ and the path lengths $l_i$ inside the channels
(Eq.~\eqref{eq:Delta_vs_l}) which in turn depend on the actual hit position
$\vec{P}$. The best hit position, and the most probable differential energy
loss will be estimated by minimizing $\chi^2$.
The (three dimensional) path lengths $l_i$ can be obtained
from the projected lengths $\lambda_i$ as
\begin{equation*}
 l_i = \frac{l}{\lambda} \lambda_i
\end{equation*}

\noindent where $l$ is the total path length and $\lambda$ is the total
projected length, both are fixed from the local trajectory direction.

\begin{figure*}[!h]

 \begin{center}
  \input{residual_perp}
  \input{residual_para}
 \end{center}

 \caption{Residuals of the reconstructed hit position in directions
perpendicular to (left) or parallel with (right) the projected trajectory using
the discussed methods.}

 \label{fig:resid_perppara}

\end{figure*}
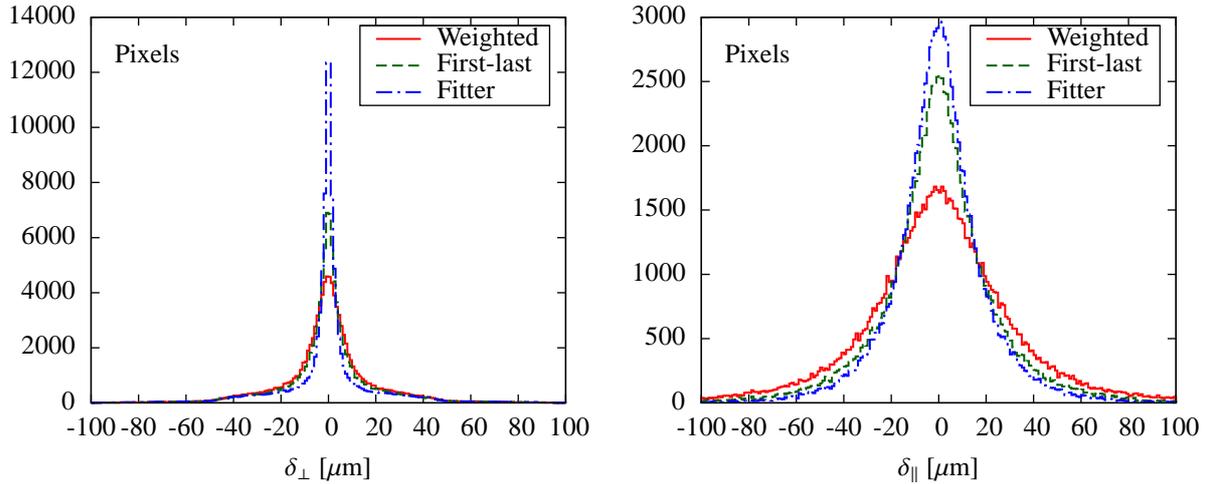

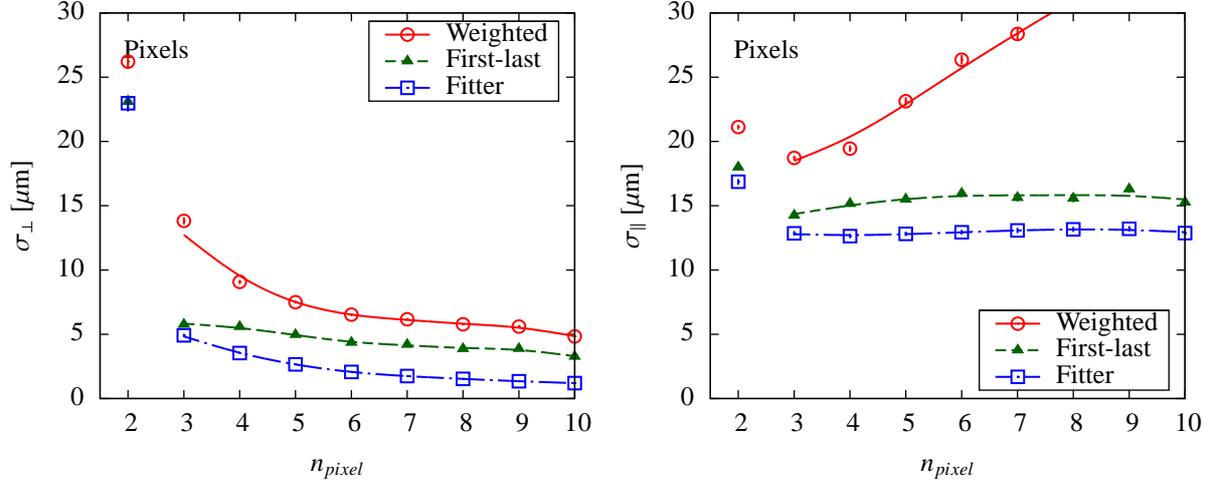
\begin{figure*}[!h]

 \begin{center}
  \input{residual_fit_perp}
  \input{residual_fit_para}
 \end{center}

 \caption{Resolution of the reconstructed hit position in directions
perpendicular to (left) or parallel with (right) the projected trajectory using
the discussed methods, as a function of the number of pixels in the cluster.
Lines are drawn to guide the eye.}

 \label{fig:resol_perppara}

\end{figure*}

In case of hit position estimation the value of $\varepsilon$ is not relevant,
hence instead of the complete relationship shown in Eq.~\eqref{eq:Delta_vs_l} a
simple linearity can be assumed
\begin{equation}
 \label{eq:Delta_vs_l_linear}
 \Delta(l) \approx \varepsilon l
\end{equation}

\noindent where $\varepsilon$ is rather the average rate of energy loss.
The first derivatives of the $\chi^2$ components with respect to
$(\varepsilon, \vec{P})$ are
\begin{align*}
 \frac{\p \chi^2}{\p \varepsilon} &= \frac{\p \chi^2}{\p \Delta} l, &
 \frac{\p \chi^2}{\p P_j}         &= \frac{\p \chi^2}{\p \Delta} 
                                      \varepsilon \frac{\p l}{\p P_j}
\end{align*}

\noindent while the second derivatives are
\begin{align*}
 \frac{\p^2 \chi^2}{\p \varepsilon^2}    &=
 \frac{\p^2 \chi^2}{\p \Delta^2} l^2, &
 \frac{\p^2 \chi^2}{\p P_j \p P_k}       &=
 \frac{\p^2 \chi^2}{\p \Delta^2} \varepsilon^2 \frac{\p l}{\p P_j}
                                                \frac{\p l}{\p P_k}
\end{align*}

\noindent where the term containing $\frac{\p^2 l}{\p \vec{P}^2} = 0$ is not
shown. The second derivative cross-term is
\begin{equation*}
 \frac{\p^2 \chi^2}{\p \varepsilon \p P_j} =
 \left(\frac{\p^2 \chi^2}{\p \Delta^2} \Delta +
       \frac{\p   \chi^2}{\p \Delta} \right) \frac{\p l}{\p P_j}. \\
\end{equation*}

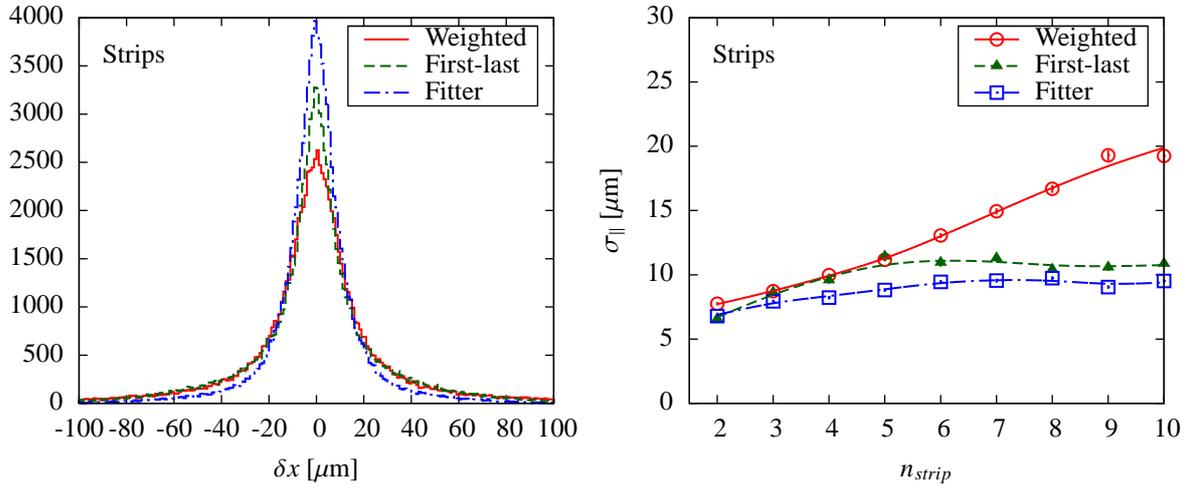
\begin{figure*}

 \begin{center}
  \input{residual_x_strips}
  \input{residual_fit_strips}
 \end{center}

 \caption{Residuals of the reconstructed hit position (left) and the resolution
of the reconstructed hit position using the discussed methods, as a function of
the number of strips in the cluster (right). Lines are drawn to guide the eye.}

 \label{fig:resid_resol_strips}

\end{figure*}

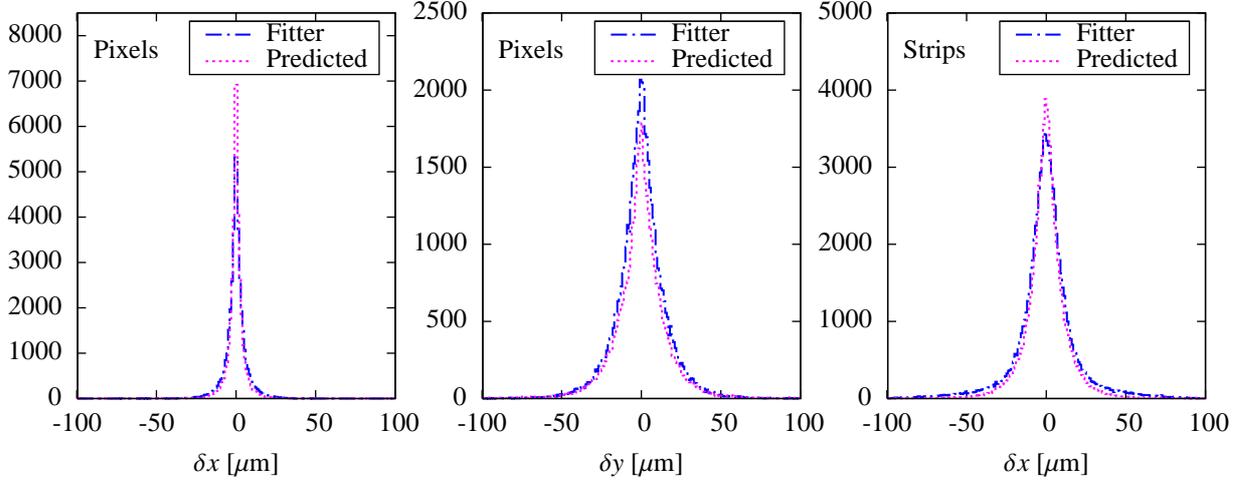
\begin{figure*}[!h]

 \begin{center}
   \input{resolution_x_pixels}
   \input{resolution_y_pixels}
   \input{resolution_x_strips}
 \end{center}

  \caption{Residuals of the reconstructed hit position for pixels ($x$ and $y$
directions) and strips, from left to right. The measured values from comparison
of reconstructed and simulated positions (Fitter, dash-dotted line) are shown
together with the distribution predicted from the errors of the fitted
parameters (Predicted, dashed line).}

 \label{fig:predicted_resol}

\end{figure*}

\noindent The partial derivative of the path length is
\begin{equation*}
 \frac{\p l}{\p \vec{P}} = \frac{l}{\lambda} \frac{\p \lambda}{\p \vec{P}}
\end{equation*}

\noindent and the last term is given in Eqs.~\eqref{eq:pderTouched} and
\eqref{eq:pderOutlier}. During minimization the positivity of $\varepsilon$ can
be assured by choosing $\log\varepsilon$ as the free parameter.  Since both the
first and second derivatives can be calculated, Newton's method
\cite{Press:1058313} can be used for fast and precise minimization.

In case of strips the observed signals are correlated. Before starting with the
position estimation the coupling has to be undone first: it will be discussed in
Sec.~\ref{sec:coupled}.

\begin{table}
 \begin{center}

 \caption{Some relevant quantities used in silicon detector simulation.}

 \label{tab:sim}

 \begin{tabular}{lc}
 \hline
 Detector thickness            & 300~$\mu$m \\
 Pitch, pixels ($x$ direction) & 100~$\mu$m \\
 Pitch, pixels ($y$ direction) & 200~$\mu$m \\
 Pitch, strips                 & 100~$\mu$m \\
 \hline
 Signal coupling, strips       & 0.1 \\
 \hline
 Channel noise      & 1.5~keV \\
 Channel threshold  & 7.5~keV \\
 Channel saturation & 150~keV \\
 \hline
 \end{tabular}

 \end{center}
\end{table}

\subsection{Results}

\label{sec:sim_hits}

For the studies in Sec.~\ref{sec:position} and \ref{sec:energy} charged
particles in the $\beta\gamma$ range of 0.56 -- 10.0 were generated with flat
distribution in $\log(\beta\gamma)$, and in pseudo-rapidity, $-2.5 < \eta <
2.5$, following a $p_T \exp(-p_T/T)$ shape for the transverse momentum
distribution, where $T = 0.15~\mathrm{GeV}/c$ was set.
Some relevant quantities used in the simulation are given in
Table~\ref{tab:sim}. The pixel and strip layers were located in a barrel
geometry at 10~cm and 50~cm radial distances from the beam axis, respectively,
in a $B = $ 4~T magnetic field. 100~000 complete pixel and strip hits were
generated, down to the level of individual channels. For simplicity, the charge
drift direction was assumed to be parallel with the electric field,
perpendicular to the silicon unit.  Hence the change in direction due to the
$\vec{E} \times \vec{B}$ effect was omitted. 

The addition of local particle direction and energy loss information will make
the residuals of the hit position measurement smaller. In case of pixels the
deviations from the true value can be decomposed into projections parallel and
perpendicular to the passage of the particle (Fig.~\ref{fig:resid_perppara}).
In both cases the proposed model (labeled as Fitter) gives better results than
the weighted or first-last methods introduced in Sec.~\ref{sec:pos_standard}.
The differences are mostly seen in the perpendicular direction. The measured
Gaussian resolutions as a function of the number of pixels in the cluster for
all three discussed methods are shown in Fig.~\ref{fig:resol_perppara}, in some
cases reaching 1~$\mu$m levels.
The corresponding plots of residuals and position resolution for strips are
shown in Fig.~\ref{fig:resid_resol_strips}, again the fitter giving the best
results.

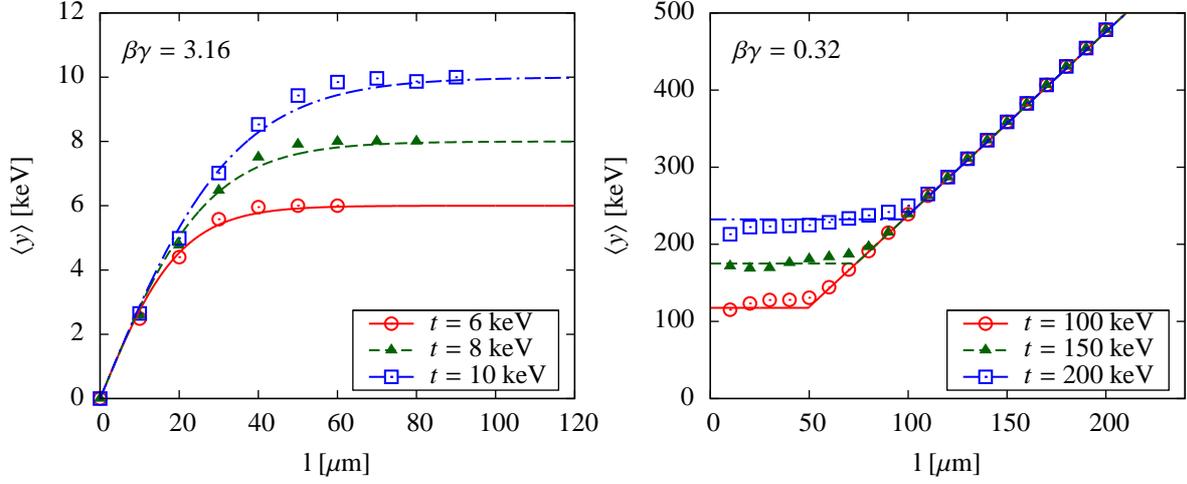
\begin{figure*}[!h]

 \begin{center}
  \input{corrected_3}
  \input{corrected_high_0}
 \end{center}

 \caption{The average lost deposit $\langle y \rangle$ as function of the path
length $l$ for several threshold. Left: deposits below threshold with $t =$ 6,
8 and 10~keV, for particles with $\beta\gamma = 3.16$. Right: deposits above
censoring level with $t = $ 100, 150 and 200~keV, for particles with
$\beta\gamma = 0.32$. The curves show the functional forms of
Eq.~\eqref{eq:corrected_below} and Eq.~\eqref{eq:corrected_above},
respectively.}

 \label{fig:corrected}

\end{figure*}

The resolution of the position estimate can be calculated hit by hit using the
observed Fisher information (Sec.~\ref{sec:fisher}).
Residuals of the reconstructed hit position for pixels and strips are shown in
Fig.~\ref{fig:predicted_resol}. The measured values from comparison of
reconstructed and simulated positions are plotted together with the
distribution {\it predicted} from the errors of the fitted parameters. There is
a very good agreement between observed and predicted values.
Note that for pixels the cross-correlation term can also be deduced giving
valuable input for track refit.

\section{Estimation of hit energy deposit}
\label{sec:energy}

The deposited energy $Q$ in the cluster could be plainly estimated with the sum
of the individual channel deposits $y_i$:
\begin{equation*}
 Q_\mathrm{Sum} = \sum_i y_i
\end{equation*}

\noindent but this approach clearly has various biases
(Sec.~\ref{sec:trunc_censor}).

The deposited energy can be lost during clustering, because the deposit can be
below the channel threshold (left truncation). The average lost deposit
$\langle y \rangle$ can be estimated using the discussed microscopical model.
Example values as function of path length $l$ for several threshold settings
($t =$ 6, 8 and 10~keV) are shown in Fig.~\ref{fig:corrected}-left, for
particles with $\beta\gamma = 3.16$. In case of large $l$ the average lost
deposit is close to the threshold, $\langle y \rangle \approx t$, because the
probability density function $p(y|l)$ is very steep for low $y$ values. If the
path length $l$ is small the average lost deposit will be simply $\langle y
\rangle \approx \varepsilon l$. These observations can be successfully
described and matched with simulated data (Fig.~\ref{fig:corrected}-left) using
the following formula:
\begin{equation}
 \langle y \rangle_{y < t} \approx t \tanh\left(\frac{\varepsilon l}{t}\right).
 \label{eq:corrected_below}
\end{equation}

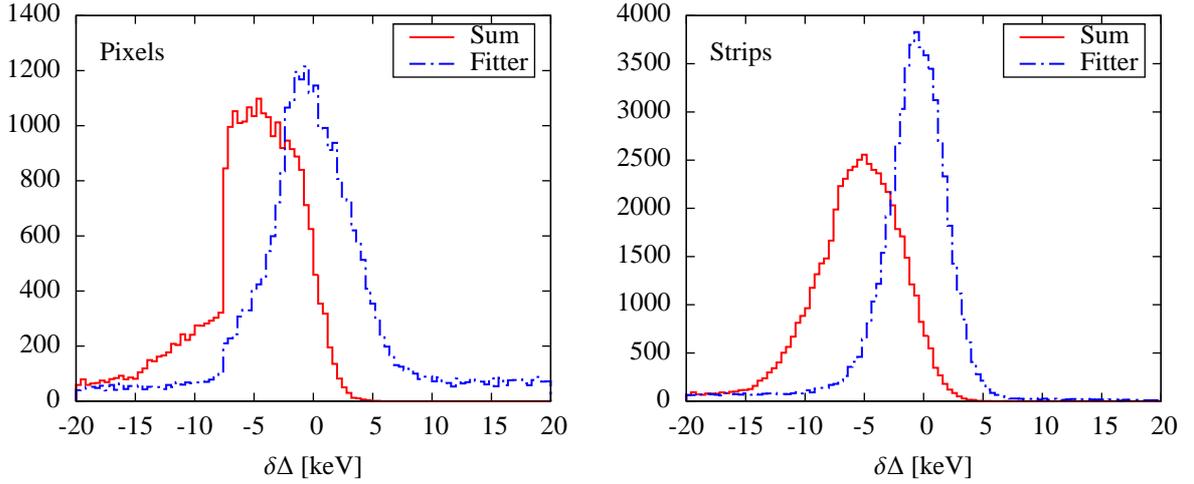
\begin{figure*}[!t]

 \begin{center}
  \input{residual_charge_pixels}
  \input{residual_charge_strips}
 \end{center}

 \caption{Residuals of the reconstructed deposited energy of the hit cluster
for pixels (left) and strips (right).  The result of the simple sum and the
proposed fitter are compared.}

 \label{fig:resid_charge}

\end{figure*}

The measured value can also be saturated, when the deposit is above censoring
level (right censoring). Example values as function of path length $l$ for
several threshold settings ($t =$ 100, 150 and 200~keV) are shown in
Fig.~\ref{fig:corrected}-right, for particles with $\beta\gamma = 0.32$. In
case of small $l$ the average lost deposit can be estimated as
\begin{equation*}
 \langle y \rangle \approx
   \frac{\int_t          y  \exp\left[-\nu y/\sigma_\Delta(t)\right]}
        {\int_t \phantom{y} \exp\left[-\nu y/\sigma_\Delta(t)\right]} =
   t + \frac{\sigma_\Delta(t)}{\nu}.
\end{equation*}

\noindent If the path length $l$ is large the average
lost deposit will be simply $\langle y \rangle \approx \varepsilon l$. These
observations can be successfully described and matched with simulated data
(Fig.~\ref{fig:corrected}-right) using the following formula:
\begin{equation}
 \langle y \rangle_{y > t} \approx
  \max\left(t + \frac{\sigma_\Delta(t)}{\nu}, \varepsilon l\right).
 \label{eq:corrected_above}
\end{equation}

\subsection{Coupled signals}

\label{sec:coupled}

In case of strips the observed signals are correlated due to the capacitive
coupling of the neighboring strips and cross-talk. This effect can be modelled
by a tridiagonal matrix:
\begin{gather*}
 C = 
 \begin{pmatrix}
 1-2\alpha & \alpha    &        & 0 \\
 \alpha    & 1-2\alpha & \ddots & \\
           & \ddots    & \ddots & \alpha \\
 0         &           & \alpha & 1-2\alpha 
 \end{pmatrix}
\intertext{where $\alpha > 0$, such that the measured deposits $y'$ can be obtained from the original ones $y$ by}
 \vec{y'} = C \vec{y}.
\end{gather*}

Since the signals are coupled, $\sum_i \chi_i^2$ would sum dependent values
making the minimization false. The coupling should be first undone by applying
$C^{-1}$ to the measured deposits. The inverse of $C$ can be exactly calculated
\cite{Lewis:1982}, furthermore if $\alpha \ll 1$ it is well approximated by
\begin{equation*}
 {C^{-1}}_{ij} = - \frac{(2 - 1/\alpha)^{|i-j|-1}}{\alpha}.
\end{equation*}

Note that due to the presence of signals below threshold and saturated values
the decoupling is not exact. The measured deposit in the $i$th channel is
either given, or subject to inequalities: 
\begin{align*}
 (C \vec{y})_i = y_i' && \text{or} &&
 (C \vec{y})_i < t    && \text{or} &&
 (C \vec{y})_i > t.
\end{align*}

In that sense a system of equalities and inequalities has to be solved for
$\vec{y}$.
The problem can be easily handled with the tools of linear programming
\cite{Press:1058313,140356}, the OOQP package \cite{Gertz:2003}, available in
ROOT \cite{Brun:1997pa} under MATH/QUADP, could be used for this purpose. The
solution is not unique, hence the minimization of a scalar product $\vec{c}
\vec{y'}$ is also required. In this case it is advantageous to set $\vec{c}
=$ (1, \dots, 1). This way those solution will be selected where $\sum_i y_i
\equiv Q_\mathrm{Fitter}$ is minimal.

In fact the situation is even more complicated due to the presence of detector
and readout noise. Assuming a Gaussian noise with standard deviation $\sigma$,
the $\chi^2$ contribution of the $i$th channel is
\begin{align*}
 \chi^2_{y'}   &= \left(\frac{y_i' - (Cy)_i}{\sigma}\right)^2 \\
 \chi^2_{y'<t} &\approx
  \begin{cases}
   \frac{1}{2} \left(\frac{t - (Cy)_i}{\sigma} - 2\right)^2, &
   \text{if $\frac{t - (Cy)_i}{\sigma} - 2 < 0$} \\
   0, & \text{otherwise}
  \end{cases} \\
 \chi^2_{y'>t} &\approx
  \begin{cases}
   \frac{1}{2} \left(\frac{t - (Cy)_i}{\sigma} + 2\right)^2, &
   \text{if $\frac{t - (Cy)_i}{\sigma} + 2 > 0$} \\
   0, & \text{otherwise}.
  \end{cases}
\end{align*}

\noindent
Here, in case of the inequalities, the exact values
\begin{gather*}
 P(y < t | m) =
  \frac{1}{2} \left[1 + \erf \left(\frac{t - m}{\sigma\sqrt{2}} \right)\right]
\\
\intertext{were approximated as}
 \chi^2 = -2 \log P \approx
  \begin{cases}
   \frac{1}{2} \left(\frac{t - m}{\sigma} - 2\right)^2, &
   \text{if $\frac{t - m}{\sigma} - 2 < 0$} \\
   0, & \text{otherwise}.
  \end{cases}
\end{gather*}

In order to keep the minimization away from unphysical negative $y_i$ values,
a penalty $(C \vec{y})_i/\sigma)^2$ is added if $(C \vec{y})_i < 0$.

The joint chi-square $\sum \chi^2_{y_i'}$ is a positive definite quadratic
function of the original deposits $y_i$. The first and second derivatives with
respect to $y_i$ components are easily calculable and the function can be
minimized with Newton's method, in few iterative steps.

If some $y_i$ values are negative at the minimum, the one with lowest value is
fixed to 0, the minimization is redone. In the end we have a minimum joint
chi-square with all the $y_i$ values being non-negative.
At the minimum the total charge and its variance are given as
\begin{align*}
 Q          &= \sum_i y_i, & 
 \sigma_Q^2 &= \sum_{i,j} \mathrm{Var}(y_i,y_j),
\end{align*}

\noindent respectively.

\subsection{Results}

The details of the simulation are the same as they were described in
Sec.~\ref{sec:sim_hits}.
Residuals of the reconstructed deposited energy of the hit cluster are shown in
Fig.~\ref{fig:resid_charge}, for pixels and strips. The result of the simple
sum and the proposed fitter are compared. The sum has a $-5$~keV shift for both
configurations which is a 5\% effect for minimum ionizing particles along a
300~$\mu$m path in silicon. At the same time the fitter has practically no bias
and a better resolution.

\section{Estimation of the most probable differential energy loss for tracks}
\label{sec:differential}

Having the proper hit energy deposits $y_i$, the next step is to estimate the
most probable differential energy loss for the whole trajectory. The task is very similar
(Eq.~\eqref{eq:joint}), but only $\varepsilon$ has to be optimized, because the
path lengths $l_i$ are given from track finding,
\begin{equation*}
 \chi^2(\varepsilon) = \sum_i \chi^2_{y_i} \left(\Delta(\varepsilon,l_i)\right).
\end{equation*}

\noindent Since the terms on the right contain linear or positive definite
quadratic functions (Eqs.~\eqref{eq:chi2_a}-\eqref{eq:chi2_c}) a fast
convergence with Newton's method is guaranteed.

The derivatives are
\begin{align*}
 \frac{\p\chi^2_y(\Delta)}{\p\Delta} &= 
  \begin{cases}
   \frac{-2\nu}{\sigma_\Delta(y)}
    ,& \hspace{0.25in} \text{if $\Delta <   y - \nu\sigma_\Delta(y)$} \\
   \frac{2(\Delta - y)}{\sigma_\Delta(y)^2} 
    ,& \hspace{0.25in} \text{if $\Delta \ge y - \nu\sigma_\Delta(y)$}
  \end{cases} \\
 \frac{\p\chi^2_{y<t}(\Delta)}{\p\Delta} &=
  \begin{cases}
    0
    ,& \text{if $\Delta <   t - \sigma_\Delta(t)$} \\
     \frac{2(\Delta - t + \sigma_\Delta(t))}{\sigma_\Delta(t)^2}
    ,& \text{if $\Delta \ge t - \sigma_\Delta(t)$}
  \end{cases} \\
 \frac{\chi^2_{y>t}(\Delta)}{\p\Delta} &=
  \begin{cases}
    -\frac{1}{\sigma_\Delta(t)}
    ,& \hspace{0.25in} \text{if $\Delta <   t + \sigma_\Delta(t)$} \\
    0
    ,& \hspace{0.25in} \text{if $\Delta \ge t + \sigma_\Delta(t)$}.
  \end{cases}
\end{align*}

Only the following second derivatives are non-zero:
\begin{align*}
 \frac{\p^2\chi^2_y(\Delta)}{\p\Delta^2} &= 
   \frac{2}{\sigma_\Delta(y)^2} 
    ,& \text{if $\Delta \ge y - \nu\sigma_\Delta(y)$} \\
 \frac{\p^2\chi^2_{y<t}(\Delta)}{\p\Delta^2} &=
    \frac{2}{\sigma_\Delta(t)^2}
    ,& \text{if $\Delta \ge t - \sigma_\Delta(t)$}.
\end{align*}

\begin{figure}
 
 \begin{center}
  \input{fisher}
 \end{center}

 \caption{The expectation value of $\sigma_\Delta^{-2}$ as function of path
length $l$ at several $\beta\gamma$ values (points). The curves show the
functional form defined in Eq.~\eqref{eq:fisher}. The corresponding fitted
powers for $\varepsilon$ and $l$ are also
indicated.}

 \label{fig:fisher}

\end{figure}
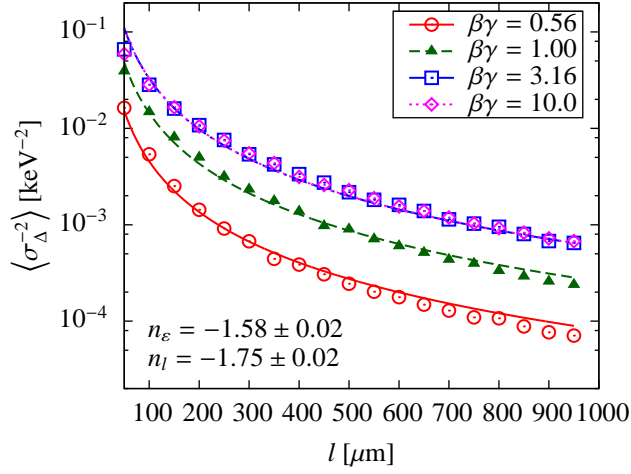

\subsection{Resolution}
\label{sec:resolution}

The dependence of the standard deviation of the $\varepsilon$ estimate can be
obtained as (Sec.~\ref{sec:fisher}):
\begin{gather*}
  {\cal I}(\varepsilon) =
    \frac{1}{2} \sum_i E \left[ \frac{\partial^2 \chi_i^2}
                                     {\partial \varepsilon^2} \right] \\
\intertext{where the second derivative is}
  \frac{1}{2} \frac{\partial^2 \chi_i^2}{\partial \varepsilon^2} = 
   \begin{cases}
   \left(\frac{l_i [1 + a\log(l_i/l_0)]}{\sigma_\Delta(y)}\right)^2,
      & \text{if in the Gaussian part}\\
   0, & \text{if in the exponential part.}\\
   \end{cases}
\end{gather*}

\noindent According to the fit shown in Fig.~\ref{fig:fisher} the
expectation value of $\sigma_\Delta^{-2}$ is a power function of
$\varepsilon$ and $l$:
\begin{equation}
 \label{eq:fisher}
 \left\langle \sigma_\Delta^{-2} \right\rangle
  \propto \varepsilon^{-1.6} l^{-1.8}.
\end{equation}

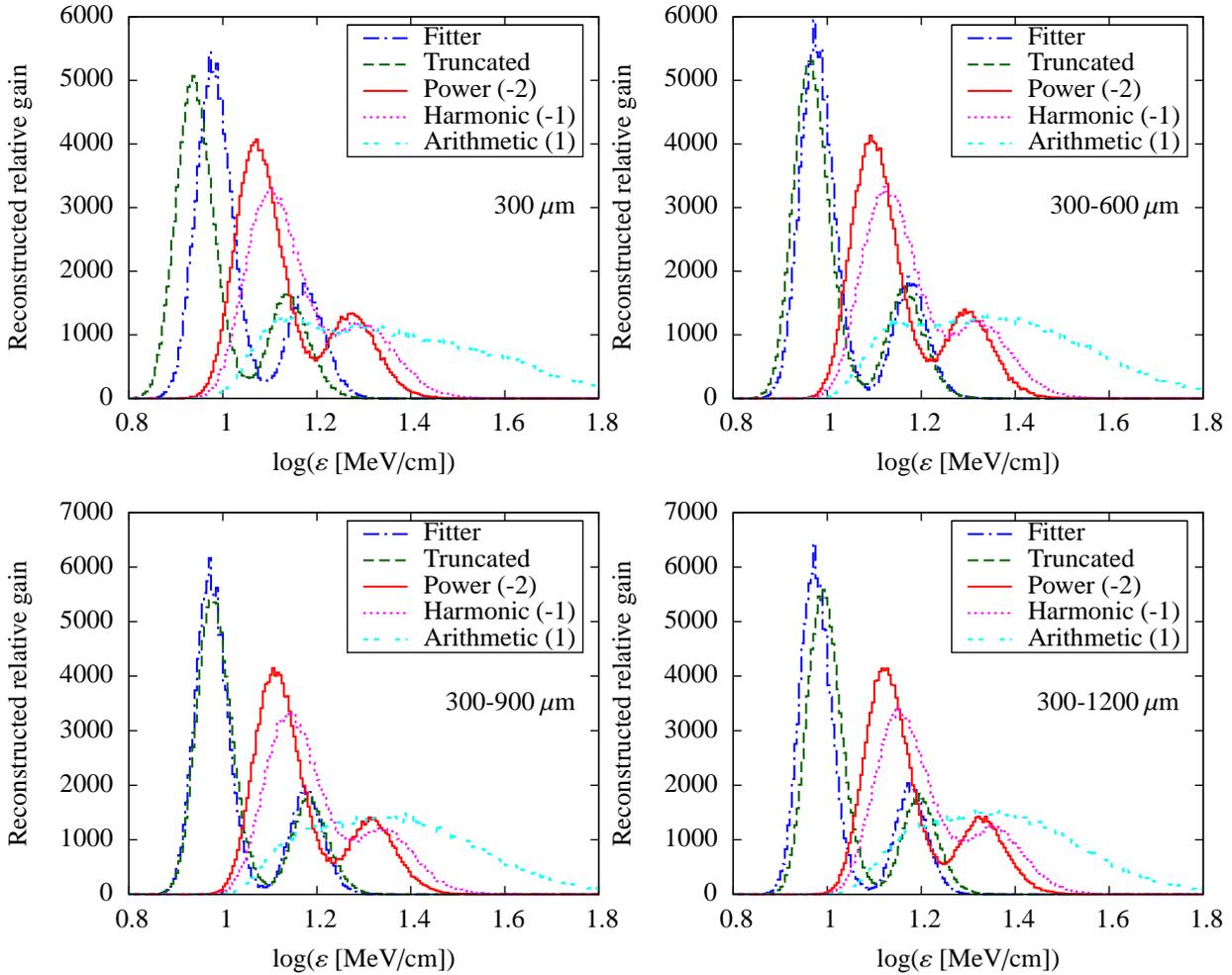
\begin{figure*}[!t]

 \begin{center}
  \input{estimates_0}
  \input{estimates_1}
 \end{center}
 
 \vspace{-0.3in}
 
 \begin{center}
  \input{estimates_2}
  \input{estimates_3}
 \end{center}

 \caption{Distribution of the estimated most probable differential energy loss
values at reference length ($\varepsilon$), obtained with the four detector
layer settings. Several methods are shown: this maximum likelihood fitter, truncated mean,
power mean (power -2), harmonic mean (-1) and arithmetic mean (1).}

 \label{fig:estimates}

\end{figure*}

\noindent
With that, also approximating the factor $a \log(l_i/l_0)$ in the range $l =$ 50
 -- 1000~$\mu$m,
\begin{equation*}
 \sigma^2(\varepsilon)
  \propto \frac{\varepsilon^{1.6}}{\sum_i l_i^{0.2}}.
\end{equation*}

\label{sec:ln_epsilon}

\noindent It is clear that the relative resolution
$\sigma(\varepsilon)/\varepsilon$ only slightly depends on $\varepsilon$, hence
$\log\varepsilon$ is a convenient and uniform estimator. Since the exponent of
the path length is also small (0.2), it is not the total thickness of the
silicon, but the {\it number of independent measurements} that matters. For a
given total path length the relative resolution is proportional to $n^{-0.4}$,
where $n$ is the number of independent measurements. The reason for that
originates in the non-Gaussian nature of the energy deposit distribution.
Grouped measurements would have worse resolution unlike in the Gaussian case
where the resolution would stay unchanged.

\subsection{False hit removal}

Since the association of hits to trajectories is not always unambiguous some
hits do not belong to the proper track. Although their measured deposit is
correct, the calculated path length can be false. Assuming that there is at
most one false hit in a trajectory it can be detected and removed: the
exclusion of the outlier hit decreases the joint chi-square of the trajectory
by a considerable amount.

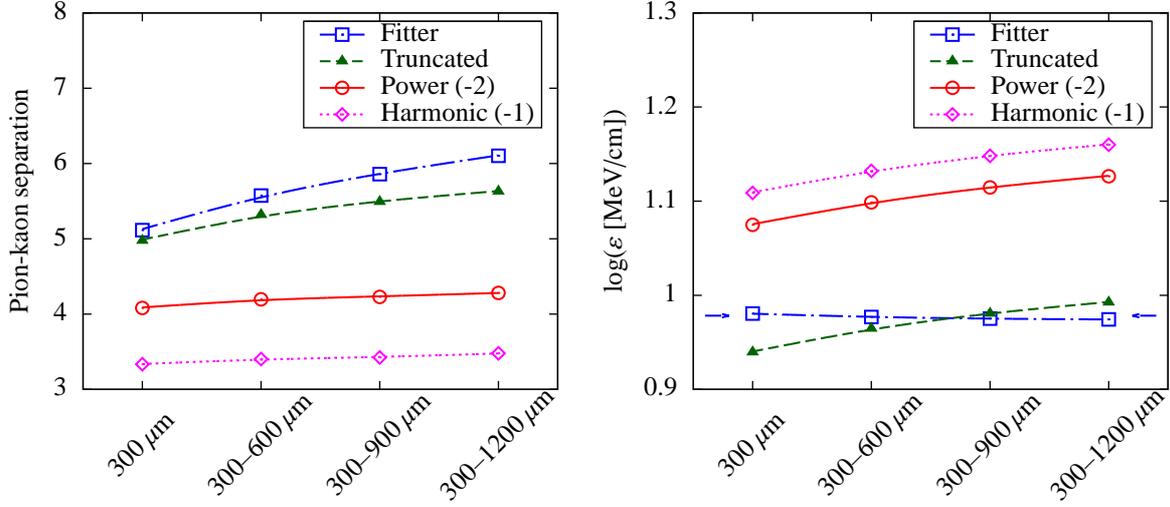
\begin{figure*}

 \begin{center}
  \input{separation}
  \input{dependence}
 \end{center}

 \caption{Left: pion-kaon separation power for four detector layer settings,
using several methods: this maximum likelihood fitter, truncated mean, power
mean (power -2), and harmonic mean (-1). Right: most probable differential
energy loss values at reference length ($\log\varepsilon$), shown for the four
detector layer settings. The horizontal arrows show the expected theoretical
value. Lines are drawn to guide the eye.}

 \label{fig:comparison}

\end{figure*}

\subsection{Distribution of the estimator}

Although the most probable value of $\varepsilon$ is estimated, it is equally
important to deduce, or also estimate, its distribution. Although the variance
of the estimate is given, the original probability density is a skewed
function: the estimator is not expected to follow a perfect Gaussian
distribution. The distribution of the joint $\chi^2$ could be determined track
by track with detailed Monte Carlo simulation, but this possibility would 
have huge computational demands and timing issues.

Are there other methods that could be used to estimate the distribution of the
estimator? One of them is the {\it statistical bootstrap}
\cite{jun1995jackknife}, that is a random
sampling with replacement from the original dataset. The so called {\it
jackknife} method \cite{jun1995jackknife} can also be considered where the estimate is systematically
recomputed leaving out one observation at a time from the sample. Both methods
can be used to estimate bias, variance and the shape of the distributions. It
can be shown that the jackknife method cancels biases with terms proportional
to $1/n$ and $1/n^2$.
A problem common to both resampling methods is that they do not work if the
sample size is small. Since here we deal sometimes with tracks with as few as
two or three measurements these approaches are not viable. For example, in case
of the jackknife method, the variance of the subsamples will have large errors
if the subsample consist of only one or two measurements.

As a solution, the shape of the $\log\varepsilon$ distribution could be
determined by energy deposit {\it regeneration}. A measured track is used to
construct the shape distribution of a given particle species (mass $m$), if its
measured $\varepsilon$ value is compatible with the corresponding value at a
given $p/m$.
While all kinematical parameters and path lengths of hits in silicon are kept,
all energy deposit values are randomly regenerated using assumed $p/m$ values,
as described in Sec.~\ref{sec:random}.
The chosen procedure ensures meaningful shape determination even for tracks
with very few hits and exploits the success of the energy loss model seen at
the hit level.

\subsection{Results}

\label{sec:sim_tracks}

In order to demonstrate track level applications a very simple detector model
was used: 16 layers of silicon at normal incidence. The orientation of layers
and the bending of particle trajectories are taken into account by defining
four different layer settings:

\begin{itemize}

 \item all layers are 300~$\mu$m thick (labeled as 300~$\mu$m);

 \item four groups of layers, each containing four 300, 400, 500 and 600~$\mu$m
thick layers (300-600~$\mu$m);

 \item four groups of layers, each containing four 300, 500, 700 and 900~$\mu$m
thick layers (300-900~$\mu$m);

 \item four groups of layers, each containing four 300, 600, 900 and 1200~$\mu$m
thick layers (300-1200~$\mu$m).

\end{itemize}

For the study the 100~000 pions and 30~000 kaons were generated at total
momentum $p = 0.8~\mathrm{GeV}/c$, and a Gaussian noise with 1~keV standard
deviation was added to each energy deposit.
 
Distribution of the estimated most probable differential energy loss values at
reference length ($\log\varepsilon$), obtained with the four detector layer
settings, are shown in Fig.~\ref{fig:estimates} (Fitter). Results of several
other methods working with the differential energy deposit values $y_i/l_i$ are
also plotted: truncated mean (average of the lowest half of the $y_i/l_i$
numbers, 0-50\% truncation) \cite{Fernow:1986,Grupen:2008}; as well as those
using all measurements, such as power mean (power -2), harmonic mean (power -1)
and arithmetic mean (power 1). While the maximum likelihood fitter gives the
best results in all the cases, power means have much worse resolution. The
truncated mean performs surprisingly well.

The resulted distributions can be fitted with sum of Gaussians.  The
performance can be further examined by plotting the pion-kaon separation power,
defined as $|m_1 - m_2| / [(\sigma_1^2 + \sigma_2^2)/2]^{1/2}$, where $m_i$ and
$\sigma_i$ are the means and standard deviations for pions and kaons,
respectively.  Comparisons including estimators listed above are shown in
Fig.~\ref{fig:comparison}-left, for the four detector layers settings.  Again
the maximum likelihood fitter gives the best result over the others. The
separation power increases if longer hit path lengths or thicker detectors are
used (compare with observations in Sec.~\ref{sec:resolution}). 

Ideally the mean of the $\log\varepsilon$ estimates should not depend on path
lengths and detector details. The dependence of the mean for the four layers
settings is given in Fig.~\ref{fig:comparison}-right. The horizontal arrows
show the expected theoretical value. It is clear that the maximum likelihood
fitter provides stable means, while the others show a pronounced increase.
Although for these latter the dependencies could be compensated, in case of
tracks with varying path length distribution only the proposed method would
perform appropriately.

\section{Detector gain calibration with tracks}
\label{sec:gain}

In order to determine the multiplicative gain correction $g$ for a
chip, terms such as
\begin{equation*}
 X_j^2(g) \equiv \chi^2_{g y_j}(\Delta) 
\end{equation*}

\noindent should be summed for collected hits and the sum minimized
by varying $g$. The partial derivatives of $\chi^2_y$ are
\begin{align*}
 \frac{\p\chi^2_{y}(\Delta)}{\p y} &= \;\; \frac{2b}{\sigma_\Delta(y)} +
  \begin{cases}
     \frac{2 \nu \sigma_\Delta(\Delta)}{\sigma_\Delta(y)^2}
    ,& \text{if $\Delta <   y - \nu\sigma_\Delta(y)$} \\
   - \frac{2(\Delta-y) \sigma_\Delta(\Delta)}{\sigma_\Delta(y)^3}
    ,& \text{if $\Delta \ge y - \nu\sigma_\Delta(y)$}
  \end{cases} \\
 \frac{\p^2\chi^2_{y}(\Delta)}{\p y^2} &= -\frac{2b^2}{\sigma_\Delta(y)^2} +
  \\ + & \begin{cases}
   - \frac{4 \nu b \sigma_\Delta(\Delta)}{\sigma_\Delta(y)^3}
    ,& \text{if $\Delta <   y - \nu\sigma_\Delta(y)$} \\
     \frac{2\left[\sigma_\Delta(y) + 3 b (\Delta-y) \right] \sigma_\Delta(\Delta)}{\sigma_\Delta(y)^4}
    ,& \text{if $\Delta \ge y - \nu\sigma_\Delta(y)$}.
  \end{cases}
\end{align*}

\noindent With that
\begin{align*}
 \frac{\p   X_j^2}{\p g  } &= \left(
  \frac{y  }{g  } \frac{\p  \chi^2}{\p y  }\right) \Bigg |_{y = g y_j} \\
 \frac{\p^2 X_j^2}{\p g^2} &= \left(
  \frac{y^2}{g^2} \frac{\p^2\chi^2}{\p y^2}\right) \Bigg |_{y = g y_j}
\end{align*}

\noindent where $y = g y_j$ should be substituted.

\subsection{Complete gain calibration}
\label{sec:cross}

The cross-calibration can be performed in the following steps.

\begin{enumerate}

 \item With help of a preliminary gain calibration estimate $\varepsilon$ for
each track. Select pion-like tracks and collect the values of $\beta\gamma$,
path length and deposit of each hit, and store them for every chip separately.
For each chip minimize the joint chi-square of all selected hits by varying the
gains. The minimization is best performed by a golden section search
\cite{MR0055639} first, in order to get near the region of the minimum,
followed by a refinement using Newton's method.

 \item Using the updated gains select only those tracks which are certainly
pions, kaons, protons or decay daughters of abundantly produced hadrons ($\PKzS
\rightarrow \Pgpp \Pgpm$, $\PgL \rightarrow \Pp \Pgpm$, $\PagL \rightarrow \Pap
\Pgpp$), or conversion products ($\Pgg \rightarrow \Pep \Pem$). Collect their
hits for every chip separately and minimize again the joint chi-square chip by
chip by varying the gains with similar methods as above.

 \item Finally, using all tracks, minimize the their joint chi-square
simultaneously.  This can be accomplished by nested minimizations. For a given
set of gains the $\varepsilon_i$ values are determined for each track
separately in the course of a local minimization.
The gains of chips can be highly correlated since tracks often traverse
detector units which are behind each other or very close in space (double-sided
units), but also due to manufacturing details. In case of $10^4$-$10^5$ chips
the covariance matrix needed for the minimization step is sparse, but huge. It
is usually numerically difficult to invert. In order to have a treatable
problem the detector units with highest correlation are identified, this way
forming a covariance matrix with a block diagonal structure. This latter is
already easily inverted and the Newtonian step can be computed.

\end{enumerate}

\subsection{Results}

\begin{figure}

 \begin{center}
  \input{gains_1}
 \end{center}

 \caption{Reconstructed relative gains as function of the real values, shown
for the 16 layers used in the detector simulation.}

 \label{fig:gainCalib}

\end{figure}
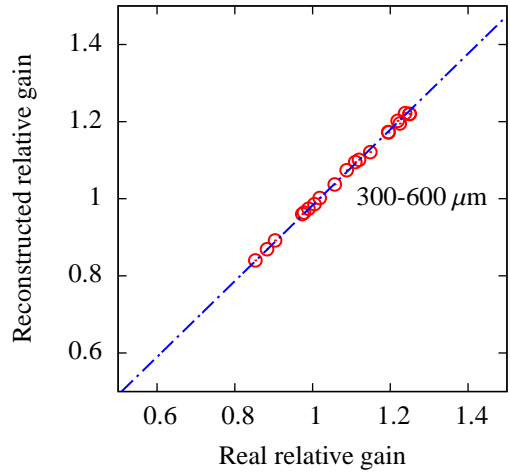

The details of the simulation are the same as they were described in
Sec.~\ref{sec:sim_tracks}, but now the initial gains of the layers were set
randomly in the range $0.8-1.2$. Only the first step in the list of
Sec.~\ref{sec:cross} was performed. The extracted relative gains as function of
the real values are shown in Fig.~\ref{fig:gainCalib}, for all the 16 layers
used. While the gains are steadily smaller by about 1\% (bias), they have an
excellent relative precision.

\section{Conclusions}

In this work a new analytical energy loss model for charged particles in
silicon was introduced. It has few parameters and it is based on the
recognition of a special connection between the distribution of the deposited
energy and the most probable energy loss.
Its use was demonstrated through several examples. With help of measured charge
deposits in pixels or strips of hit clusters their position and energy can be
estimated with better accuracy and much less bias. Deposits below threshold and
saturated values are treated properly, resulting in a wider dynamic range.
The model was successfully applied to track differential energy loss estimation
and to detector gain calibration tasks, again showing a performance superior to
standard methods.

\section*{Acknowledgements}
 
The author wishes to thank to S\'andor Hegyi, Andr\'as L\'aszl\'o and D\'aniel
Barna for helpful discussions. This work was supported by the Hungarian
Scientific Research Fund with the National Office for Research and Technology
(K 81614, H07-B 74296), and the Swiss National Science Foundation (128079).

\bibliographystyle{elsarticle-num}
\bibliography{newModel}

\end{document}

%% file: density_1.tex
\begingroup
  \makeatletter
  \providecommand\color[2][]{%
    \GenericError{(gnuplot) \space\space\space\@spaces}{%
      Package color not loaded in conjunction with
      terminal option `colourtext'%
    }{See the gnuplot documentation for explanation.%
    }{Either use 'blacktext' in gnuplot or load the package
      color.sty in LaTeX.}%
    \renewcommand\color[2][]{}%
  }%
  \providecommand\includegraphics[2][]{%
    \GenericError{(gnuplot) \space\space\space\@spaces}{%
      Package graphicx or graphics not loaded%
    }{See the gnuplot documentation for explanation.%
    }{The gnuplot epslatex terminal needs graphicx.sty or graphics.sty.}%
    \renewcommand\includegraphics[2][]{}%
  }%
  \providecommand\rotatebox[2]{#2}%
  \@ifundefined{ifGPcolor}{%
    \newif\ifGPcolor
    \GPcolortrue
  }{}%
  \@ifundefined{ifGPblacktext}{%
    \newif\ifGPblacktext
    \GPblacktexttrue
  }{}%
  \let\gplgaddtomacro\g@addto@macro
  \gdef\gplbacktext{}%
  \gdef\gplfronttext{}%
  \makeatother
  \ifGPblacktext
    \def\colorrgb#1{}%
    \def\colorgray#1{}%
  \else
    \ifGPcolor
      \def\colorrgb#1{\color[rgb]{#1}}%
      \def\colorgray#1{\color[gray]{#1}}%
      \expandafter\def\csname LTw\endcsname{\color{white}}%
      \expandafter\def\csname LTb\endcsname{\color{black}}%
      \expandafter\def\csname LTa\endcsname{\color{black}}%
      \expandafter\def\csname LT0\endcsname{\color[rgb]{1,0,0}}%
      \expandafter\def\csname LT1\endcsname{\color[rgb]{0,1,0}}%
      \expandafter\def\csname LT2\endcsname{\color[rgb]{0,0,1}}%
      \expandafter\def\csname LT3\endcsname{\color[rgb]{1,0,1}}%
      \expandafter\def\csname LT4\endcsname{\color[rgb]{0,1,1}}%
      \expandafter\def\csname LT5\endcsname{\color[rgb]{1,1,0}}%
      \expandafter\def\csname LT6\endcsname{\color[rgb]{0,0,0}}%
      \expandafter\def\csname LT7\endcsname{\color[rgb]{1,0.3,0}}%
      \expandafter\def\csname LT8\endcsname{\color[rgb]{0.5,0.5,0.5}}%
    \else
      \def\colorrgb#1{\color{black}}%
      \def\colorgray#1{\color[gray]{#1}}%
      \expandafter\def\csname LTw\endcsname{\color{white}}%
      \expandafter\def\csname LTb\endcsname{\color{black}}%
      \expandafter\def\csname LTa\endcsname{\color{black}}%
      \expandafter\def\csname LT0\endcsname{\color{black}}%
      \expandafter\def\csname LT1\endcsname{\color{black}}%
      \expandafter\def\csname LT2\endcsname{\color{black}}%
      \expandafter\def\csname LT3\endcsname{\color{black}}%
      \expandafter\def\csname LT4\endcsname{\color{black}}%
      \expandafter\def\csname LT5\endcsname{\color{black}}%
      \expandafter\def\csname LT6\endcsname{\color{black}}%
      \expandafter\def\csname LT7\endcsname{\color{black}}%
      \expandafter\def\csname LT8\endcsname{\color{black}}%
    \fi
  \fi
  \setlength{\unitlength}{0.0500bp}%
  \begin{picture}(4500.00,3780.00)%
    \gplgaddtomacro\gplbacktext{%
      \csname LTb\endcsname%
      \put(600,640){\makebox(0,0)[r]{\strut{}$10^{-5}$}}%
      \put(600,1220){\makebox(0,0)[r]{\strut{}$10^{-4}$}}%
      \put(600,1800){\makebox(0,0)[r]{\strut{}$10^{-3}$}}%
      \put(600,2379){\makebox(0,0)[r]{\strut{}$10^{-2}$}}%
      \put(600,2959){\makebox(0,0)[r]{\strut{}$10^{-1}$}}%
      \put(600,3539){\makebox(0,0)[r]{\strut{}$10^{0}$}}%
      \put(967,440){\makebox(0,0){\strut{} 0}}%
      \put(1790,440){\makebox(0,0){\strut{} 50}}%
      \put(2613,440){\makebox(0,0){\strut{} 100}}%
      \put(3436,440){\makebox(0,0){\strut{} 150}}%
      \put(4259,440){\makebox(0,0){\strut{} 200}}%
      \put(20,2089){\rotatebox{-270}{\makebox(0,0){\strut{}$p(y|l)$}}}%
      \put(2489,140){\makebox(0,0){\strut{}$y$ [keV]}}%
      \put(4082,2090){\makebox(0,0)[r]{\strut{}$\beta\gamma = 1.00$}}%
    }%
    \gplgaddtomacro\gplfronttext{%
      \csname LTb\endcsname%
      \put(3179,3376){\makebox(0,0)[l]{\strut{}$l = ~20$ $\mu$m}}%
      \csname LTb\endcsname%
      \put(3179,3176){\makebox(0,0)[l]{\strut{}$l = ~50$ $\mu$m}}%
      \csname LTb\endcsname%
      \put(3179,2976){\makebox(0,0)[l]{\strut{}$l = 100$ $\mu$m}}%
      \csname LTb\endcsname%
      \put(3179,2776){\makebox(0,0)[l]{\strut{}$l = 200$ $\mu$m}}%
    }%
    \gplbacktext
    \put(0,0){\includegraphics{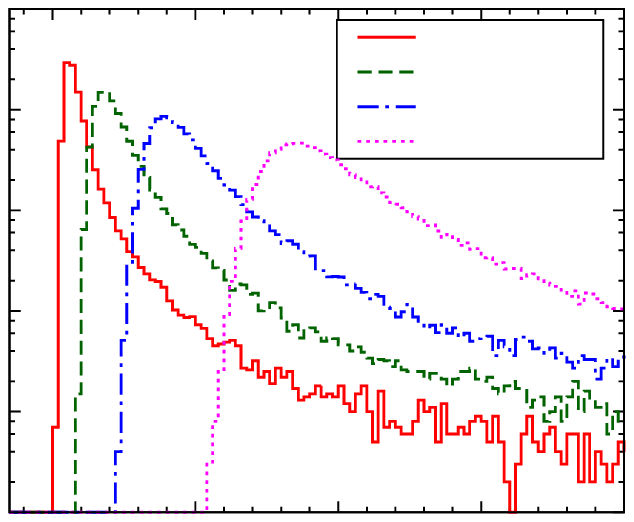}}%
    \gplfronttext
  \end{picture}%
\endgroup

%% file: bg_3.16_all.tex
\begingroup
  \makeatletter
  \providecommand\color[2][]{%
    \GenericError{(gnuplot) \space\space\space\@spaces}{%
      Package color not loaded in conjunction with
      terminal option `colourtext'%
    }{See the gnuplot documentation for explanation.%
    }{Either use 'blacktext' in gnuplot or load the package
      color.sty in LaTeX.}%
    \renewcommand\color[2][]{}%
  }%
  \providecommand\includegraphics[2][]{%
    \GenericError{(gnuplot) \space\space\space\@spaces}{%
      Package graphicx or graphics not loaded%
    }{See the gnuplot documentation for explanation.%
    }{The gnuplot epslatex terminal needs graphicx.sty or graphics.sty.}%
    \renewcommand\includegraphics[2][]{}%
  }%
  \providecommand\rotatebox[2]{#2}%
  \@ifundefined{ifGPcolor}{%
    \newif\ifGPcolor
    \GPcolortrue
  }{}%
  \@ifundefined{ifGPblacktext}{%
    \newif\ifGPblacktext
    \GPblacktexttrue
  }{}%
  \let\gplgaddtomacro\g@addto@macro
  \gdef\gplbacktext{}%
  \gdef\gplfronttext{}%
  \makeatother
  \ifGPblacktext
    \def\colorrgb#1{}%
    \def\colorgray#1{}%
  \else
    \ifGPcolor
      \def\colorrgb#1{\color[rgb]{#1}}%
      \def\colorgray#1{\color[gray]{#1}}%
      \expandafter\def\csname LTw\endcsname{\color{white}}%
      \expandafter\def\csname LTb\endcsname{\color{black}}%
      \expandafter\def\csname LTa\endcsname{\color{black}}%
      \expandafter\def\csname LT0\endcsname{\color[rgb]{1,0,0}}%
      \expandafter\def\csname LT1\endcsname{\color[rgb]{0,1,0}}%
      \expandafter\def\csname LT2\endcsname{\color[rgb]{0,0,1}}%
      \expandafter\def\csname LT3\endcsname{\color[rgb]{1,0,1}}%
      \expandafter\def\csname LT4\endcsname{\color[rgb]{0,1,1}}%
      \expandafter\def\csname LT5\endcsname{\color[rgb]{1,1,0}}%
      \expandafter\def\csname LT6\endcsname{\color[rgb]{0,0,0}}%
      \expandafter\def\csname LT7\endcsname{\color[rgb]{1,0.3,0}}%
      \expandafter\def\csname LT8\endcsname{\color[rgb]{0.5,0.5,0.5}}%
    \else
      \def\colorrgb#1{\color{black}}%
      \def\colorgray#1{\color[gray]{#1}}%
      \expandafter\def\csname LTw\endcsname{\color{white}}%
      \expandafter\def\csname LTb\endcsname{\color{black}}%
      \expandafter\def\csname LTa\endcsname{\color{black}}%
      \expandafter\def\csname LT0\endcsname{\color{black}}%
      \expandafter\def\csname LT1\endcsname{\color{black}}%
      \expandafter\def\csname LT2\endcsname{\color{black}}%
      \expandafter\def\csname LT3\endcsname{\color{black}}%
      \expandafter\def\csname LT4\endcsname{\color{black}}%
      \expandafter\def\csname LT5\endcsname{\color{black}}%
      \expandafter\def\csname LT6\endcsname{\color{black}}%
      \expandafter\def\csname LT7\endcsname{\color{black}}%
      \expandafter\def\csname LT8\endcsname{\color{black}}%
    \fi
  \fi
  \setlength{\unitlength}{0.0500bp}%
  \begin{picture}(4500.00,3780.00)%
    \gplgaddtomacro\gplbacktext{%
      \csname LTb\endcsname%
      \put(600,640){\makebox(0,0)[r]{\strut{}$10^{-5}$}}%
      \put(600,1220){\makebox(0,0)[r]{\strut{}$10^{-4}$}}%
      \put(600,1800){\makebox(0,0)[r]{\strut{}$10^{-3}$}}%
      \put(600,2379){\makebox(0,0)[r]{\strut{}$10^{-2}$}}%
      \put(600,2959){\makebox(0,0)[r]{\strut{}$10^{-1}$}}%
      \put(600,3539){\makebox(0,0)[r]{\strut{}$10^{0}$}}%
      \put(956,440){\makebox(0,0){\strut{} 0}}%
      \put(1428,440){\makebox(0,0){\strut{} 200}}%
      \put(1900,440){\makebox(0,0){\strut{} 400}}%
      \put(2372,440){\makebox(0,0){\strut{} 600}}%
      \put(2843,440){\makebox(0,0){\strut{} 800}}%
      \put(3315,440){\makebox(0,0){\strut{} 1000}}%
      \put(3787,440){\makebox(0,0){\strut{} 1200}}%
      \put(4259,440){\makebox(0,0){\strut{} 1400}}%
      \put(20,2089){\rotatebox{-270}{\makebox(0,0){\strut{}$f_y(l)$}}}%
      \put(2489,140){\makebox(0,0){\strut{}$l$ [$\mu$m]}}%
      \put(4082,3249){\makebox(0,0)[r]{\strut{}$\beta\gamma =$ 3.16}}%
    }%
    \gplgaddtomacro\gplfronttext{%
    }%
    \gplgaddtomacro\gplbacktext{%
      \csname LTb\endcsname%
      \put(600,640){\makebox(0,0)[r]{\strut{}$10^{-5}$}}%
      \put(600,1220){\makebox(0,0)[r]{\strut{}$10^{-4}$}}%
      \put(600,1800){\makebox(0,0)[r]{\strut{}$10^{-3}$}}%
      \put(600,2379){\makebox(0,0)[r]{\strut{}$10^{-2}$}}%
      \put(600,2959){\makebox(0,0)[r]{\strut{}$10^{-1}$}}%
      \put(600,3539){\makebox(0,0)[r]{\strut{}$10^{0}$}}%
      \put(956,440){\makebox(0,0){\strut{} 0}}%
      \put(1428,440){\makebox(0,0){\strut{} 200}}%
      \put(1900,440){\makebox(0,0){\strut{} 400}}%
      \put(2372,440){\makebox(0,0){\strut{} 600}}%
      \put(2843,440){\makebox(0,0){\strut{} 800}}%
      \put(3315,440){\makebox(0,0){\strut{} 1000}}%
      \put(3787,440){\makebox(0,0){\strut{} 1200}}%
      \put(4259,440){\makebox(0,0){\strut{} 1400}}%
      \put(20,2089){\rotatebox{-270}{\makebox(0,0){\strut{}$f_y(l)$}}}%
      \put(2489,140){\makebox(0,0){\strut{}$l$ [$\mu$m]}}%
    }%
    \gplgaddtomacro\gplfronttext{%
    }%
    \gplgaddtomacro\gplbacktext{%
      \csname LTb\endcsname%
      \put(600,640){\makebox(0,0)[r]{\strut{}$10^{-5}$}}%
      \put(600,1220){\makebox(0,0)[r]{\strut{}$10^{-4}$}}%
      \put(600,1800){\makebox(0,0)[r]{\strut{}$10^{-3}$}}%
      \put(600,2379){\makebox(0,0)[r]{\strut{}$10^{-2}$}}%
      \put(600,2959){\makebox(0,0)[r]{\strut{}$10^{-1}$}}%
      \put(600,3539){\makebox(0,0)[r]{\strut{}$10^{0}$}}%
      \put(956,440){\makebox(0,0){\strut{} 0}}%
      \put(1428,440){\makebox(0,0){\strut{} 200}}%
      \put(1900,440){\makebox(0,0){\strut{} 400}}%
      \put(2372,440){\makebox(0,0){\strut{} 600}}%
      \put(2843,440){\makebox(0,0){\strut{} 800}}%
      \put(3315,440){\makebox(0,0){\strut{} 1000}}%
      \put(3787,440){\makebox(0,0){\strut{} 1200}}%
      \put(4259,440){\makebox(0,0){\strut{} 1400}}%
      \put(20,2089){\rotatebox{-270}{\makebox(0,0){\strut{}$f_y(l)$}}}%
      \put(2489,140){\makebox(0,0){\strut{}$l$ [$\mu$m]}}%
    }%
    \gplgaddtomacro\gplfronttext{%
    }%
    \gplgaddtomacro\gplbacktext{%
      \csname LTb\endcsname%
      \put(600,640){\makebox(0,0)[r]{\strut{}$10^{-5}$}}%
      \put(600,1220){\makebox(0,0)[r]{\strut{}$10^{-4}$}}%
      \put(600,1800){\makebox(0,0)[r]{\strut{}$10^{-3}$}}%
      \put(600,2379){\makebox(0,0)[r]{\strut{}$10^{-2}$}}%
      \put(600,2959){\makebox(0,0)[r]{\strut{}$10^{-1}$}}%
      \put(600,3539){\makebox(0,0)[r]{\strut{}$10^{0}$}}%
      \put(956,440){\makebox(0,0){\strut{} 0}}%
      \put(1428,440){\makebox(0,0){\strut{} 200}}%
      \put(1900,440){\makebox(0,0){\strut{} 400}}%
      \put(2372,440){\makebox(0,0){\strut{} 600}}%
      \put(2843,440){\makebox(0,0){\strut{} 800}}%
      \put(3315,440){\makebox(0,0){\strut{} 1000}}%
      \put(3787,440){\makebox(0,0){\strut{} 1200}}%
      \put(4259,440){\makebox(0,0){\strut{} 1400}}%
      \put(20,2089){\rotatebox{-270}{\makebox(0,0){\strut{}$f_y(l)$}}}%
      \put(2489,140){\makebox(0,0){\strut{}$l$ [$\mu$m]}}%
    }%
    \gplgaddtomacro\gplfronttext{%
    }%
    \gplbacktext
    \put(0,0){\includegraphics{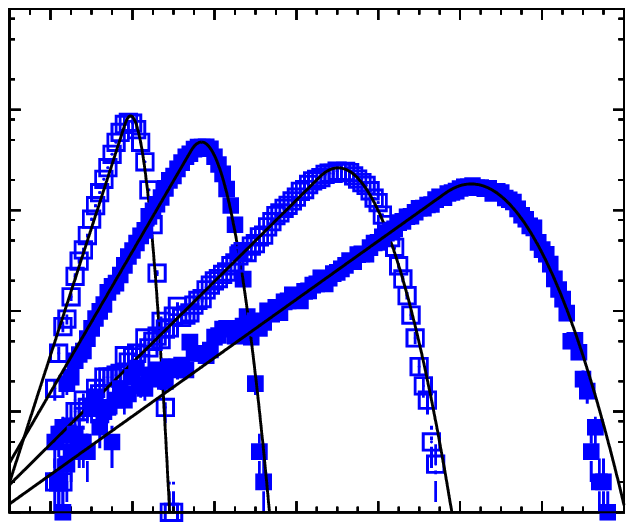}}%
    \gplfronttext
  \end{picture}%
\endgroup

%% file: ylp_fit.tex
\begingroup
  \makeatletter
  \providecommand\color[2][]{%
    \GenericError{(gnuplot) \space\space\space\@spaces}{%
      Package color not loaded in conjunction with
      terminal option `colourtext'%
    }{See the gnuplot documentation for explanation.%
    }{Either use 'blacktext' in gnuplot or load the package
      color.sty in LaTeX.}%
    \renewcommand\color[2][]{}%
  }%
  \providecommand\includegraphics[2][]{%
    \GenericError{(gnuplot) \space\space\space\@spaces}{%
      Package graphicx or graphics not loaded%
    }{See the gnuplot documentation for explanation.%
    }{The gnuplot epslatex terminal needs graphicx.sty or graphics.sty.}%
    \renewcommand\includegraphics[2][]{}%
  }%
  \providecommand\rotatebox[2]{#2}%
  \@ifundefined{ifGPcolor}{%
    \newif\ifGPcolor
    \GPcolortrue
  }{}%
  \@ifundefined{ifGPblacktext}{%
    \newif\ifGPblacktext
    \GPblacktexttrue
  }{}%
  \let\gplgaddtomacro\g@addto@macro
  \gdef\gplbacktext{}%
  \gdef\gplfronttext{}%
  \makeatother
  \ifGPblacktext
    \def\colorrgb#1{}%
    \def\colorgray#1{}%
  \else
    \ifGPcolor
      \def\colorrgb#1{\color[rgb]{#1}}%
      \def\colorgray#1{\color[gray]{#1}}%
      \expandafter\def\csname LTw\endcsname{\color{white}}%
      \expandafter\def\csname LTb\endcsname{\color{black}}%
      \expandafter\def\csname LTa\endcsname{\color{black}}%
      \expandafter\def\csname LT0\endcsname{\color[rgb]{1,0,0}}%
      \expandafter\def\csname LT1\endcsname{\color[rgb]{0,1,0}}%
      \expandafter\def\csname LT2\endcsname{\color[rgb]{0,0,1}}%
      \expandafter\def\csname LT3\endcsname{\color[rgb]{1,0,1}}%
      \expandafter\def\csname LT4\endcsname{\color[rgb]{0,1,1}}%
      \expandafter\def\csname LT5\endcsname{\color[rgb]{1,1,0}}%
      \expandafter\def\csname LT6\endcsname{\color[rgb]{0,0,0}}%
      \expandafter\def\csname LT7\endcsname{\color[rgb]{1,0.3,0}}%
      \expandafter\def\csname LT8\endcsname{\color[rgb]{0.5,0.5,0.5}}%
    \else
      \def\colorrgb#1{\color{black}}%
      \def\colorgray#1{\color[gray]{#1}}%
      \expandafter\def\csname LTw\endcsname{\color{white}}%
      \expandafter\def\csname LTb\endcsname{\color{black}}%
      \expandafter\def\csname LTa\endcsname{\color{black}}%
      \expandafter\def\csname LT0\endcsname{\color{black}}%
      \expandafter\def\csname LT1\endcsname{\color{black}}%
      \expandafter\def\csname LT2\endcsname{\color{black}}%
      \expandafter\def\csname LT3\endcsname{\color{black}}%
      \expandafter\def\csname LT4\endcsname{\color{black}}%
      \expandafter\def\csname LT5\endcsname{\color{black}}%
      \expandafter\def\csname LT6\endcsname{\color{black}}%
      \expandafter\def\csname LT7\endcsname{\color{black}}%
      \expandafter\def\csname LT8\endcsname{\color{black}}%
    \fi
  \fi
  \setlength{\unitlength}{0.0500bp}%
  \begin{picture}(4500.00,3780.00)%
    \gplgaddtomacro\gplbacktext{%
      \csname LTb\endcsname%
      \put(600,640){\makebox(0,0)[r]{\strut{} 0}}%
      \put(600,1220){\makebox(0,0)[r]{\strut{} 0.2}}%
      \put(600,1800){\makebox(0,0)[r]{\strut{} 0.4}}%
      \put(600,2379){\makebox(0,0)[r]{\strut{} 0.6}}%
      \put(600,2959){\makebox(0,0)[r]{\strut{} 0.8}}%
      \put(600,3539){\makebox(0,0)[r]{\strut{} 1}}%
      \put(720,440){\makebox(0,0){\strut{} 0}}%
      \put(1605,440){\makebox(0,0){\strut{} 200}}%
      \put(2490,440){\makebox(0,0){\strut{} 400}}%
      \put(3374,440){\makebox(0,0){\strut{} 600}}%
      \put(4259,440){\makebox(0,0){\strut{} 800}}%
      \put(140,2089){\rotatebox{-270}{\makebox(0,0){\strut{}$y/l_P$ [keV/$\mu$m]}}}%
      \put(2489,140){\makebox(0,0){\strut{}$l_P$ [$\mu$m]}}%
    }%
    \gplgaddtomacro\gplfronttext{%
      \csname LTb\endcsname%
      \put(3313,3236){\makebox(0,0)[l]{\strut{}$\beta\gamma = 0.56$}}%
      \csname LTb\endcsname%
      \put(3313,3036){\makebox(0,0)[l]{\strut{}$\beta\gamma = 1.00$}}%
      \csname LTb\endcsname%
      \put(3313,2836){\makebox(0,0)[l]{\strut{}$\beta\gamma = 3.16$}}%
      \csname LTb\endcsname%
      \put(3313,2636){\makebox(0,0)[l]{\strut{}$\beta\gamma = 10.0$}}%
    }%
    \gplbacktext
    \put(0,0){\includegraphics{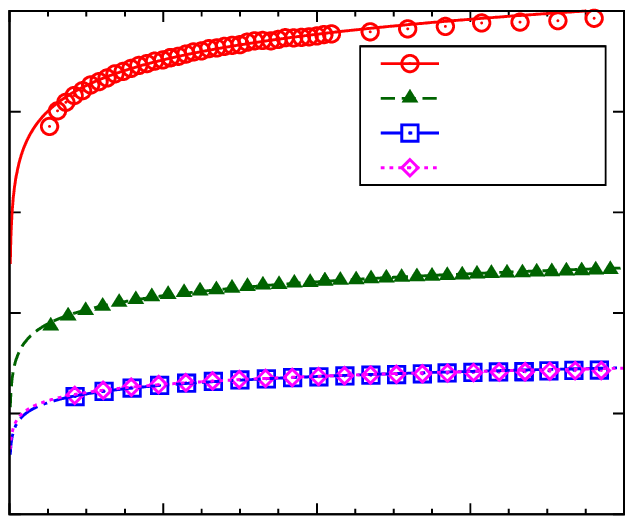}}%
    \gplfronttext
  \end{picture}%
\endgroup

%% file: sllp_fit.tex
\begingroup
  \makeatletter
  \providecommand\color[2][]{%
    \GenericError{(gnuplot) \space\space\space\@spaces}{%
      Package color not loaded in conjunction with
      terminal option `colourtext'%
    }{See the gnuplot documentation for explanation.%
    }{Either use 'blacktext' in gnuplot or load the package
      color.sty in LaTeX.}%
    \renewcommand\color[2][]{}%
  }%
  \providecommand\includegraphics[2][]{%
    \GenericError{(gnuplot) \space\space\space\@spaces}{%
      Package graphicx or graphics not loaded%
    }{See the gnuplot documentation for explanation.%
    }{The gnuplot epslatex terminal needs graphicx.sty or graphics.sty.}%
    \renewcommand\includegraphics[2][]{}%
  }%
  \providecommand\rotatebox[2]{#2}%
  \@ifundefined{ifGPcolor}{%
    \newif\ifGPcolor
    \GPcolortrue
  }{}%
  \@ifundefined{ifGPblacktext}{%
    \newif\ifGPblacktext
    \GPblacktexttrue
  }{}%
  \let\gplgaddtomacro\g@addto@macro
  \gdef\gplbacktext{}%
  \gdef\gplfronttext{}%
  \makeatother
  \ifGPblacktext
    \def\colorrgb#1{}%
    \def\colorgray#1{}%
  \else
    \ifGPcolor
      \def\colorrgb#1{\color[rgb]{#1}}%
      \def\colorgray#1{\color[gray]{#1}}%
      \expandafter\def\csname LTw\endcsname{\color{white}}%
      \expandafter\def\csname LTb\endcsname{\color{black}}%
      \expandafter\def\csname LTa\endcsname{\color{black}}%
      \expandafter\def\csname LT0\endcsname{\color[rgb]{1,0,0}}%
      \expandafter\def\csname LT1\endcsname{\color[rgb]{0,1,0}}%
      \expandafter\def\csname LT2\endcsname{\color[rgb]{0,0,1}}%
      \expandafter\def\csname LT3\endcsname{\color[rgb]{1,0,1}}%
      \expandafter\def\csname LT4\endcsname{\color[rgb]{0,1,1}}%
      \expandafter\def\csname LT5\endcsname{\color[rgb]{1,1,0}}%
      \expandafter\def\csname LT6\endcsname{\color[rgb]{0,0,0}}%
      \expandafter\def\csname LT7\endcsname{\color[rgb]{1,0.3,0}}%
      \expandafter\def\csname LT8\endcsname{\color[rgb]{0.5,0.5,0.5}}%
    \else
      \def\colorrgb#1{\color{black}}%
      \def\colorgray#1{\color[gray]{#1}}%
      \expandafter\def\csname LTw\endcsname{\color{white}}%
      \expandafter\def\csname LTb\endcsname{\color{black}}%
      \expandafter\def\csname LTa\endcsname{\color{black}}%
      \expandafter\def\csname LT0\endcsname{\color{black}}%
      \expandafter\def\csname LT1\endcsname{\color{black}}%
      \expandafter\def\csname LT2\endcsname{\color{black}}%
      \expandafter\def\csname LT3\endcsname{\color{black}}%
      \expandafter\def\csname LT4\endcsname{\color{black}}%
      \expandafter\def\csname LT5\endcsname{\color{black}}%
      \expandafter\def\csname LT6\endcsname{\color{black}}%
      \expandafter\def\csname LT7\endcsname{\color{black}}%
      \expandafter\def\csname LT8\endcsname{\color{black}}%
    \fi
  \fi
  \setlength{\unitlength}{0.0500bp}%
  \begin{picture}(4500.00,3780.00)%
    \gplgaddtomacro\gplbacktext{%
      \csname LTb\endcsname%
      \put(600,640){\makebox(0,0)[r]{\strut{} 0}}%
      \put(600,962){\makebox(0,0)[r]{\strut{} 5}}%
      \put(600,1284){\makebox(0,0)[r]{\strut{} 10}}%
      \put(600,1606){\makebox(0,0)[r]{\strut{} 15}}%
      \put(600,1928){\makebox(0,0)[r]{\strut{} 20}}%
      \put(600,2251){\makebox(0,0)[r]{\strut{} 25}}%
      \put(600,2573){\makebox(0,0)[r]{\strut{} 30}}%
      \put(600,2895){\makebox(0,0)[r]{\strut{} 35}}%
      \put(600,3217){\makebox(0,0)[r]{\strut{} 40}}%
      \put(600,3539){\makebox(0,0)[r]{\strut{} 45}}%
      \put(720,440){\makebox(0,0){\strut{} 0}}%
      \put(1162,440){\makebox(0,0){\strut{} 50}}%
      \put(1605,440){\makebox(0,0){\strut{} 100}}%
      \put(2047,440){\makebox(0,0){\strut{} 150}}%
      \put(2490,440){\makebox(0,0){\strut{} 200}}%
      \put(2932,440){\makebox(0,0){\strut{} 250}}%
      \put(3374,440){\makebox(0,0){\strut{} 300}}%
      \put(3817,440){\makebox(0,0){\strut{} 350}}%
      \put(4259,440){\makebox(0,0){\strut{} 400}}%
      \put(140,2089){\rotatebox{-270}{\makebox(0,0){\strut{}$\sigma_l$ [$\mu$m]}}}%
      \put(2489,140){\makebox(0,0){\strut{}$y$ [keV]}}%
    }%
    \gplgaddtomacro\gplfronttext{%
      \csname LTb\endcsname%
      \put(3299,1403){\makebox(0,0)[l]{\strut{}$\beta\gamma = 0.56$}}%
      \csname LTb\endcsname%
      \put(3299,1203){\makebox(0,0)[l]{\strut{}$\beta\gamma = 1.00$}}%
      \csname LTb\endcsname%
      \put(3299,1003){\makebox(0,0)[l]{\strut{}$\beta\gamma = 3.16$}}%
      \csname LTb\endcsname%
      \put(3299,803){\makebox(0,0)[l]{\strut{}$\beta\gamma = 10.0$}}%
    }%
    \gplbacktext
    \put(0,0){\includegraphics{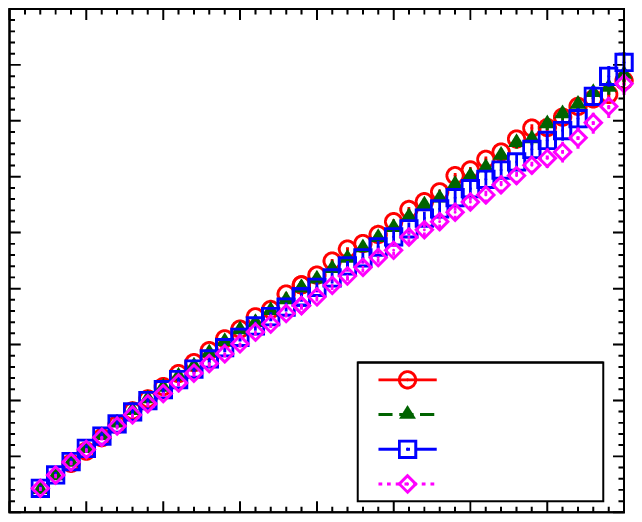}}%
    \gplfronttext
  \end{picture}%
\endgroup

%% file: ylp_fit_2.tex
\begingroup
  \makeatletter
  \providecommand\color[2][]{%
    \GenericError{(gnuplot) \space\space\space\@spaces}{%
      Package color not loaded in conjunction with
      terminal option `colourtext'%
    }{See the gnuplot documentation for explanation.%
    }{Either use 'blacktext' in gnuplot or load the package
      color.sty in LaTeX.}%
    \renewcommand\color[2][]{}%
  }%
  \providecommand\includegraphics[2][]{%
    \GenericError{(gnuplot) \space\space\space\@spaces}{%
      Package graphicx or graphics not loaded%
    }{See the gnuplot documentation for explanation.%
    }{The gnuplot epslatex terminal needs graphicx.sty or graphics.sty.}%
    \renewcommand\includegraphics[2][]{}%
  }%
  \providecommand\rotatebox[2]{#2}%
  \@ifundefined{ifGPcolor}{%
    \newif\ifGPcolor
    \GPcolortrue
  }{}%
  \@ifundefined{ifGPblacktext}{%
    \newif\ifGPblacktext
    \GPblacktexttrue
  }{}%
  \let\gplgaddtomacro\g@addto@macro
  \gdef\gplbacktext{}%
  \gdef\gplfronttext{}%
  \makeatother
  \ifGPblacktext
    \def\colorrgb#1{}%
    \def\colorgray#1{}%
  \else
    \ifGPcolor
      \def\colorrgb#1{\color[rgb]{#1}}%
      \def\colorgray#1{\color[gray]{#1}}%
      \expandafter\def\csname LTw\endcsname{\color{white}}%
      \expandafter\def\csname LTb\endcsname{\color{black}}%
      \expandafter\def\csname LTa\endcsname{\color{black}}%
      \expandafter\def\csname LT0\endcsname{\color[rgb]{1,0,0}}%
      \expandafter\def\csname LT1\endcsname{\color[rgb]{0,1,0}}%
      \expandafter\def\csname LT2\endcsname{\color[rgb]{0,0,1}}%
      \expandafter\def\csname LT3\endcsname{\color[rgb]{1,0,1}}%
      \expandafter\def\csname LT4\endcsname{\color[rgb]{0,1,1}}%
      \expandafter\def\csname LT5\endcsname{\color[rgb]{1,1,0}}%
      \expandafter\def\csname LT6\endcsname{\color[rgb]{0,0,0}}%
      \expandafter\def\csname LT7\endcsname{\color[rgb]{1,0.3,0}}%
      \expandafter\def\csname LT8\endcsname{\color[rgb]{0.5,0.5,0.5}}%
    \else
      \def\colorrgb#1{\color{black}}%
      \def\colorgray#1{\color[gray]{#1}}%
      \expandafter\def\csname LTw\endcsname{\color{white}}%
      \expandafter\def\csname LTb\endcsname{\color{black}}%
      \expandafter\def\csname LTa\endcsname{\color{black}}%
      \expandafter\def\csname LT0\endcsname{\color{black}}%
      \expandafter\def\csname LT1\endcsname{\color{black}}%
      \expandafter\def\csname LT2\endcsname{\color{black}}%
      \expandafter\def\csname LT3\endcsname{\color{black}}%
      \expandafter\def\csname LT4\endcsname{\color{black}}%
      \expandafter\def\csname LT5\endcsname{\color{black}}%
      \expandafter\def\csname LT6\endcsname{\color{black}}%
      \expandafter\def\csname LT7\endcsname{\color{black}}%
      \expandafter\def\csname LT8\endcsname{\color{black}}%
    \fi
  \fi
  \setlength{\unitlength}{0.0500bp}%
  \begin{picture}(4500.00,3780.00)%
    \gplgaddtomacro\gplbacktext{%
      \csname LTb\endcsname%
      \put(600,640){\makebox(0,0)[r]{\strut{} 0}}%
      \put(600,1220){\makebox(0,0)[r]{\strut{} 0.2}}%
      \put(600,1800){\makebox(0,0)[r]{\strut{} 0.4}}%
      \put(600,2379){\makebox(0,0)[r]{\strut{} 0.6}}%
      \put(600,2959){\makebox(0,0)[r]{\strut{} 0.8}}%
      \put(600,3539){\makebox(0,0)[r]{\strut{} 1}}%
      \put(720,440){\makebox(0,0){\strut{} 0}}%
      \put(1605,440){\makebox(0,0){\strut{} 200}}%
      \put(2490,440){\makebox(0,0){\strut{} 400}}%
      \put(3374,440){\makebox(0,0){\strut{} 600}}%
      \put(4259,440){\makebox(0,0){\strut{} 800}}%
      \put(140,2089){\rotatebox{-270}{\makebox(0,0){\strut{}$\Delta/l$ [keV/$\mu$m]}}}%
      \put(2489,140){\makebox(0,0){\strut{}$l$ [$\mu$m]}}%
    }%
    \gplgaddtomacro\gplfronttext{%
      \csname LTb\endcsname%
      \put(3313,3236){\makebox(0,0)[l]{\strut{}$\beta\gamma = 0.56$}}%
      \csname LTb\endcsname%
      \put(3313,3036){\makebox(0,0)[l]{\strut{}$\beta\gamma = 1.00$}}%
      \csname LTb\endcsname%
      \put(3313,2836){\makebox(0,0)[l]{\strut{}$\beta\gamma = 3.16$}}%
      \csname LTb\endcsname%
      \put(3313,2636){\makebox(0,0)[l]{\strut{}$\beta\gamma = 10.0$}}%
    }%
    \gplbacktext
    \put(0,0){\includegraphics{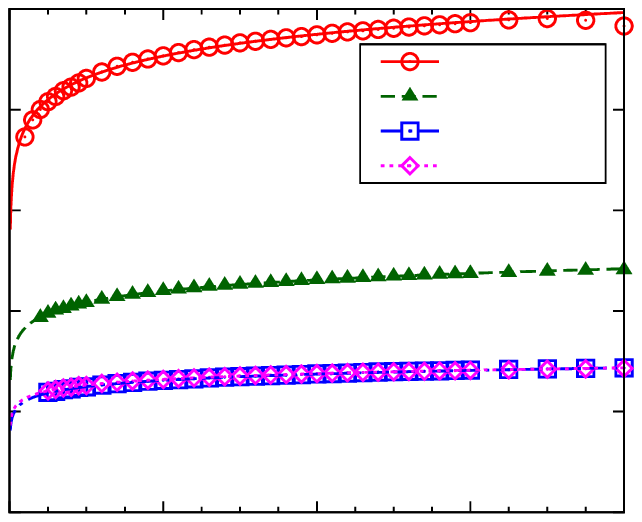}}%
    \gplfronttext
  \end{picture}%
\endgroup

%% file: bg_0.56_all_2.tex
\begingroup
  \makeatletter
  \providecommand\color[2][]{%
    \GenericError{(gnuplot) \space\space\space\@spaces}{%
      Package color not loaded in conjunction with
      terminal option `colourtext'%
    }{See the gnuplot documentation for explanation.%
    }{Either use 'blacktext' in gnuplot or load the package
      color.sty in LaTeX.}%
    \renewcommand\color[2][]{}%
  }%
  \providecommand\includegraphics[2][]{%
    \GenericError{(gnuplot) \space\space\space\@spaces}{%
      Package graphicx or graphics not loaded%
    }{See the gnuplot documentation for explanation.%
    }{The gnuplot epslatex terminal needs graphicx.sty or graphics.sty.}%
    \renewcommand\includegraphics[2][]{}%
  }%
  \providecommand\rotatebox[2]{#2}%
  \@ifundefined{ifGPcolor}{%
    \newif\ifGPcolor
    \GPcolortrue
  }{}%
  \@ifundefined{ifGPblacktext}{%
    \newif\ifGPblacktext
    \GPblacktexttrue
  }{}%
  \let\gplgaddtomacro\g@addto@macro
  \gdef\gplbacktext{}%
  \gdef\gplfronttext{}%
  \makeatother
  \ifGPblacktext
    \def\colorrgb#1{}%
    \def\colorgray#1{}%
  \else
    \ifGPcolor
      \def\colorrgb#1{\color[rgb]{#1}}%
      \def\colorgray#1{\color[gray]{#1}}%
      \expandafter\def\csname LTw\endcsname{\color{white}}%
      \expandafter\def\csname LTb\endcsname{\color{black}}%
      \expandafter\def\csname LTa\endcsname{\color{black}}%
      \expandafter\def\csname LT0\endcsname{\color[rgb]{1,0,0}}%
      \expandafter\def\csname LT1\endcsname{\color[rgb]{0,1,0}}%
      \expandafter\def\csname LT2\endcsname{\color[rgb]{0,0,1}}%
      \expandafter\def\csname LT3\endcsname{\color[rgb]{1,0,1}}%
      \expandafter\def\csname LT4\endcsname{\color[rgb]{0,1,1}}%
      \expandafter\def\csname LT5\endcsname{\color[rgb]{1,1,0}}%
      \expandafter\def\csname LT6\endcsname{\color[rgb]{0,0,0}}%
      \expandafter\def\csname LT7\endcsname{\color[rgb]{1,0.3,0}}%
      \expandafter\def\csname LT8\endcsname{\color[rgb]{0.5,0.5,0.5}}%
    \else
      \def\colorrgb#1{\color{black}}%
      \def\colorgray#1{\color[gray]{#1}}%
      \expandafter\def\csname LTw\endcsname{\color{white}}%
      \expandafter\def\csname LTb\endcsname{\color{black}}%
      \expandafter\def\csname LTa\endcsname{\color{black}}%
      \expandafter\def\csname LT0\endcsname{\color{black}}%
      \expandafter\def\csname LT1\endcsname{\color{black}}%
      \expandafter\def\csname LT2\endcsname{\color{black}}%
      \expandafter\def\csname LT3\endcsname{\color{black}}%
      \expandafter\def\csname LT4\endcsname{\color{black}}%
      \expandafter\def\csname LT5\endcsname{\color{black}}%
      \expandafter\def\csname LT6\endcsname{\color{black}}%
      \expandafter\def\csname LT7\endcsname{\color{black}}%
      \expandafter\def\csname LT8\endcsname{\color{black}}%
    \fi
  \fi
  \setlength{\unitlength}{0.0500bp}%
  \begin{picture}(4500.00,3780.00)%
    \gplgaddtomacro\gplbacktext{%
      \csname LTb\endcsname%
      \put(600,640){\makebox(0,0)[r]{\strut{}$10^{-5}$}}%
      \put(600,1220){\makebox(0,0)[r]{\strut{}$10^{-4}$}}%
      \put(600,1800){\makebox(0,0)[r]{\strut{}$10^{-3}$}}%
      \put(600,2379){\makebox(0,0)[r]{\strut{}$10^{-2}$}}%
      \put(600,2959){\makebox(0,0)[r]{\strut{}$10^{-1}$}}%
      \put(600,3539){\makebox(0,0)[r]{\strut{}$10^{0}$}}%
      \put(1363,440){\makebox(0,0){\strut{} 0}}%
      \put(2650,440){\makebox(0,0){\strut{} 200}}%
      \put(3937,440){\makebox(0,0){\strut{} 400}}%
      \put(20,2089){\rotatebox{-270}{\makebox(0,0){\strut{}$f_y(\Delta)$}}}%
      \put(2489,140){\makebox(0,0){\strut{}$\Delta$ [keV]}}%
      \put(4082,3249){\makebox(0,0)[r]{\strut{}$\beta\gamma =$ 0.56}}%
    }%
    \gplgaddtomacro\gplfronttext{%
    }%
    \gplgaddtomacro\gplbacktext{%
      \csname LTb\endcsname%
      \put(600,640){\makebox(0,0)[r]{\strut{}$10^{-5}$}}%
      \put(600,1220){\makebox(0,0)[r]{\strut{}$10^{-4}$}}%
      \put(600,1800){\makebox(0,0)[r]{\strut{}$10^{-3}$}}%
      \put(600,2379){\makebox(0,0)[r]{\strut{}$10^{-2}$}}%
      \put(600,2959){\makebox(0,0)[r]{\strut{}$10^{-1}$}}%
      \put(600,3539){\makebox(0,0)[r]{\strut{}$10^{0}$}}%
      \put(1363,440){\makebox(0,0){\strut{} 0}}%
      \put(2650,440){\makebox(0,0){\strut{} 200}}%
      \put(3937,440){\makebox(0,0){\strut{} 400}}%
      \put(20,2089){\rotatebox{-270}{\makebox(0,0){\strut{}$f_y(\Delta)$}}}%
      \put(2489,140){\makebox(0,0){\strut{}$\Delta$ [keV]}}%
    }%
    \gplgaddtomacro\gplfronttext{%
    }%
    \gplgaddtomacro\gplbacktext{%
      \csname LTb\endcsname%
      \put(600,640){\makebox(0,0)[r]{\strut{}$10^{-5}$}}%
      \put(600,1220){\makebox(0,0)[r]{\strut{}$10^{-4}$}}%
      \put(600,1800){\makebox(0,0)[r]{\strut{}$10^{-3}$}}%
      \put(600,2379){\makebox(0,0)[r]{\strut{}$10^{-2}$}}%
      \put(600,2959){\makebox(0,0)[r]{\strut{}$10^{-1}$}}%
      \put(600,3539){\makebox(0,0)[r]{\strut{}$10^{0}$}}%
      \put(1363,440){\makebox(0,0){\strut{} 0}}%
      \put(2650,440){\makebox(0,0){\strut{} 200}}%
      \put(3937,440){\makebox(0,0){\strut{} 400}}%
      \put(20,2089){\rotatebox{-270}{\makebox(0,0){\strut{}$f_y(\Delta)$}}}%
      \put(2489,140){\makebox(0,0){\strut{}$\Delta$ [keV]}}%
    }%
    \gplgaddtomacro\gplfronttext{%
    }%
    \gplgaddtomacro\gplbacktext{%
      \csname LTb\endcsname%
      \put(600,640){\makebox(0,0)[r]{\strut{}$10^{-5}$}}%
      \put(600,1220){\makebox(0,0)[r]{\strut{}$10^{-4}$}}%
      \put(600,1800){\makebox(0,0)[r]{\strut{}$10^{-3}$}}%
      \put(600,2379){\makebox(0,0)[r]{\strut{}$10^{-2}$}}%
      \put(600,2959){\makebox(0,0)[r]{\strut{}$10^{-1}$}}%
      \put(600,3539){\makebox(0,0)[r]{\strut{}$10^{0}$}}%
      \put(1363,440){\makebox(0,0){\strut{} 0}}%
      \put(2650,440){\makebox(0,0){\strut{} 200}}%
      \put(3937,440){\makebox(0,0){\strut{} 400}}%
      \put(20,2089){\rotatebox{-270}{\makebox(0,0){\strut{}$f_y(\Delta)$}}}%
      \put(2489,140){\makebox(0,0){\strut{}$\Delta$ [keV]}}%
    }%
    \gplgaddtomacro\gplfronttext{%
    }%
    \gplbacktext
    \put(0,0){\includegraphics{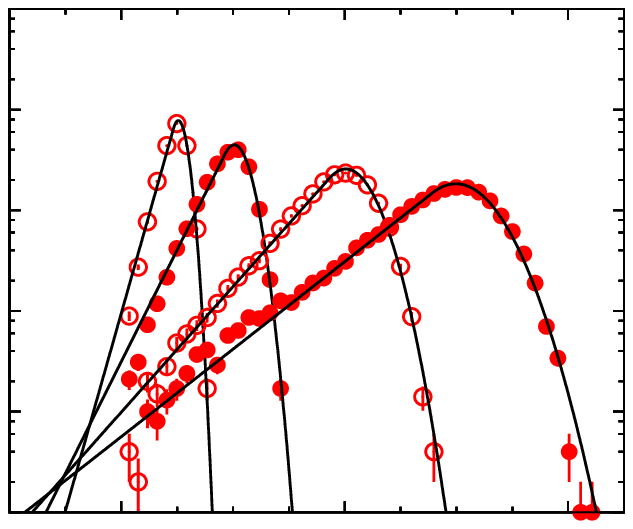}}%
    \gplfronttext
  \end{picture}%
\endgroup

%% file: bg_1.00_all_2.tex
\begingroup
  \makeatletter
  \providecommand\color[2][]{%
    \GenericError{(gnuplot) \space\space\space\@spaces}{%
      Package color not loaded in conjunction with
      terminal option `colourtext'%
    }{See the gnuplot documentation for explanation.%
    }{Either use 'blacktext' in gnuplot or load the package
      color.sty in LaTeX.}%
    \renewcommand\color[2][]{}%
  }%
  \providecommand\includegraphics[2][]{%
    \GenericError{(gnuplot) \space\space\space\@spaces}{%
      Package graphicx or graphics not loaded%
    }{See the gnuplot documentation for explanation.%
    }{The gnuplot epslatex terminal needs graphicx.sty or graphics.sty.}%
    \renewcommand\includegraphics[2][]{}%
  }%
  \providecommand\rotatebox[2]{#2}%
  \@ifundefined{ifGPcolor}{%
    \newif\ifGPcolor
    \GPcolortrue
  }{}%
  \@ifundefined{ifGPblacktext}{%
    \newif\ifGPblacktext
    \GPblacktexttrue
  }{}%
  \let\gplgaddtomacro\g@addto@macro
  \gdef\gplbacktext{}%
  \gdef\gplfronttext{}%
  \makeatother
  \ifGPblacktext
    \def\colorrgb#1{}%
    \def\colorgray#1{}%
  \else
    \ifGPcolor
      \def\colorrgb#1{\color[rgb]{#1}}%
      \def\colorgray#1{\color[gray]{#1}}%
      \expandafter\def\csname LTw\endcsname{\color{white}}%
      \expandafter\def\csname LTb\endcsname{\color{black}}%
      \expandafter\def\csname LTa\endcsname{\color{black}}%
      \expandafter\def\csname LT0\endcsname{\color[rgb]{1,0,0}}%
      \expandafter\def\csname LT1\endcsname{\color[rgb]{0,1,0}}%
      \expandafter\def\csname LT2\endcsname{\color[rgb]{0,0,1}}%
      \expandafter\def\csname LT3\endcsname{\color[rgb]{1,0,1}}%
      \expandafter\def\csname LT4\endcsname{\color[rgb]{0,1,1}}%
      \expandafter\def\csname LT5\endcsname{\color[rgb]{1,1,0}}%
      \expandafter\def\csname LT6\endcsname{\color[rgb]{0,0,0}}%
      \expandafter\def\csname LT7\endcsname{\color[rgb]{1,0.3,0}}%
      \expandafter\def\csname LT8\endcsname{\color[rgb]{0.5,0.5,0.5}}%
    \else
      \def\colorrgb#1{\color{black}}%
      \def\colorgray#1{\color[gray]{#1}}%
      \expandafter\def\csname LTw\endcsname{\color{white}}%
      \expandafter\def\csname LTb\endcsname{\color{black}}%
      \expandafter\def\csname LTa\endcsname{\color{black}}%
      \expandafter\def\csname LT0\endcsname{\color{black}}%
      \expandafter\def\csname LT1\endcsname{\color{black}}%
      \expandafter\def\csname LT2\endcsname{\color{black}}%
      \expandafter\def\csname LT3\endcsname{\color{black}}%
      \expandafter\def\csname LT4\endcsname{\color{black}}%
      \expandafter\def\csname LT5\endcsname{\color{black}}%
      \expandafter\def\csname LT6\endcsname{\color{black}}%
      \expandafter\def\csname LT7\endcsname{\color{black}}%
      \expandafter\def\csname LT8\endcsname{\color{black}}%
    \fi
  \fi
  \setlength{\unitlength}{0.0500bp}%
  \begin{picture}(4500.00,3780.00)%
    \gplgaddtomacro\gplbacktext{%
      \csname LTb\endcsname%
      \put(600,640){\makebox(0,0)[r]{\strut{}$10^{-5}$}}%
      \put(600,1220){\makebox(0,0)[r]{\strut{}$10^{-4}$}}%
      \put(600,1800){\makebox(0,0)[r]{\strut{}$10^{-3}$}}%
      \put(600,2379){\makebox(0,0)[r]{\strut{}$10^{-2}$}}%
      \put(600,2959){\makebox(0,0)[r]{\strut{}$10^{-1}$}}%
      \put(600,3539){\makebox(0,0)[r]{\strut{}$10^{0}$}}%
      \put(1363,440){\makebox(0,0){\strut{} 0}}%
      \put(2650,440){\makebox(0,0){\strut{} 200}}%
      \put(3937,440){\makebox(0,0){\strut{} 400}}%
      \put(20,2089){\rotatebox{-270}{\makebox(0,0){\strut{}$f_y(\Delta)$}}}%
      \put(2489,140){\makebox(0,0){\strut{}$\Delta$ [keV]}}%
      \put(4082,3249){\makebox(0,0)[r]{\strut{}$\beta\gamma =$ 1.00}}%
    }%
    \gplgaddtomacro\gplfronttext{%
    }%
    \gplgaddtomacro\gplbacktext{%
      \csname LTb\endcsname%
      \put(600,640){\makebox(0,0)[r]{\strut{}$10^{-5}$}}%
      \put(600,1220){\makebox(0,0)[r]{\strut{}$10^{-4}$}}%
      \put(600,1800){\makebox(0,0)[r]{\strut{}$10^{-3}$}}%
      \put(600,2379){\makebox(0,0)[r]{\strut{}$10^{-2}$}}%
      \put(600,2959){\makebox(0,0)[r]{\strut{}$10^{-1}$}}%
      \put(600,3539){\makebox(0,0)[r]{\strut{}$10^{0}$}}%
      \put(1363,440){\makebox(0,0){\strut{} 0}}%
      \put(2650,440){\makebox(0,0){\strut{} 200}}%
      \put(3937,440){\makebox(0,0){\strut{} 400}}%
      \put(20,2089){\rotatebox{-270}{\makebox(0,0){\strut{}$f_y(\Delta)$}}}%
      \put(2489,140){\makebox(0,0){\strut{}$\Delta$ [keV]}}%
    }%
    \gplgaddtomacro\gplfronttext{%
    }%
    \gplgaddtomacro\gplbacktext{%
      \csname LTb\endcsname%
      \put(600,640){\makebox(0,0)[r]{\strut{}$10^{-5}$}}%
      \put(600,1220){\makebox(0,0)[r]{\strut{}$10^{-4}$}}%
      \put(600,1800){\makebox(0,0)[r]{\strut{}$10^{-3}$}}%
      \put(600,2379){\makebox(0,0)[r]{\strut{}$10^{-2}$}}%
      \put(600,2959){\makebox(0,0)[r]{\strut{}$10^{-1}$}}%
      \put(600,3539){\makebox(0,0)[r]{\strut{}$10^{0}$}}%
      \put(1363,440){\makebox(0,0){\strut{} 0}}%
      \put(2650,440){\makebox(0,0){\strut{} 200}}%
      \put(3937,440){\makebox(0,0){\strut{} 400}}%
      \put(20,2089){\rotatebox{-270}{\makebox(0,0){\strut{}$f_y(\Delta)$}}}%
      \put(2489,140){\makebox(0,0){\strut{}$\Delta$ [keV]}}%
    }%
    \gplgaddtomacro\gplfronttext{%
    }%
    \gplgaddtomacro\gplbacktext{%
      \csname LTb\endcsname%
      \put(600,640){\makebox(0,0)[r]{\strut{}$10^{-5}$}}%
      \put(600,1220){\makebox(0,0)[r]{\strut{}$10^{-4}$}}%
      \put(600,1800){\makebox(0,0)[r]{\strut{}$10^{-3}$}}%
      \put(600,2379){\makebox(0,0)[r]{\strut{}$10^{-2}$}}%
      \put(600,2959){\makebox(0,0)[r]{\strut{}$10^{-1}$}}%
      \put(600,3539){\makebox(0,0)[r]{\strut{}$10^{0}$}}%
      \put(1363,440){\makebox(0,0){\strut{} 0}}%
      \put(2650,440){\makebox(0,0){\strut{} 200}}%
      \put(3937,440){\makebox(0,0){\strut{} 400}}%
      \put(20,2089){\rotatebox{-270}{\makebox(0,0){\strut{}$f_y(\Delta)$}}}%
      \put(2489,140){\makebox(0,0){\strut{}$\Delta$ [keV]}}%
    }%
    \gplgaddtomacro\gplfronttext{%
    }%
    \gplbacktext
    \put(0,0){\includegraphics{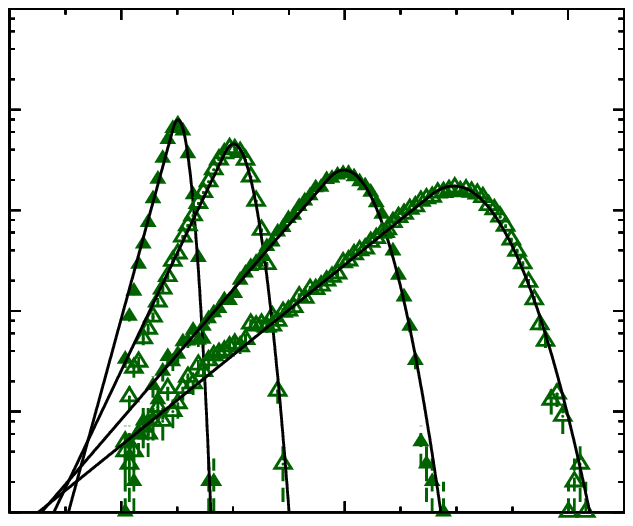}}%
    \gplfronttext
  \end{picture}%
\endgroup

%% file: bg_3.16_all_2.tex
\begingroup
  \makeatletter
  \providecommand\color[2][]{%
    \GenericError{(gnuplot) \space\space\space\@spaces}{%
      Package color not loaded in conjunction with
      terminal option `colourtext'%
    }{See the gnuplot documentation for explanation.%
    }{Either use 'blacktext' in gnuplot or load the package
      color.sty in LaTeX.}%
    \renewcommand\color[2][]{}%
  }%
  \providecommand\includegraphics[2][]{%
    \GenericError{(gnuplot) \space\space\space\@spaces}{%
      Package graphicx or graphics not loaded%
    }{See the gnuplot documentation for explanation.%
    }{The gnuplot epslatex terminal needs graphicx.sty or graphics.sty.}%
    \renewcommand\includegraphics[2][]{}%
  }%
  \providecommand\rotatebox[2]{#2}%
  \@ifundefined{ifGPcolor}{%
    \newif\ifGPcolor
    \GPcolortrue
  }{}%
  \@ifundefined{ifGPblacktext}{%
    \newif\ifGPblacktext
    \GPblacktexttrue
  }{}%
  \let\gplgaddtomacro\g@addto@macro
  \gdef\gplbacktext{}%
  \gdef\gplfronttext{}%
  \makeatother
  \ifGPblacktext
    \def\colorrgb#1{}%
    \def\colorgray#1{}%
  \else
    \ifGPcolor
      \def\colorrgb#1{\color[rgb]{#1}}%
      \def\colorgray#1{\color[gray]{#1}}%
      \expandafter\def\csname LTw\endcsname{\color{white}}%
      \expandafter\def\csname LTb\endcsname{\color{black}}%
      \expandafter\def\csname LTa\endcsname{\color{black}}%
      \expandafter\def\csname LT0\endcsname{\color[rgb]{1,0,0}}%
      \expandafter\def\csname LT1\endcsname{\color[rgb]{0,1,0}}%
      \expandafter\def\csname LT2\endcsname{\color[rgb]{0,0,1}}%
      \expandafter\def\csname LT3\endcsname{\color[rgb]{1,0,1}}%
      \expandafter\def\csname LT4\endcsname{\color[rgb]{0,1,1}}%
      \expandafter\def\csname LT5\endcsname{\color[rgb]{1,1,0}}%
      \expandafter\def\csname LT6\endcsname{\color[rgb]{0,0,0}}%
      \expandafter\def\csname LT7\endcsname{\color[rgb]{1,0.3,0}}%
      \expandafter\def\csname LT8\endcsname{\color[rgb]{0.5,0.5,0.5}}%
    \else
      \def\colorrgb#1{\color{black}}%
      \def\colorgray#1{\color[gray]{#1}}%
      \expandafter\def\csname LTw\endcsname{\color{white}}%
      \expandafter\def\csname LTb\endcsname{\color{black}}%
      \expandafter\def\csname LTa\endcsname{\color{black}}%
      \expandafter\def\csname LT0\endcsname{\color{black}}%
      \expandafter\def\csname LT1\endcsname{\color{black}}%
      \expandafter\def\csname LT2\endcsname{\color{black}}%
      \expandafter\def\csname LT3\endcsname{\color{black}}%
      \expandafter\def\csname LT4\endcsname{\color{black}}%
      \expandafter\def\csname LT5\endcsname{\color{black}}%
      \expandafter\def\csname LT6\endcsname{\color{black}}%
      \expandafter\def\csname LT7\endcsname{\color{black}}%
      \expandafter\def\csname LT8\endcsname{\color{black}}%
    \fi
  \fi
  \setlength{\unitlength}{0.0500bp}%
  \begin{picture}(4500.00,3780.00)%
    \gplgaddtomacro\gplbacktext{%
      \csname LTb\endcsname%
      \put(600,640){\makebox(0,0)[r]{\strut{}$10^{-5}$}}%
      \put(600,1220){\makebox(0,0)[r]{\strut{}$10^{-4}$}}%
      \put(600,1800){\makebox(0,0)[r]{\strut{}$10^{-3}$}}%
      \put(600,2379){\makebox(0,0)[r]{\strut{}$10^{-2}$}}%
      \put(600,2959){\makebox(0,0)[r]{\strut{}$10^{-1}$}}%
      \put(600,3539){\makebox(0,0)[r]{\strut{}$10^{0}$}}%
      \put(1363,440){\makebox(0,0){\strut{} 0}}%
      \put(2650,440){\makebox(0,0){\strut{} 200}}%
      \put(3937,440){\makebox(0,0){\strut{} 400}}%
      \put(20,2089){\rotatebox{-270}{\makebox(0,0){\strut{}$f_y(\Delta)$}}}%
      \put(2489,140){\makebox(0,0){\strut{}$\Delta$ [keV]}}%
      \put(4082,3249){\makebox(0,0)[r]{\strut{}$\beta\gamma =$ 3.16}}%
    }%
    \gplgaddtomacro\gplfronttext{%
    }%
    \gplgaddtomacro\gplbacktext{%
      \csname LTb\endcsname%
      \put(600,640){\makebox(0,0)[r]{\strut{}$10^{-5}$}}%
      \put(600,1220){\makebox(0,0)[r]{\strut{}$10^{-4}$}}%
      \put(600,1800){\makebox(0,0)[r]{\strut{}$10^{-3}$}}%
      \put(600,2379){\makebox(0,0)[r]{\strut{}$10^{-2}$}}%
      \put(600,2959){\makebox(0,0)[r]{\strut{}$10^{-1}$}}%
      \put(600,3539){\makebox(0,0)[r]{\strut{}$10^{0}$}}%
      \put(1363,440){\makebox(0,0){\strut{} 0}}%
      \put(2650,440){\makebox(0,0){\strut{} 200}}%
      \put(3937,440){\makebox(0,0){\strut{} 400}}%
      \put(20,2089){\rotatebox{-270}{\makebox(0,0){\strut{}$f_y(\Delta)$}}}%
      \put(2489,140){\makebox(0,0){\strut{}$\Delta$ [keV]}}%
    }%
    \gplgaddtomacro\gplfronttext{%
    }%
    \gplgaddtomacro\gplbacktext{%
      \csname LTb\endcsname%
      \put(600,640){\makebox(0,0)[r]{\strut{}$10^{-5}$}}%
      \put(600,1220){\makebox(0,0)[r]{\strut{}$10^{-4}$}}%
      \put(600,1800){\makebox(0,0)[r]{\strut{}$10^{-3}$}}%
      \put(600,2379){\makebox(0,0)[r]{\strut{}$10^{-2}$}}%
      \put(600,2959){\makebox(0,0)[r]{\strut{}$10^{-1}$}}%
      \put(600,3539){\makebox(0,0)[r]{\strut{}$10^{0}$}}%
      \put(1363,440){\makebox(0,0){\strut{} 0}}%
      \put(2650,440){\makebox(0,0){\strut{} 200}}%
      \put(3937,440){\makebox(0,0){\strut{} 400}}%
      \put(20,2089){\rotatebox{-270}{\makebox(0,0){\strut{}$f_y(\Delta)$}}}%
      \put(2489,140){\makebox(0,0){\strut{}$\Delta$ [keV]}}%
    }%
    \gplgaddtomacro\gplfronttext{%
    }%
    \gplgaddtomacro\gplbacktext{%
      \csname LTb\endcsname%
      \put(600,640){\makebox(0,0)[r]{\strut{}$10^{-5}$}}%
      \put(600,1220){\makebox(0,0)[r]{\strut{}$10^{-4}$}}%
      \put(600,1800){\makebox(0,0)[r]{\strut{}$10^{-3}$}}%
      \put(600,2379){\makebox(0,0)[r]{\strut{}$10^{-2}$}}%
      \put(600,2959){\makebox(0,0)[r]{\strut{}$10^{-1}$}}%
      \put(600,3539){\makebox(0,0)[r]{\strut{}$10^{0}$}}%
      \put(1363,440){\makebox(0,0){\strut{} 0}}%
      \put(2650,440){\makebox(0,0){\strut{} 200}}%
      \put(3937,440){\makebox(0,0){\strut{} 400}}%
      \put(20,2089){\rotatebox{-270}{\makebox(0,0){\strut{}$f_y(\Delta)$}}}%
      \put(2489,140){\makebox(0,0){\strut{}$\Delta$ [keV]}}%
    }%
    \gplgaddtomacro\gplfronttext{%
    }%
    \gplbacktext
    \put(0,0){\includegraphics{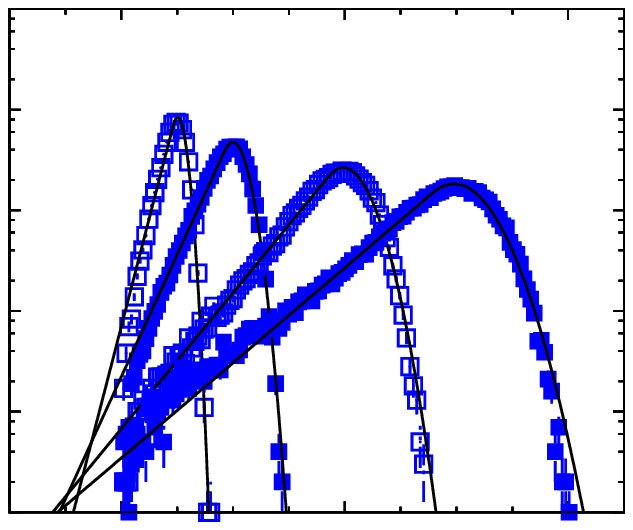}}%
    \gplfronttext
  \end{picture}%
\endgroup

%% file: bg_10.00_all_2.tex
\begingroup
  \makeatletter
  \providecommand\color[2][]{%
    \GenericError{(gnuplot) \space\space\space\@spaces}{%
      Package color not loaded in conjunction with
      terminal option `colourtext'%
    }{See the gnuplot documentation for explanation.%
    }{Either use 'blacktext' in gnuplot or load the package
      color.sty in LaTeX.}%
    \renewcommand\color[2][]{}%
  }%
  \providecommand\includegraphics[2][]{%
    \GenericError{(gnuplot) \space\space\space\@spaces}{%
      Package graphicx or graphics not loaded%
    }{See the gnuplot documentation for explanation.%
    }{The gnuplot epslatex terminal needs graphicx.sty or graphics.sty.}%
    \renewcommand\includegraphics[2][]{}%
  }%
  \providecommand\rotatebox[2]{#2}%
  \@ifundefined{ifGPcolor}{%
    \newif\ifGPcolor
    \GPcolortrue
  }{}%
  \@ifundefined{ifGPblacktext}{%
    \newif\ifGPblacktext
    \GPblacktexttrue
  }{}%
  \let\gplgaddtomacro\g@addto@macro
  \gdef\gplbacktext{}%
  \gdef\gplfronttext{}%
  \makeatother
  \ifGPblacktext
    \def\colorrgb#1{}%
    \def\colorgray#1{}%
  \else
    \ifGPcolor
      \def\colorrgb#1{\color[rgb]{#1}}%
      \def\colorgray#1{\color[gray]{#1}}%
      \expandafter\def\csname LTw\endcsname{\color{white}}%
      \expandafter\def\csname LTb\endcsname{\color{black}}%
      \expandafter\def\csname LTa\endcsname{\color{black}}%
      \expandafter\def\csname LT0\endcsname{\color[rgb]{1,0,0}}%
      \expandafter\def\csname LT1\endcsname{\color[rgb]{0,1,0}}%
      \expandafter\def\csname LT2\endcsname{\color[rgb]{0,0,1}}%
      \expandafter\def\csname LT3\endcsname{\color[rgb]{1,0,1}}%
      \expandafter\def\csname LT4\endcsname{\color[rgb]{0,1,1}}%
      \expandafter\def\csname LT5\endcsname{\color[rgb]{1,1,0}}%
      \expandafter\def\csname LT6\endcsname{\color[rgb]{0,0,0}}%
      \expandafter\def\csname LT7\endcsname{\color[rgb]{1,0.3,0}}%
      \expandafter\def\csname LT8\endcsname{\color[rgb]{0.5,0.5,0.5}}%
    \else
      \def\colorrgb#1{\color{black}}%
      \def\colorgray#1{\color[gray]{#1}}%
      \expandafter\def\csname LTw\endcsname{\color{white}}%
      \expandafter\def\csname LTb\endcsname{\color{black}}%
      \expandafter\def\csname LTa\endcsname{\color{black}}%
      \expandafter\def\csname LT0\endcsname{\color{black}}%
      \expandafter\def\csname LT1\endcsname{\color{black}}%
      \expandafter\def\csname LT2\endcsname{\color{black}}%
      \expandafter\def\csname LT3\endcsname{\color{black}}%
      \expandafter\def\csname LT4\endcsname{\color{black}}%
      \expandafter\def\csname LT5\endcsname{\color{black}}%
      \expandafter\def\csname LT6\endcsname{\color{black}}%
      \expandafter\def\csname LT7\endcsname{\color{black}}%
      \expandafter\def\csname LT8\endcsname{\color{black}}%
    \fi
  \fi
  \setlength{\unitlength}{0.0500bp}%
  \begin{picture}(4500.00,3780.00)%
    \gplgaddtomacro\gplbacktext{%
      \csname LTb\endcsname%
      \put(600,640){\makebox(0,0)[r]{\strut{}$10^{-5}$}}%
      \put(600,1220){\makebox(0,0)[r]{\strut{}$10^{-4}$}}%
      \put(600,1800){\makebox(0,0)[r]{\strut{}$10^{-3}$}}%
      \put(600,2379){\makebox(0,0)[r]{\strut{}$10^{-2}$}}%
      \put(600,2959){\makebox(0,0)[r]{\strut{}$10^{-1}$}}%
      \put(600,3539){\makebox(0,0)[r]{\strut{}$10^{0}$}}%
      \put(1363,440){\makebox(0,0){\strut{} 0}}%
      \put(2650,440){\makebox(0,0){\strut{} 200}}%
      \put(3937,440){\makebox(0,0){\strut{} 400}}%
      \put(20,2089){\rotatebox{-270}{\makebox(0,0){\strut{}$f_y(\Delta)$}}}%
      \put(2489,140){\makebox(0,0){\strut{}$\Delta$ [keV]}}%
      \put(4082,3249){\makebox(0,0)[r]{\strut{}$\beta\gamma =$ 10.00}}%
    }%
    \gplgaddtomacro\gplfronttext{%
    }%
    \gplgaddtomacro\gplbacktext{%
      \csname LTb\endcsname%
      \put(600,640){\makebox(0,0)[r]{\strut{}$10^{-5}$}}%
      \put(600,1220){\makebox(0,0)[r]{\strut{}$10^{-4}$}}%
      \put(600,1800){\makebox(0,0)[r]{\strut{}$10^{-3}$}}%
      \put(600,2379){\makebox(0,0)[r]{\strut{}$10^{-2}$}}%
      \put(600,2959){\makebox(0,0)[r]{\strut{}$10^{-1}$}}%
      \put(600,3539){\makebox(0,0)[r]{\strut{}$10^{0}$}}%
      \put(1363,440){\makebox(0,0){\strut{} 0}}%
      \put(2650,440){\makebox(0,0){\strut{} 200}}%
      \put(3937,440){\makebox(0,0){\strut{} 400}}%
      \put(20,2089){\rotatebox{-270}{\makebox(0,0){\strut{}$f_y(\Delta)$}}}%
      \put(2489,140){\makebox(0,0){\strut{}$\Delta$ [keV]}}%
    }%
    \gplgaddtomacro\gplfronttext{%
    }%
    \gplgaddtomacro\gplbacktext{%
      \csname LTb\endcsname%
      \put(600,640){\makebox(0,0)[r]{\strut{}$10^{-5}$}}%
      \put(600,1220){\makebox(0,0)[r]{\strut{}$10^{-4}$}}%
      \put(600,1800){\makebox(0,0)[r]{\strut{}$10^{-3}$}}%
      \put(600,2379){\makebox(0,0)[r]{\strut{}$10^{-2}$}}%
      \put(600,2959){\makebox(0,0)[r]{\strut{}$10^{-1}$}}%
      \put(600,3539){\makebox(0,0)[r]{\strut{}$10^{0}$}}%
      \put(1363,440){\makebox(0,0){\strut{} 0}}%
      \put(2650,440){\makebox(0,0){\strut{} 200}}%
      \put(3937,440){\makebox(0,0){\strut{} 400}}%
      \put(20,2089){\rotatebox{-270}{\makebox(0,0){\strut{}$f_y(\Delta)$}}}%
      \put(2489,140){\makebox(0,0){\strut{}$\Delta$ [keV]}}%
    }%
    \gplgaddtomacro\gplfronttext{%
    }%
    \gplgaddtomacro\gplbacktext{%
      \csname LTb\endcsname%
      \put(600,640){\makebox(0,0)[r]{\strut{}$10^{-5}$}}%
      \put(600,1220){\makebox(0,0)[r]{\strut{}$10^{-4}$}}%
      \put(600,1800){\makebox(0,0)[r]{\strut{}$10^{-3}$}}%
      \put(600,2379){\makebox(0,0)[r]{\strut{}$10^{-2}$}}%
      \put(600,2959){\makebox(0,0)[r]{\strut{}$10^{-1}$}}%
      \put(600,3539){\makebox(0,0)[r]{\strut{}$10^{0}$}}%
      \put(1363,440){\makebox(0,0){\strut{} 0}}%
      \put(2650,440){\makebox(0,0){\strut{} 200}}%
      \put(3937,440){\makebox(0,0){\strut{} 400}}%
      \put(20,2089){\rotatebox{-270}{\makebox(0,0){\strut{}$f_y(\Delta)$}}}%
      \put(2489,140){\makebox(0,0){\strut{}$\Delta$ [keV]}}%
    }%
    \gplgaddtomacro\gplfronttext{%
    }%
    \gplbacktext
    \put(0,0){\includegraphics{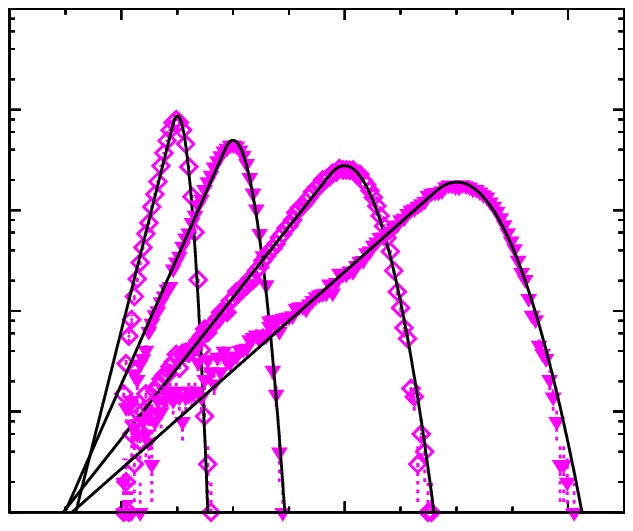}}%
    \gplfronttext
  \end{picture}%
\endgroup

%% file: sllp_fit_2.tex
\begingroup
  \makeatletter
  \providecommand\color[2][]{%
    \GenericError{(gnuplot) \space\space\space\@spaces}{%
      Package color not loaded in conjunction with
      terminal option `colourtext'%
    }{See the gnuplot documentation for explanation.%
    }{Either use 'blacktext' in gnuplot or load the package
      color.sty in LaTeX.}%
    \renewcommand\color[2][]{}%
  }%
  \providecommand\includegraphics[2][]{%
    \GenericError{(gnuplot) \space\space\space\@spaces}{%
      Package graphicx or graphics not loaded%
    }{See the gnuplot documentation for explanation.%
    }{The gnuplot epslatex terminal needs graphicx.sty or graphics.sty.}%
    \renewcommand\includegraphics[2][]{}%
  }%
  \providecommand\rotatebox[2]{#2}%
  \@ifundefined{ifGPcolor}{%
    \newif\ifGPcolor
    \GPcolortrue
  }{}%
  \@ifundefined{ifGPblacktext}{%
    \newif\ifGPblacktext
    \GPblacktexttrue
  }{}%
  \let\gplgaddtomacro\g@addto@macro
  \gdef\gplbacktext{}%
  \gdef\gplfronttext{}%
  \makeatother
  \ifGPblacktext
    \def\colorrgb#1{}%
    \def\colorgray#1{}%
  \else
    \ifGPcolor
      \def\colorrgb#1{\color[rgb]{#1}}%
      \def\colorgray#1{\color[gray]{#1}}%
      \expandafter\def\csname LTw\endcsname{\color{white}}%
      \expandafter\def\csname LTb\endcsname{\color{black}}%
      \expandafter\def\csname LTa\endcsname{\color{black}}%
      \expandafter\def\csname LT0\endcsname{\color[rgb]{1,0,0}}%
      \expandafter\def\csname LT1\endcsname{\color[rgb]{0,1,0}}%
      \expandafter\def\csname LT2\endcsname{\color[rgb]{0,0,1}}%
      \expandafter\def\csname LT3\endcsname{\color[rgb]{1,0,1}}%
      \expandafter\def\csname LT4\endcsname{\color[rgb]{0,1,1}}%
      \expandafter\def\csname LT5\endcsname{\color[rgb]{1,1,0}}%
      \expandafter\def\csname LT6\endcsname{\color[rgb]{0,0,0}}%
      \expandafter\def\csname LT7\endcsname{\color[rgb]{1,0.3,0}}%
      \expandafter\def\csname LT8\endcsname{\color[rgb]{0.5,0.5,0.5}}%
    \else
      \def\colorrgb#1{\color{black}}%
      \def\colorgray#1{\color[gray]{#1}}%
      \expandafter\def\csname LTw\endcsname{\color{white}}%
      \expandafter\def\csname LTb\endcsname{\color{black}}%
      \expandafter\def\csname LTa\endcsname{\color{black}}%
      \expandafter\def\csname LT0\endcsname{\color{black}}%
      \expandafter\def\csname LT1\endcsname{\color{black}}%
      \expandafter\def\csname LT2\endcsname{\color{black}}%
      \expandafter\def\csname LT3\endcsname{\color{black}}%
      \expandafter\def\csname LT4\endcsname{\color{black}}%
      \expandafter\def\csname LT5\endcsname{\color{black}}%
      \expandafter\def\csname LT6\endcsname{\color{black}}%
      \expandafter\def\csname LT7\endcsname{\color{black}}%
      \expandafter\def\csname LT8\endcsname{\color{black}}%
    \fi
  \fi
  \setlength{\unitlength}{0.0500bp}%
  \begin{picture}(4500.00,3780.00)%
    \gplgaddtomacro\gplbacktext{%
      \csname LTb\endcsname%
      \put(600,640){\makebox(0,0)[r]{\strut{} 0}}%
      \put(600,962){\makebox(0,0)[r]{\strut{} 5}}%
      \put(600,1284){\makebox(0,0)[r]{\strut{} 10}}%
      \put(600,1606){\makebox(0,0)[r]{\strut{} 15}}%
      \put(600,1928){\makebox(0,0)[r]{\strut{} 20}}%
      \put(600,2251){\makebox(0,0)[r]{\strut{} 25}}%
      \put(600,2573){\makebox(0,0)[r]{\strut{} 30}}%
      \put(600,2895){\makebox(0,0)[r]{\strut{} 35}}%
      \put(600,3217){\makebox(0,0)[r]{\strut{} 40}}%
      \put(600,3539){\makebox(0,0)[r]{\strut{} 45}}%
      \put(720,440){\makebox(0,0){\strut{} 0}}%
      \put(1162,440){\makebox(0,0){\strut{} 50}}%
      \put(1605,440){\makebox(0,0){\strut{} 100}}%
      \put(2047,440){\makebox(0,0){\strut{} 150}}%
      \put(2490,440){\makebox(0,0){\strut{} 200}}%
      \put(2932,440){\makebox(0,0){\strut{} 250}}%
      \put(3374,440){\makebox(0,0){\strut{} 300}}%
      \put(3817,440){\makebox(0,0){\strut{} 350}}%
      \put(4259,440){\makebox(0,0){\strut{} 400}}%
      \put(140,2089){\rotatebox{-270}{\makebox(0,0){\strut{}$\sigma_\Delta$ [keV]}}}%
      \put(2489,140){\makebox(0,0){\strut{}$y$ [keV]}}%
      \put(897,3249){\makebox(0,0)[l]{\strut{}$\sigma_0 \approx 2.1$ keV}}%
      \put(897,2959){\makebox(0,0)[l]{\strut{}$b \approx 0.094$}}%
    }%
    \gplgaddtomacro\gplfronttext{%
      \csname LTb\endcsname%
      \put(3299,1403){\makebox(0,0)[l]{\strut{}$\beta\gamma = 0.56$}}%
      \csname LTb\endcsname%
      \put(3299,1203){\makebox(0,0)[l]{\strut{}$\beta\gamma = 1.00$}}%
      \csname LTb\endcsname%
      \put(3299,1003){\makebox(0,0)[l]{\strut{}$\beta\gamma = 3.16$}}%
      \csname LTb\endcsname%
      \put(3299,803){\makebox(0,0)[l]{\strut{}$\beta\gamma = 10.0$}}%
    }%
    \gplbacktext
    \put(0,0){\includegraphics{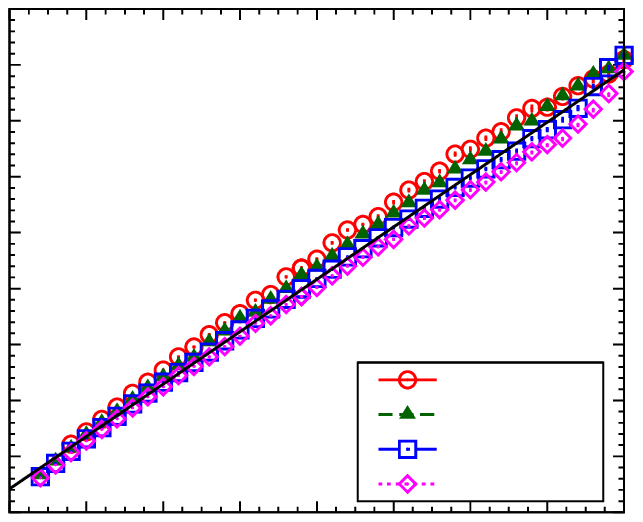}}%
    \gplfronttext
  \end{picture}%
\endgroup

%% file: exponent.tex
\begingroup
  \makeatletter
  \providecommand\color[2][]{%
    \GenericError{(gnuplot) \space\space\space\@spaces}{%
      Package color not loaded in conjunction with
      terminal option `colourtext'%
    }{See the gnuplot documentation for explanation.%
    }{Either use 'blacktext' in gnuplot or load the package
      color.sty in LaTeX.}%
    \renewcommand\color[2][]{}%
  }%
  \providecommand\includegraphics[2][]{%
    \GenericError{(gnuplot) \space\space\space\@spaces}{%
      Package graphicx or graphics not loaded%
    }{See the gnuplot documentation for explanation.%
    }{The gnuplot epslatex terminal needs graphicx.sty or graphics.sty.}%
    \renewcommand\includegraphics[2][]{}%
  }%
  \providecommand\rotatebox[2]{#2}%
  \@ifundefined{ifGPcolor}{%
    \newif\ifGPcolor
    \GPcolortrue
  }{}%
  \@ifundefined{ifGPblacktext}{%
    \newif\ifGPblacktext
    \GPblacktexttrue
  }{}%
  \let\gplgaddtomacro\g@addto@macro
  \gdef\gplbacktext{}%
  \gdef\gplfronttext{}%
  \makeatother
  \ifGPblacktext
    \def\colorrgb#1{}%
    \def\colorgray#1{}%
  \else
    \ifGPcolor
      \def\colorrgb#1{\color[rgb]{#1}}%
      \def\colorgray#1{\color[gray]{#1}}%
      \expandafter\def\csname LTw\endcsname{\color{white}}%
      \expandafter\def\csname LTb\endcsname{\color{black}}%
      \expandafter\def\csname LTa\endcsname{\color{black}}%
      \expandafter\def\csname LT0\endcsname{\color[rgb]{1,0,0}}%
      \expandafter\def\csname LT1\endcsname{\color[rgb]{0,1,0}}%
      \expandafter\def\csname LT2\endcsname{\color[rgb]{0,0,1}}%
      \expandafter\def\csname LT3\endcsname{\color[rgb]{1,0,1}}%
      \expandafter\def\csname LT4\endcsname{\color[rgb]{0,1,1}}%
      \expandafter\def\csname LT5\endcsname{\color[rgb]{1,1,0}}%
      \expandafter\def\csname LT6\endcsname{\color[rgb]{0,0,0}}%
      \expandafter\def\csname LT7\endcsname{\color[rgb]{1,0.3,0}}%
      \expandafter\def\csname LT8\endcsname{\color[rgb]{0.5,0.5,0.5}}%
    \else
      \def\colorrgb#1{\color{black}}%
      \def\colorgray#1{\color[gray]{#1}}%
      \expandafter\def\csname LTw\endcsname{\color{white}}%
      \expandafter\def\csname LTb\endcsname{\color{black}}%
      \expandafter\def\csname LTa\endcsname{\color{black}}%
      \expandafter\def\csname LT0\endcsname{\color{black}}%
      \expandafter\def\csname LT1\endcsname{\color{black}}%
      \expandafter\def\csname LT2\endcsname{\color{black}}%
      \expandafter\def\csname LT3\endcsname{\color{black}}%
      \expandafter\def\csname LT4\endcsname{\color{black}}%
      \expandafter\def\csname LT5\endcsname{\color{black}}%
      \expandafter\def\csname LT6\endcsname{\color{black}}%
      \expandafter\def\csname LT7\endcsname{\color{black}}%
      \expandafter\def\csname LT8\endcsname{\color{black}}%
    \fi
  \fi
  \setlength{\unitlength}{0.0500bp}%
  \begin{picture}(4500.00,3780.00)%
    \gplgaddtomacro\gplbacktext{%
      \csname LTb\endcsname%
      \put(600,640){\makebox(0,0)[r]{\strut{}$10^{-3}$}}%
      \put(600,1606){\makebox(0,0)[r]{\strut{}$10^{-2}$}}%
      \put(600,2573){\makebox(0,0)[r]{\strut{}$10^{-1}$}}%
      \put(600,3539){\makebox(0,0)[r]{\strut{}$10^{0}$}}%
      \put(720,440){\makebox(0,0){\strut{} 0}}%
      \put(1162,440){\makebox(0,0){\strut{} 50}}%
      \put(1605,440){\makebox(0,0){\strut{} 100}}%
      \put(2047,440){\makebox(0,0){\strut{} 150}}%
      \put(2490,440){\makebox(0,0){\strut{} 200}}%
      \put(2932,440){\makebox(0,0){\strut{} 250}}%
      \put(3374,440){\makebox(0,0){\strut{} 300}}%
      \put(3817,440){\makebox(0,0){\strut{} 350}}%
      \put(4259,440){\makebox(0,0){\strut{} 400}}%
      \put(380,2089){\rotatebox{-270}{\makebox(0,0){\strut{}C(y)}}}%
      \put(2489,140){\makebox(0,0){\strut{}y [keV]}}%
    }%
    \gplgaddtomacro\gplfronttext{%
      \csname LTb\endcsname%
      \put(3299,3376){\makebox(0,0)[l]{\strut{}$\beta\gamma =$ 0.56}}%
      \csname LTb\endcsname%
      \put(3299,3176){\makebox(0,0)[l]{\strut{}$\beta\gamma =$ 1.00}}%
      \csname LTb\endcsname%
      \put(3299,2976){\makebox(0,0)[l]{\strut{}$\beta\gamma =$ 3.16}}%
      \csname LTb\endcsname%
      \put(3299,2776){\makebox(0,0)[l]{\strut{}$\beta\gamma =$ 10.0}}%
    }%
    \gplgaddtomacro\gplbacktext{%
      \csname LTb\endcsname%
      \put(600,640){\makebox(0,0)[r]{\strut{}$10^{-3}$}}%
      \put(600,1606){\makebox(0,0)[r]{\strut{}$10^{-2}$}}%
      \put(600,2573){\makebox(0,0)[r]{\strut{}$10^{-1}$}}%
      \put(600,3539){\makebox(0,0)[r]{\strut{}$10^{0}$}}%
      \put(720,440){\makebox(0,0){\strut{} 0}}%
      \put(1162,440){\makebox(0,0){\strut{} 50}}%
      \put(1605,440){\makebox(0,0){\strut{} 100}}%
      \put(2047,440){\makebox(0,0){\strut{} 150}}%
      \put(2490,440){\makebox(0,0){\strut{} 200}}%
      \put(2932,440){\makebox(0,0){\strut{} 250}}%
      \put(3374,440){\makebox(0,0){\strut{} 300}}%
      \put(3817,440){\makebox(0,0){\strut{} 350}}%
      \put(4259,440){\makebox(0,0){\strut{} 400}}%
      \put(380,2089){\rotatebox{-270}{\makebox(0,0){\strut{}C(y)}}}%
      \put(2489,140){\makebox(0,0){\strut{}y [keV]}}%
    }%
    \gplgaddtomacro\gplfronttext{%
    }%
    \gplgaddtomacro\gplbacktext{%
      \csname LTb\endcsname%
      \put(600,640){\makebox(0,0)[r]{\strut{}$10^{-3}$}}%
      \put(600,1606){\makebox(0,0)[r]{\strut{}$10^{-2}$}}%
      \put(600,2573){\makebox(0,0)[r]{\strut{}$10^{-1}$}}%
      \put(600,3539){\makebox(0,0)[r]{\strut{}$10^{0}$}}%
      \put(720,440){\makebox(0,0){\strut{} 0}}%
      \put(1162,440){\makebox(0,0){\strut{} 50}}%
      \put(1605,440){\makebox(0,0){\strut{} 100}}%
      \put(2047,440){\makebox(0,0){\strut{} 150}}%
      \put(2490,440){\makebox(0,0){\strut{} 200}}%
      \put(2932,440){\makebox(0,0){\strut{} 250}}%
      \put(3374,440){\makebox(0,0){\strut{} 300}}%
      \put(3817,440){\makebox(0,0){\strut{} 350}}%
      \put(4259,440){\makebox(0,0){\strut{} 400}}%
      \put(380,2089){\rotatebox{-270}{\makebox(0,0){\strut{}C(y)}}}%
      \put(2489,140){\makebox(0,0){\strut{}y [keV]}}%
    }%
    \gplgaddtomacro\gplfronttext{%
    }%
    \gplgaddtomacro\gplbacktext{%
      \csname LTb\endcsname%
      \put(600,640){\makebox(0,0)[r]{\strut{}$10^{-3}$}}%
      \put(600,1606){\makebox(0,0)[r]{\strut{}$10^{-2}$}}%
      \put(600,2573){\makebox(0,0)[r]{\strut{}$10^{-1}$}}%
      \put(600,3539){\makebox(0,0)[r]{\strut{}$10^{0}$}}%
      \put(720,440){\makebox(0,0){\strut{} 0}}%
      \put(1162,440){\makebox(0,0){\strut{} 50}}%
      \put(1605,440){\makebox(0,0){\strut{} 100}}%
      \put(2047,440){\makebox(0,0){\strut{} 150}}%
      \put(2490,440){\makebox(0,0){\strut{} 200}}%
      \put(2932,440){\makebox(0,0){\strut{} 250}}%
      \put(3374,440){\makebox(0,0){\strut{} 300}}%
      \put(3817,440){\makebox(0,0){\strut{} 350}}%
      \put(4259,440){\makebox(0,0){\strut{} 400}}%
      \put(380,2089){\rotatebox{-270}{\makebox(0,0){\strut{}C(y)}}}%
      \put(2489,140){\makebox(0,0){\strut{}y [keV]}}%
    }%
    \gplgaddtomacro\gplfronttext{%
    }%
    \gplgaddtomacro\gplbacktext{%
      \csname LTb\endcsname%
      \put(600,640){\makebox(0,0)[r]{\strut{}$10^{-3}$}}%
      \put(600,1606){\makebox(0,0)[r]{\strut{}$10^{-2}$}}%
      \put(600,2573){\makebox(0,0)[r]{\strut{}$10^{-1}$}}%
      \put(600,3539){\makebox(0,0)[r]{\strut{}$10^{0}$}}%
      \put(720,440){\makebox(0,0){\strut{} 0}}%
      \put(1162,440){\makebox(0,0){\strut{} 50}}%
      \put(1605,440){\makebox(0,0){\strut{} 100}}%
      \put(2047,440){\makebox(0,0){\strut{} 150}}%
      \put(2490,440){\makebox(0,0){\strut{} 200}}%
      \put(2932,440){\makebox(0,0){\strut{} 250}}%
      \put(3374,440){\makebox(0,0){\strut{} 300}}%
      \put(3817,440){\makebox(0,0){\strut{} 350}}%
      \put(4259,440){\makebox(0,0){\strut{} 400}}%
      \put(380,2089){\rotatebox{-270}{\makebox(0,0){\strut{}C(y)}}}%
      \put(2489,140){\makebox(0,0){\strut{}y [keV]}}%
    }%
    \gplgaddtomacro\gplfronttext{%
    }%
    \gplbacktext
    \put(0,0){\includegraphics{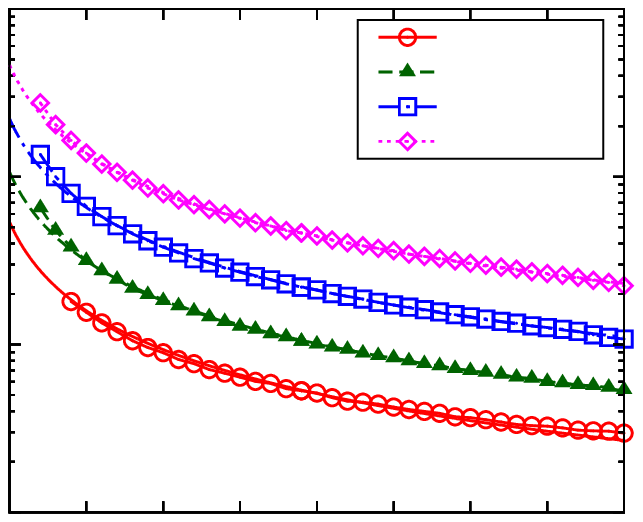}}%
    \gplfronttext
  \end{picture}%
\endgroup

%% file: threshold.tex
\begingroup
  \makeatletter
  \providecommand\color[2][]{%
    \GenericError{(gnuplot) \space\space\space\@spaces}{%
      Package color not loaded in conjunction with
      terminal option `colourtext'%
    }{See the gnuplot documentation for explanation.%
    }{Either use 'blacktext' in gnuplot or load the package
      color.sty in LaTeX.}%
    \renewcommand\color[2][]{}%
  }%
  \providecommand\includegraphics[2][]{%
    \GenericError{(gnuplot) \space\space\space\@spaces}{%
      Package graphicx or graphics not loaded%
    }{See the gnuplot documentation for explanation.%
    }{The gnuplot epslatex terminal needs graphicx.sty or graphics.sty.}%
    \renewcommand\includegraphics[2][]{}%
  }%
  \providecommand\rotatebox[2]{#2}%
  \@ifundefined{ifGPcolor}{%
    \newif\ifGPcolor
    \GPcolortrue
  }{}%
  \@ifundefined{ifGPblacktext}{%
    \newif\ifGPblacktext
    \GPblacktexttrue
  }{}%
  \let\gplgaddtomacro\g@addto@macro
  \gdef\gplbacktext{}%
  \gdef\gplfronttext{}%
  \makeatother
  \ifGPblacktext
    \def\colorrgb#1{}%
    \def\colorgray#1{}%
  \else
    \ifGPcolor
      \def\colorrgb#1{\color[rgb]{#1}}%
      \def\colorgray#1{\color[gray]{#1}}%
      \expandafter\def\csname LTw\endcsname{\color{white}}%
      \expandafter\def\csname LTb\endcsname{\color{black}}%
      \expandafter\def\csname LTa\endcsname{\color{black}}%
      \expandafter\def\csname LT0\endcsname{\color[rgb]{1,0,0}}%
      \expandafter\def\csname LT1\endcsname{\color[rgb]{0,1,0}}%
      \expandafter\def\csname LT2\endcsname{\color[rgb]{0,0,1}}%
      \expandafter\def\csname LT3\endcsname{\color[rgb]{1,0,1}}%
      \expandafter\def\csname LT4\endcsname{\color[rgb]{0,1,1}}%
      \expandafter\def\csname LT5\endcsname{\color[rgb]{1,1,0}}%
      \expandafter\def\csname LT6\endcsname{\color[rgb]{0,0,0}}%
      \expandafter\def\csname LT7\endcsname{\color[rgb]{1,0.3,0}}%
      \expandafter\def\csname LT8\endcsname{\color[rgb]{0.5,0.5,0.5}}%
    \else
      \def\colorrgb#1{\color{black}}%
      \def\colorgray#1{\color[gray]{#1}}%
      \expandafter\def\csname LTw\endcsname{\color{white}}%
      \expandafter\def\csname LTb\endcsname{\color{black}}%
      \expandafter\def\csname LTa\endcsname{\color{black}}%
      \expandafter\def\csname LT0\endcsname{\color{black}}%
      \expandafter\def\csname LT1\endcsname{\color{black}}%
      \expandafter\def\csname LT2\endcsname{\color{black}}%
      \expandafter\def\csname LT3\endcsname{\color{black}}%
      \expandafter\def\csname LT4\endcsname{\color{black}}%
      \expandafter\def\csname LT5\endcsname{\color{black}}%
      \expandafter\def\csname LT6\endcsname{\color{black}}%
      \expandafter\def\csname LT7\endcsname{\color{black}}%
      \expandafter\def\csname LT8\endcsname{\color{black}}%
    \fi
  \fi
  \setlength{\unitlength}{0.0500bp}%
  \begin{picture}(4500.00,3780.00)%
    \gplgaddtomacro\gplbacktext{%
      \csname LTb\endcsname%
      \put(600,640){\makebox(0,0)[r]{\strut{}$10^{-5}$}}%
      \put(600,1220){\makebox(0,0)[r]{\strut{}$10^{-4}$}}%
      \put(600,1800){\makebox(0,0)[r]{\strut{}$10^{-3}$}}%
      \put(600,2379){\makebox(0,0)[r]{\strut{}$10^{-2}$}}%
      \put(600,2959){\makebox(0,0)[r]{\strut{}$10^{-1}$}}%
      \put(600,3539){\makebox(0,0)[r]{\strut{}$10^{0}$}}%
      \put(720,440){\makebox(0,0){\strut{}-5}}%
      \put(1226,440){\makebox(0,0){\strut{} 0}}%
      \put(1731,440){\makebox(0,0){\strut{} 5}}%
      \put(2237,440){\makebox(0,0){\strut{} 10}}%
      \put(2742,440){\makebox(0,0){\strut{} 15}}%
      \put(3248,440){\makebox(0,0){\strut{} 20}}%
      \put(3753,440){\makebox(0,0){\strut{} 25}}%
      \put(4259,440){\makebox(0,0){\strut{} 30}}%
      \put(20,2089){\rotatebox{-270}{\makebox(0,0){\strut{}$f_{y<t}(\Delta)$}}}%
      \put(2489,140){\makebox(0,0){\strut{}$\Delta$ [keV]}}%
    }%
    \gplgaddtomacro\gplfronttext{%
      \csname LTb\endcsname%
      \put(1414,1243){\makebox(0,0)[l]{\strut{}$f_{y < 5~\mathrm{keV}}$}}%
      \csname LTb\endcsname%
      \put(1414,1043){\makebox(0,0)[l]{\strut{}$f_{y < 9~\mathrm{keV}}$}}%
      \csname LTb\endcsname%
      \put(1414,843){\makebox(0,0)[l]{\strut{}$f_{y < 13~\mathrm{keV}}$}}%
    }%
    \gplbacktext
    \put(0,0){\includegraphics{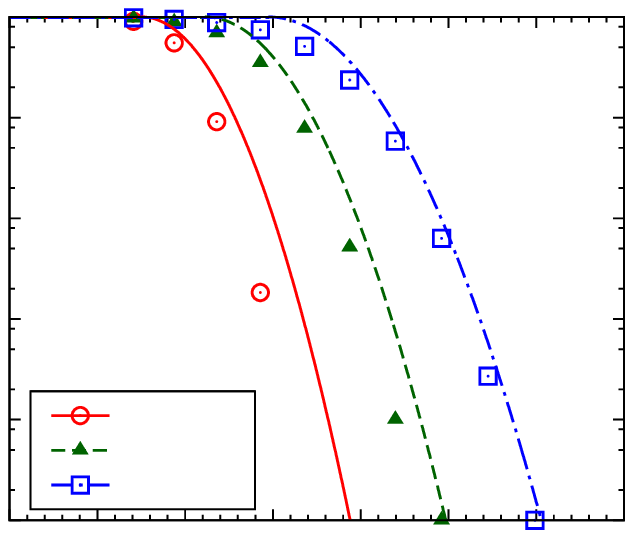}}%
    \gplfronttext
  \end{picture}%
\endgroup

%% file: saturation.tex
\begingroup
  \makeatletter
  \providecommand\color[2][]{%
    \GenericError{(gnuplot) \space\space\space\@spaces}{%
      Package color not loaded in conjunction with
      terminal option `colourtext'%
    }{See the gnuplot documentation for explanation.%
    }{Either use 'blacktext' in gnuplot or load the package
      color.sty in LaTeX.}%
    \renewcommand\color[2][]{}%
  }%
  \providecommand\includegraphics[2][]{%
    \GenericError{(gnuplot) \space\space\space\@spaces}{%
      Package graphicx or graphics not loaded%
    }{See the gnuplot documentation for explanation.%
    }{The gnuplot epslatex terminal needs graphicx.sty or graphics.sty.}%
    \renewcommand\includegraphics[2][]{}%
  }%
  \providecommand\rotatebox[2]{#2}%
  \@ifundefined{ifGPcolor}{%
    \newif\ifGPcolor
    \GPcolortrue
  }{}%
  \@ifundefined{ifGPblacktext}{%
    \newif\ifGPblacktext
    \GPblacktexttrue
  }{}%
  \let\gplgaddtomacro\g@addto@macro
  \gdef\gplbacktext{}%
  \gdef\gplfronttext{}%
  \makeatother
  \ifGPblacktext
    \def\colorrgb#1{}%
    \def\colorgray#1{}%
  \else
    \ifGPcolor
      \def\colorrgb#1{\color[rgb]{#1}}%
      \def\colorgray#1{\color[gray]{#1}}%
      \expandafter\def\csname LTw\endcsname{\color{white}}%
      \expandafter\def\csname LTb\endcsname{\color{black}}%
      \expandafter\def\csname LTa\endcsname{\color{black}}%
      \expandafter\def\csname LT0\endcsname{\color[rgb]{1,0,0}}%
      \expandafter\def\csname LT1\endcsname{\color[rgb]{0,1,0}}%
      \expandafter\def\csname LT2\endcsname{\color[rgb]{0,0,1}}%
      \expandafter\def\csname LT3\endcsname{\color[rgb]{1,0,1}}%
      \expandafter\def\csname LT4\endcsname{\color[rgb]{0,1,1}}%
      \expandafter\def\csname LT5\endcsname{\color[rgb]{1,1,0}}%
      \expandafter\def\csname LT6\endcsname{\color[rgb]{0,0,0}}%
      \expandafter\def\csname LT7\endcsname{\color[rgb]{1,0.3,0}}%
      \expandafter\def\csname LT8\endcsname{\color[rgb]{0.5,0.5,0.5}}%
    \else
      \def\colorrgb#1{\color{black}}%
      \def\colorgray#1{\color[gray]{#1}}%
      \expandafter\def\csname LTw\endcsname{\color{white}}%
      \expandafter\def\csname LTb\endcsname{\color{black}}%
      \expandafter\def\csname LTa\endcsname{\color{black}}%
      \expandafter\def\csname LT0\endcsname{\color{black}}%
      \expandafter\def\csname LT1\endcsname{\color{black}}%
      \expandafter\def\csname LT2\endcsname{\color{black}}%
      \expandafter\def\csname LT3\endcsname{\color{black}}%
      \expandafter\def\csname LT4\endcsname{\color{black}}%
      \expandafter\def\csname LT5\endcsname{\color{black}}%
      \expandafter\def\csname LT6\endcsname{\color{black}}%
      \expandafter\def\csname LT7\endcsname{\color{black}}%
      \expandafter\def\csname LT8\endcsname{\color{black}}%
    \fi
  \fi
  \setlength{\unitlength}{0.0500bp}%
  \begin{picture}(4500.00,3780.00)%
    \gplgaddtomacro\gplbacktext{%
      \csname LTb\endcsname%
      \put(600,640){\makebox(0,0)[r]{\strut{}$10^{-5}$}}%
      \put(600,1220){\makebox(0,0)[r]{\strut{}$10^{-4}$}}%
      \put(600,1800){\makebox(0,0)[r]{\strut{}$10^{-3}$}}%
      \put(600,2379){\makebox(0,0)[r]{\strut{}$10^{-2}$}}%
      \put(600,2959){\makebox(0,0)[r]{\strut{}$10^{-1}$}}%
      \put(600,3539){\makebox(0,0)[r]{\strut{}$10^{0}$}}%
      \put(992,440){\makebox(0,0){\strut{} 0}}%
      \put(1673,440){\makebox(0,0){\strut{} 50}}%
      \put(2353,440){\makebox(0,0){\strut{} 100}}%
      \put(3034,440){\makebox(0,0){\strut{} 150}}%
      \put(3715,440){\makebox(0,0){\strut{} 200}}%
      \put(20,2089){\rotatebox{-270}{\makebox(0,0){\strut{}$f_{y>t}(\Delta)$}}}%
      \put(2489,140){\makebox(0,0){\strut{}$\Delta$ [keV]}}%
    }%
    \gplgaddtomacro\gplfronttext{%
      \csname LTb\endcsname%
      \put(3299,1243){\makebox(0,0)[l]{\strut{}$f_{y >  80~\mathrm{keV}}$}}%
      \csname LTb\endcsname%
      \put(3299,1043){\makebox(0,0)[l]{\strut{}$f_{y > 135~\mathrm{keV}}$}}%
      \csname LTb\endcsname%
      \put(3299,843){\makebox(0,0)[l]{\strut{}$f_{y > 180~\mathrm{keV}}$}}%
    }%
    \gplbacktext
    \put(0,0){\includegraphics{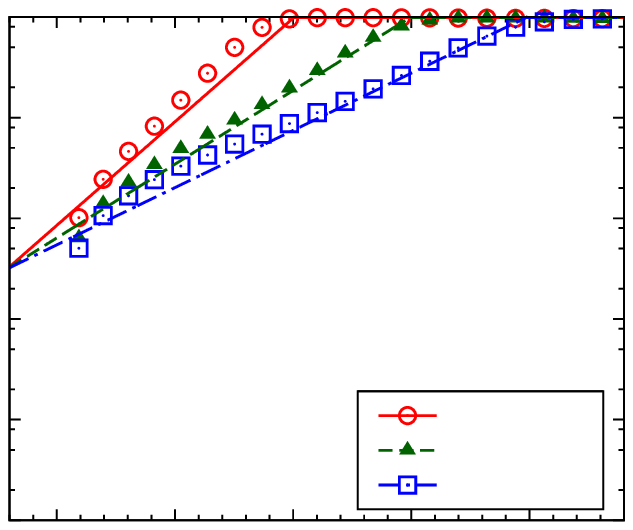}}%
    \gplfronttext
  \end{picture}%
\endgroup

%% file: drawing2.tex
\begingroup
  \makeatletter
  \providecommand\color[2][]{%
    \GenericError{(gnuplot) \space\space\space\@spaces}{%
      Package color not loaded in conjunction with
      terminal option `colourtext'%
    }{See the gnuplot documentation for explanation.%
    }{Either use 'blacktext' in gnuplot or load the package
      color.sty in LaTeX.}%
    \renewcommand\color[2][]{}%
  }%
  \providecommand\includegraphics[2][]{%
    \GenericError{(gnuplot) \space\space\space\@spaces}{%
      Package graphicx or graphics not loaded%
    }{See the gnuplot documentation for explanation.%
    }{The gnuplot epslatex terminal needs graphicx.sty or graphics.sty.}%
    \renewcommand\includegraphics[2][]{}%
  }%
  \providecommand\rotatebox[2]{#2}%
  \@ifundefined{ifGPcolor}{%
    \newif\ifGPcolor
    \GPcolortrue
  }{}%
  \@ifundefined{ifGPblacktext}{%
    \newif\ifGPblacktext
    \GPblacktexttrue
  }{}%
  \let\gplgaddtomacro\g@addto@macro
  \gdef\gplbacktext{}%
  \gdef\gplfronttext{}%
  \makeatother
  \ifGPblacktext
    \def\colorrgb#1{}%
    \def\colorgray#1{}%
  \else
    \ifGPcolor
      \def\colorrgb#1{\color[rgb]{#1}}%
      \def\colorgray#1{\color[gray]{#1}}%
      \expandafter\def\csname LTw\endcsname{\color{white}}%
      \expandafter\def\csname LTb\endcsname{\color{black}}%
      \expandafter\def\csname LTa\endcsname{\color{black}}%
      \expandafter\def\csname LT0\endcsname{\color[rgb]{1,0,0}}%
      \expandafter\def\csname LT1\endcsname{\color[rgb]{0,1,0}}%
      \expandafter\def\csname LT2\endcsname{\color[rgb]{0,0,1}}%
      \expandafter\def\csname LT3\endcsname{\color[rgb]{1,0,1}}%
      \expandafter\def\csname LT4\endcsname{\color[rgb]{0,1,1}}%
      \expandafter\def\csname LT5\endcsname{\color[rgb]{1,1,0}}%
      \expandafter\def\csname LT6\endcsname{\color[rgb]{0,0,0}}%
      \expandafter\def\csname LT7\endcsname{\color[rgb]{1,0.3,0}}%
      \expandafter\def\csname LT8\endcsname{\color[rgb]{0.5,0.5,0.5}}%
    \else
      \def\colorrgb#1{\color{black}}%
      \def\colorgray#1{\color[gray]{#1}}%
      \expandafter\def\csname LTw\endcsname{\color{white}}%
      \expandafter\def\csname LTb\endcsname{\color{black}}%
      \expandafter\def\csname LTa\endcsname{\color{black}}%
      \expandafter\def\csname LT0\endcsname{\color{black}}%
      \expandafter\def\csname LT1\endcsname{\color{black}}%
      \expandafter\def\csname LT2\endcsname{\color{black}}%
      \expandafter\def\csname LT3\endcsname{\color{black}}%
      \expandafter\def\csname LT4\endcsname{\color{black}}%
      \expandafter\def\csname LT5\endcsname{\color{black}}%
      \expandafter\def\csname LT6\endcsname{\color{black}}%
      \expandafter\def\csname LT7\endcsname{\color{black}}%
      \expandafter\def\csname LT8\endcsname{\color{black}}%
    \fi
  \fi
  \setlength{\unitlength}{0.0500bp}%
  \begin{picture}(4500.00,3780.00)%
    \gplgaddtomacro\gplbacktext{%
    }%
    \gplgaddtomacro\gplfronttext{%
      \csname LT2\endcsname%
      \put(1190,1302){\makebox(0,0){\strut{}below}}%
      \put(1190,1134){\makebox(0,0){\strut{}thresh}}%
      \put(1974,1554){\makebox(0,0){\strut{}$\vec{\lambda}$}}%
      \csname LTb\endcsname%
      \put(1750,1890){\makebox(0,0){\strut{}27}}%
      \put(1750,1050){\makebox(0,0){\strut{}38}}%
      \put(2309,1890){\makebox(0,0){\strut{}44}}%
      \put(2869,2729){\makebox(0,0){\strut{}30}}%
      \put(3429,2729){\makebox(0,0){\strut{}12}}%
      \csname LT2\endcsname%
      \put(2309,2981){\makebox(0,0){\strut{}below}}%
      \put(2309,2813){\makebox(0,0){\strut{}thresh}}%
      \csname LT0\endcsname%
      \put(3429,2981){\makebox(0,0){\strut{}outlier}}%
    }%
    \gplbacktext
    \put(0,0){\includegraphics{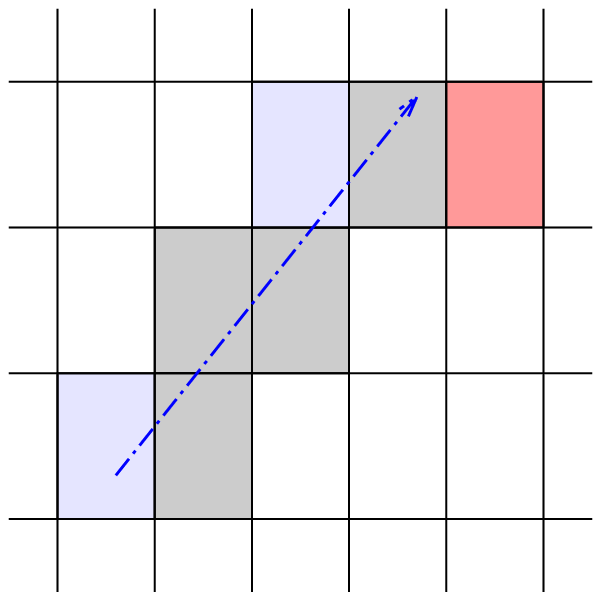}}%
    \gplfronttext
  \end{picture}%
\endgroup

%% file: pathLength_drawing.tex
\begingroup
  \makeatletter
  \providecommand\color[2][]{%
    \GenericError{(gnuplot) \space\space\space\@spaces}{%
      Package color not loaded in conjunction with
      terminal option `colourtext'%
    }{See the gnuplot documentation for explanation.%
    }{Either use 'blacktext' in gnuplot or load the package
      color.sty in LaTeX.}%
    \renewcommand\color[2][]{}%
  }%
  \providecommand\includegraphics[2][]{%
    \GenericError{(gnuplot) \space\space\space\@spaces}{%
      Package graphicx or graphics not loaded%
    }{See the gnuplot documentation for explanation.%
    }{The gnuplot epslatex terminal needs graphicx.sty or graphics.sty.}%
    \renewcommand\includegraphics[2][]{}%
  }%
  \providecommand\rotatebox[2]{#2}%
  \@ifundefined{ifGPcolor}{%
    \newif\ifGPcolor
    \GPcolortrue
  }{}%
  \@ifundefined{ifGPblacktext}{%
    \newif\ifGPblacktext
    \GPblacktexttrue
  }{}%
  \let\gplgaddtomacro\g@addto@macro
  \gdef\gplbacktext{}%
  \gdef\gplfronttext{}%
  \makeatother
  \ifGPblacktext
    \def\colorrgb#1{}%
    \def\colorgray#1{}%
  \else
    \ifGPcolor
      \def\colorrgb#1{\color[rgb]{#1}}%
      \def\colorgray#1{\color[gray]{#1}}%
      \expandafter\def\csname LTw\endcsname{\color{white}}%
      \expandafter\def\csname LTb\endcsname{\color{black}}%
      \expandafter\def\csname LTa\endcsname{\color{black}}%
      \expandafter\def\csname LT0\endcsname{\color[rgb]{1,0,0}}%
      \expandafter\def\csname LT1\endcsname{\color[rgb]{0,1,0}}%
      \expandafter\def\csname LT2\endcsname{\color[rgb]{0,0,1}}%
      \expandafter\def\csname LT3\endcsname{\color[rgb]{1,0,1}}%
      \expandafter\def\csname LT4\endcsname{\color[rgb]{0,1,1}}%
      \expandafter\def\csname LT5\endcsname{\color[rgb]{1,1,0}}%
      \expandafter\def\csname LT6\endcsname{\color[rgb]{0,0,0}}%
      \expandafter\def\csname LT7\endcsname{\color[rgb]{1,0.3,0}}%
      \expandafter\def\csname LT8\endcsname{\color[rgb]{0.5,0.5,0.5}}%
    \else
      \def\colorrgb#1{\color{black}}%
      \def\colorgray#1{\color[gray]{#1}}%
      \expandafter\def\csname LTw\endcsname{\color{white}}%
      \expandafter\def\csname LTb\endcsname{\color{black}}%
      \expandafter\def\csname LTa\endcsname{\color{black}}%
      \expandafter\def\csname LT0\endcsname{\color{black}}%
      \expandafter\def\csname LT1\endcsname{\color{black}}%
      \expandafter\def\csname LT2\endcsname{\color{black}}%
      \expandafter\def\csname LT3\endcsname{\color{black}}%
      \expandafter\def\csname LT4\endcsname{\color{black}}%
      \expandafter\def\csname LT5\endcsname{\color{black}}%
      \expandafter\def\csname LT6\endcsname{\color{black}}%
      \expandafter\def\csname LT7\endcsname{\color{black}}%
      \expandafter\def\csname LT8\endcsname{\color{black}}%
    \fi
  \fi
  \setlength{\unitlength}{0.0500bp}%
  \begin{picture}(4090.00,3436.00)%
    \gplgaddtomacro\gplbacktext{%
      \csname LTb\endcsname%
      \put(511,511){\makebox(0,0){\strut{}a}}%
      \put(1329,511){\makebox(0,0){\strut{}b}}%
      \put(3578,1022){\makebox(0,0){\strut{}c}}%
      \csname LT0\endcsname%
      \put(1431,1226){\makebox(0,0)[l]{\strut{}$\vec{\lambda_a}$}}%
      \put(2045,1226){\makebox(0,0)[l]{\strut{}$\vec{\lambda_b}$}}%
      \put(2965,1226){\makebox(0,0)[l]{\strut{}$\vec{\lambda_c}$}}%
    }%
    \gplgaddtomacro\gplfronttext{%
    }%
    \gplbacktext
    \put(0,0){\includegraphics{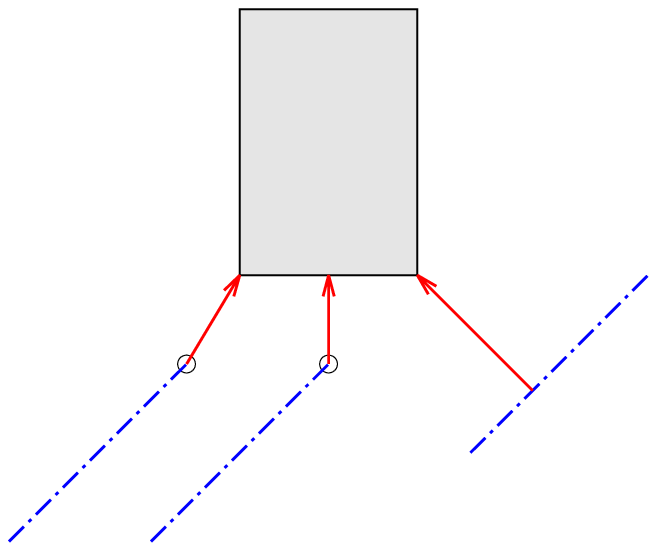}}%
    \gplfronttext
  \end{picture}%
\endgroup

%% file: drawing.tex
\begingroup
  \makeatletter
  \providecommand\color[2][]{%
    \GenericError{(gnuplot) \space\space\space\@spaces}{%
      Package color not loaded in conjunction with
      terminal option `colourtext'%
    }{See the gnuplot documentation for explanation.%
    }{Either use 'blacktext' in gnuplot or load the package
      color.sty in LaTeX.}%
    \renewcommand\color[2][]{}%
  }%
  \providecommand\includegraphics[2][]{%
    \GenericError{(gnuplot) \space\space\space\@spaces}{%
      Package graphicx or graphics not loaded%
    }{See the gnuplot documentation for explanation.%
    }{The gnuplot epslatex terminal needs graphicx.sty or graphics.sty.}%
    \renewcommand\includegraphics[2][]{}%
  }%
  \providecommand\rotatebox[2]{#2}%
  \@ifundefined{ifGPcolor}{%
    \newif\ifGPcolor
    \GPcolortrue
  }{}%
  \@ifundefined{ifGPblacktext}{%
    \newif\ifGPblacktext
    \GPblacktexttrue
  }{}%
  \let\gplgaddtomacro\g@addto@macro
  \gdef\gplbacktext{}%
  \gdef\gplfronttext{}%
  \makeatother
  \ifGPblacktext
    \def\colorrgb#1{}%
    \def\colorgray#1{}%
  \else
    \ifGPcolor
      \def\colorrgb#1{\color[rgb]{#1}}%
      \def\colorgray#1{\color[gray]{#1}}%
      \expandafter\def\csname LTw\endcsname{\color{white}}%
      \expandafter\def\csname LTb\endcsname{\color{black}}%
      \expandafter\def\csname LTa\endcsname{\color{black}}%
      \expandafter\def\csname LT0\endcsname{\color[rgb]{1,0,0}}%
      \expandafter\def\csname LT1\endcsname{\color[rgb]{0,1,0}}%
      \expandafter\def\csname LT2\endcsname{\color[rgb]{0,0,1}}%
      \expandafter\def\csname LT3\endcsname{\color[rgb]{1,0,1}}%
      \expandafter\def\csname LT4\endcsname{\color[rgb]{0,1,1}}%
      \expandafter\def\csname LT5\endcsname{\color[rgb]{1,1,0}}%
      \expandafter\def\csname LT6\endcsname{\color[rgb]{0,0,0}}%
      \expandafter\def\csname LT7\endcsname{\color[rgb]{1,0.3,0}}%
      \expandafter\def\csname LT8\endcsname{\color[rgb]{0.5,0.5,0.5}}%
    \else
      \def\colorrgb#1{\color{black}}%
      \def\colorgray#1{\color[gray]{#1}}%
      \expandafter\def\csname LTw\endcsname{\color{white}}%
      \expandafter\def\csname LTb\endcsname{\color{black}}%
      \expandafter\def\csname LTa\endcsname{\color{black}}%
      \expandafter\def\csname LT0\endcsname{\color{black}}%
      \expandafter\def\csname LT1\endcsname{\color{black}}%
      \expandafter\def\csname LT2\endcsname{\color{black}}%
      \expandafter\def\csname LT3\endcsname{\color{black}}%
      \expandafter\def\csname LT4\endcsname{\color{black}}%
      \expandafter\def\csname LT5\endcsname{\color{black}}%
      \expandafter\def\csname LT6\endcsname{\color{black}}%
      \expandafter\def\csname LT7\endcsname{\color{black}}%
      \expandafter\def\csname LT8\endcsname{\color{black}}%
    \fi
  \fi
  \setlength{\unitlength}{0.0500bp}%
  \begin{picture}(4500.00,1890.00)%
    \gplgaddtomacro\gplbacktext{%
      \csname LTb\endcsname%
      \put(120,1727){\makebox(0,0)[r]{\strut{}a}}%
      \put(1684,1727){\makebox(0,0)[r]{\strut{}b}}%
      \put(3248,1727){\makebox(0,0)[r]{\strut{}c}}%
    }%
    \gplgaddtomacro\gplfronttext{%
      \csname LTb\endcsname%
      \put(3665,580){\makebox(0,0){\strut{}$\vec{n_1}$}}%
      \put(3926,1206){\makebox(0,0)[l]{\strut{}$\vec{n_2}$}}%
      \csname LT0\endcsname%
      \put(3561,1101){\makebox(0,0)[l]{\strut{}$\vec{\lambda}$}}%
    }%
    \gplbacktext
    \put(0,0){\includegraphics{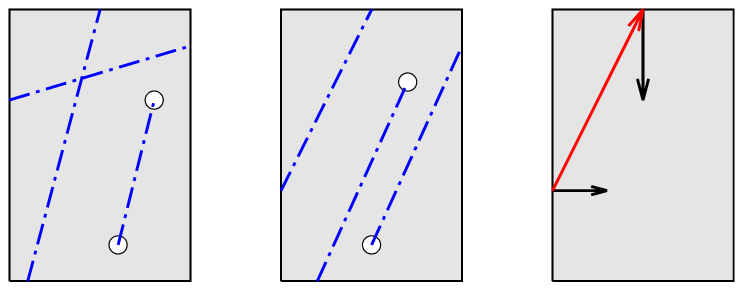}}%
    \gplfronttext
  \end{picture}%
\endgroup

%% file: residual_perp.tex
\begingroup
  \makeatletter
  \providecommand\color[2][]{%
    \GenericError{(gnuplot) \space\space\space\@spaces}{%
      Package color not loaded in conjunction with
      terminal option `colourtext'%
    }{See the gnuplot documentation for explanation.%
    }{Either use 'blacktext' in gnuplot or load the package
      color.sty in LaTeX.}%
    \renewcommand\color[2][]{}%
  }%
  \providecommand\includegraphics[2][]{%
    \GenericError{(gnuplot) \space\space\space\@spaces}{%
      Package graphicx or graphics not loaded%
    }{See the gnuplot documentation for explanation.%
    }{The gnuplot epslatex terminal needs graphicx.sty or graphics.sty.}%
    \renewcommand\includegraphics[2][]{}%
  }%
  \providecommand\rotatebox[2]{#2}%
  \@ifundefined{ifGPcolor}{%
    \newif\ifGPcolor
    \GPcolortrue
  }{}%
  \@ifundefined{ifGPblacktext}{%
    \newif\ifGPblacktext
    \GPblacktexttrue
  }{}%
  \let\gplgaddtomacro\g@addto@macro
  \gdef\gplbacktext{}%
  \gdef\gplfronttext{}%
  \makeatother
  \ifGPblacktext
    \def\colorrgb#1{}%
    \def\colorgray#1{}%
  \else
    \ifGPcolor
      \def\colorrgb#1{\color[rgb]{#1}}%
      \def\colorgray#1{\color[gray]{#1}}%
      \expandafter\def\csname LTw\endcsname{\color{white}}%
      \expandafter\def\csname LTb\endcsname{\color{black}}%
      \expandafter\def\csname LTa\endcsname{\color{black}}%
      \expandafter\def\csname LT0\endcsname{\color[rgb]{1,0,0}}%
      \expandafter\def\csname LT1\endcsname{\color[rgb]{0,1,0}}%
      \expandafter\def\csname LT2\endcsname{\color[rgb]{0,0,1}}%
      \expandafter\def\csname LT3\endcsname{\color[rgb]{1,0,1}}%
      \expandafter\def\csname LT4\endcsname{\color[rgb]{0,1,1}}%
      \expandafter\def\csname LT5\endcsname{\color[rgb]{1,1,0}}%
      \expandafter\def\csname LT6\endcsname{\color[rgb]{0,0,0}}%
      \expandafter\def\csname LT7\endcsname{\color[rgb]{1,0.3,0}}%
      \expandafter\def\csname LT8\endcsname{\color[rgb]{0.5,0.5,0.5}}%
    \else
      \def\colorrgb#1{\color{black}}%
      \def\colorgray#1{\color[gray]{#1}}%
      \expandafter\def\csname LTw\endcsname{\color{white}}%
      \expandafter\def\csname LTb\endcsname{\color{black}}%
      \expandafter\def\csname LTa\endcsname{\color{black}}%
      \expandafter\def\csname LT0\endcsname{\color{black}}%
      \expandafter\def\csname LT1\endcsname{\color{black}}%
      \expandafter\def\csname LT2\endcsname{\color{black}}%
      \expandafter\def\csname LT3\endcsname{\color{black}}%
      \expandafter\def\csname LT4\endcsname{\color{black}}%
      \expandafter\def\csname LT5\endcsname{\color{black}}%
      \expandafter\def\csname LT6\endcsname{\color{black}}%
      \expandafter\def\csname LT7\endcsname{\color{black}}%
      \expandafter\def\csname LT8\endcsname{\color{black}}%
    \fi
  \fi
  \setlength{\unitlength}{0.0500bp}%
  \begin{picture}(4500.00,3780.00)%
    \gplgaddtomacro\gplbacktext{%
      \csname LTb\endcsname%
      \put(600,640){\makebox(0,0)[r]{\strut{} 0}}%
      \put(600,1054){\makebox(0,0)[r]{\strut{} 2000}}%
      \put(600,1468){\makebox(0,0)[r]{\strut{} 4000}}%
      \put(600,1882){\makebox(0,0)[r]{\strut{} 6000}}%
      \put(600,2297){\makebox(0,0)[r]{\strut{} 8000}}%
      \put(600,2711){\makebox(0,0)[r]{\strut{} 10000}}%
      \put(600,3125){\makebox(0,0)[r]{\strut{} 12000}}%
      \put(600,3539){\makebox(0,0)[r]{\strut{} 14000}}%
      \put(720,440){\makebox(0,0){\strut{}-100}}%
      \put(1074,440){\makebox(0,0){\strut{}-80}}%
      \put(1428,440){\makebox(0,0){\strut{}-60}}%
      \put(1782,440){\makebox(0,0){\strut{}-40}}%
      \put(2136,440){\makebox(0,0){\strut{}-20}}%
      \put(2490,440){\makebox(0,0){\strut{} 0}}%
      \put(2843,440){\makebox(0,0){\strut{} 20}}%
      \put(3197,440){\makebox(0,0){\strut{} 40}}%
      \put(3551,440){\makebox(0,0){\strut{} 60}}%
      \put(3905,440){\makebox(0,0){\strut{} 80}}%
      \put(4259,440){\makebox(0,0){\strut{} 100}}%
      \put(2489,140){\makebox(0,0){\strut{}$\delta_\perp$ [$\mu$m]}}%
      \put(897,3249){\makebox(0,0)[l]{\strut{}Pixels}}%
    }%
    \gplgaddtomacro\gplfronttext{%
      \csname LTb\endcsname%
      \put(3299,3376){\makebox(0,0)[l]{\strut{}Weighted}}%
      \csname LTb\endcsname%
      \put(3299,3176){\makebox(0,0)[l]{\strut{}First-last}}%
      \csname LTb\endcsname%
      \put(3299,2976){\makebox(0,0)[l]{\strut{}Fitter}}%
    }%
    \gplbacktext
    \put(0,0){\includegraphics{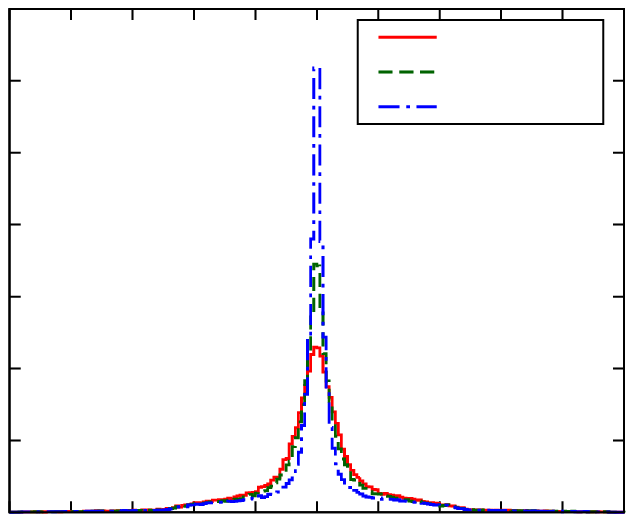}}%
    \gplfronttext
  \end{picture}%
\endgroup

%% file: residual_para.tex
\begingroup
  \makeatletter
  \providecommand\color[2][]{%
    \GenericError{(gnuplot) \space\space\space\@spaces}{%
      Package color not loaded in conjunction with
      terminal option `colourtext'%
    }{See the gnuplot documentation for explanation.%
    }{Either use 'blacktext' in gnuplot or load the package
      color.sty in LaTeX.}%
    \renewcommand\color[2][]{}%
  }%
  \providecommand\includegraphics[2][]{%
    \GenericError{(gnuplot) \space\space\space\@spaces}{%
      Package graphicx or graphics not loaded%
    }{See the gnuplot documentation for explanation.%
    }{The gnuplot epslatex terminal needs graphicx.sty or graphics.sty.}%
    \renewcommand\includegraphics[2][]{}%
  }%
  \providecommand\rotatebox[2]{#2}%
  \@ifundefined{ifGPcolor}{%
    \newif\ifGPcolor
    \GPcolortrue
  }{}%
  \@ifundefined{ifGPblacktext}{%
    \newif\ifGPblacktext
    \GPblacktexttrue
  }{}%
  \let\gplgaddtomacro\g@addto@macro
  \gdef\gplbacktext{}%
  \gdef\gplfronttext{}%
  \makeatother
  \ifGPblacktext
    \def\colorrgb#1{}%
    \def\colorgray#1{}%
  \else
    \ifGPcolor
      \def\colorrgb#1{\color[rgb]{#1}}%
      \def\colorgray#1{\color[gray]{#1}}%
      \expandafter\def\csname LTw\endcsname{\color{white}}%
      \expandafter\def\csname LTb\endcsname{\color{black}}%
      \expandafter\def\csname LTa\endcsname{\color{black}}%
      \expandafter\def\csname LT0\endcsname{\color[rgb]{1,0,0}}%
      \expandafter\def\csname LT1\endcsname{\color[rgb]{0,1,0}}%
      \expandafter\def\csname LT2\endcsname{\color[rgb]{0,0,1}}%
      \expandafter\def\csname LT3\endcsname{\color[rgb]{1,0,1}}%
      \expandafter\def\csname LT4\endcsname{\color[rgb]{0,1,1}}%
      \expandafter\def\csname LT5\endcsname{\color[rgb]{1,1,0}}%
      \expandafter\def\csname LT6\endcsname{\color[rgb]{0,0,0}}%
      \expandafter\def\csname LT7\endcsname{\color[rgb]{1,0.3,0}}%
      \expandafter\def\csname LT8\endcsname{\color[rgb]{0.5,0.5,0.5}}%
    \else
      \def\colorrgb#1{\color{black}}%
      \def\colorgray#1{\color[gray]{#1}}%
      \expandafter\def\csname LTw\endcsname{\color{white}}%
      \expandafter\def\csname LTb\endcsname{\color{black}}%
      \expandafter\def\csname LTa\endcsname{\color{black}}%
      \expandafter\def\csname LT0\endcsname{\color{black}}%
      \expandafter\def\csname LT1\endcsname{\color{black}}%
      \expandafter\def\csname LT2\endcsname{\color{black}}%
      \expandafter\def\csname LT3\endcsname{\color{black}}%
      \expandafter\def\csname LT4\endcsname{\color{black}}%
      \expandafter\def\csname LT5\endcsname{\color{black}}%
      \expandafter\def\csname LT6\endcsname{\color{black}}%
      \expandafter\def\csname LT7\endcsname{\color{black}}%
      \expandafter\def\csname LT8\endcsname{\color{black}}%
    \fi
  \fi
  \setlength{\unitlength}{0.0500bp}%
  \begin{picture}(4500.00,3780.00)%
    \gplgaddtomacro\gplbacktext{%
      \csname LTb\endcsname%
      \put(600,640){\makebox(0,0)[r]{\strut{} 0}}%
      \put(600,1123){\makebox(0,0)[r]{\strut{} 500}}%
      \put(600,1606){\makebox(0,0)[r]{\strut{} 1000}}%
      \put(600,2090){\makebox(0,0)[r]{\strut{} 1500}}%
      \put(600,2573){\makebox(0,0)[r]{\strut{} 2000}}%
      \put(600,3056){\makebox(0,0)[r]{\strut{} 2500}}%
      \put(600,3539){\makebox(0,0)[r]{\strut{} 3000}}%
      \put(720,440){\makebox(0,0){\strut{}-100}}%
      \put(1074,440){\makebox(0,0){\strut{}-80}}%
      \put(1428,440){\makebox(0,0){\strut{}-60}}%
      \put(1782,440){\makebox(0,0){\strut{}-40}}%
      \put(2136,440){\makebox(0,0){\strut{}-20}}%
      \put(2490,440){\makebox(0,0){\strut{} 0}}%
      \put(2843,440){\makebox(0,0){\strut{} 20}}%
      \put(3197,440){\makebox(0,0){\strut{} 40}}%
      \put(3551,440){\makebox(0,0){\strut{} 60}}%
      \put(3905,440){\makebox(0,0){\strut{} 80}}%
      \put(4259,440){\makebox(0,0){\strut{} 100}}%
      \put(2489,140){\makebox(0,0){\strut{}$\delta_\parallel$ [$\mu$m]}}%
      \put(897,3249){\makebox(0,0)[l]{\strut{}Pixels}}%
    }%
    \gplgaddtomacro\gplfronttext{%
      \csname LTb\endcsname%
      \put(3299,3376){\makebox(0,0)[l]{\strut{}Weighted}}%
      \csname LTb\endcsname%
      \put(3299,3176){\makebox(0,0)[l]{\strut{}First-last}}%
      \csname LTb\endcsname%
      \put(3299,2976){\makebox(0,0)[l]{\strut{}Fitter}}%
    }%
    \gplbacktext
    \put(0,0){\includegraphics{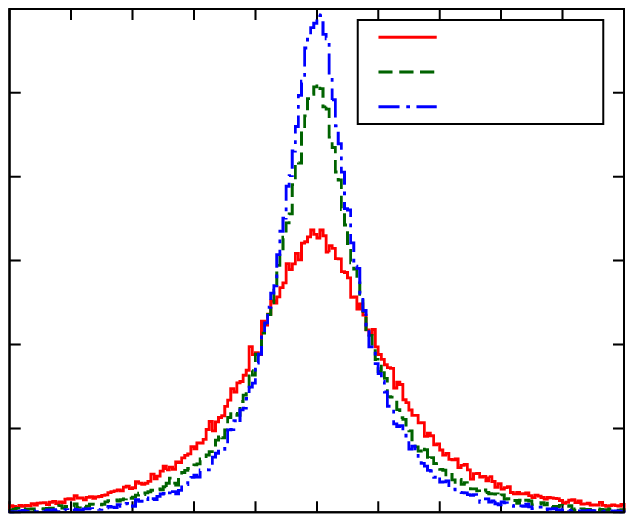}}%
    \gplfronttext
  \end{picture}%
\endgroup

%% file: residual_fit_perp.tex
\begingroup
  \makeatletter
  \providecommand\color[2][]{%
    \GenericError{(gnuplot) \space\space\space\@spaces}{%
      Package color not loaded in conjunction with
      terminal option `colourtext'%
    }{See the gnuplot documentation for explanation.%
    }{Either use 'blacktext' in gnuplot or load the package
      color.sty in LaTeX.}%
    \renewcommand\color[2][]{}%
  }%
  \providecommand\includegraphics[2][]{%
    \GenericError{(gnuplot) \space\space\space\@spaces}{%
      Package graphicx or graphics not loaded%
    }{See the gnuplot documentation for explanation.%
    }{The gnuplot epslatex terminal needs graphicx.sty or graphics.sty.}%
    \renewcommand\includegraphics[2][]{}%
  }%
  \providecommand\rotatebox[2]{#2}%
  \@ifundefined{ifGPcolor}{%
    \newif\ifGPcolor
    \GPcolortrue
  }{}%
  \@ifundefined{ifGPblacktext}{%
    \newif\ifGPblacktext
    \GPblacktexttrue
  }{}%
  \let\gplgaddtomacro\g@addto@macro
  \gdef\gplbacktext{}%
  \gdef\gplfronttext{}%
  \makeatother
  \ifGPblacktext
    \def\colorrgb#1{}%
    \def\colorgray#1{}%
  \else
    \ifGPcolor
      \def\colorrgb#1{\color[rgb]{#1}}%
      \def\colorgray#1{\color[gray]{#1}}%
      \expandafter\def\csname LTw\endcsname{\color{white}}%
      \expandafter\def\csname LTb\endcsname{\color{black}}%
      \expandafter\def\csname LTa\endcsname{\color{black}}%
      \expandafter\def\csname LT0\endcsname{\color[rgb]{1,0,0}}%
      \expandafter\def\csname LT1\endcsname{\color[rgb]{0,1,0}}%
      \expandafter\def\csname LT2\endcsname{\color[rgb]{0,0,1}}%
      \expandafter\def\csname LT3\endcsname{\color[rgb]{1,0,1}}%
      \expandafter\def\csname LT4\endcsname{\color[rgb]{0,1,1}}%
      \expandafter\def\csname LT5\endcsname{\color[rgb]{1,1,0}}%
      \expandafter\def\csname LT6\endcsname{\color[rgb]{0,0,0}}%
      \expandafter\def\csname LT7\endcsname{\color[rgb]{1,0.3,0}}%
      \expandafter\def\csname LT8\endcsname{\color[rgb]{0.5,0.5,0.5}}%
    \else
      \def\colorrgb#1{\color{black}}%
      \def\colorgray#1{\color[gray]{#1}}%
      \expandafter\def\csname LTw\endcsname{\color{white}}%
      \expandafter\def\csname LTb\endcsname{\color{black}}%
      \expandafter\def\csname LTa\endcsname{\color{black}}%
      \expandafter\def\csname LT0\endcsname{\color{black}}%
      \expandafter\def\csname LT1\endcsname{\color{black}}%
      \expandafter\def\csname LT2\endcsname{\color{black}}%
      \expandafter\def\csname LT3\endcsname{\color{black}}%
      \expandafter\def\csname LT4\endcsname{\color{black}}%
      \expandafter\def\csname LT5\endcsname{\color{black}}%
      \expandafter\def\csname LT6\endcsname{\color{black}}%
      \expandafter\def\csname LT7\endcsname{\color{black}}%
      \expandafter\def\csname LT8\endcsname{\color{black}}%
    \fi
  \fi
  \setlength{\unitlength}{0.0500bp}%
  \begin{picture}(4500.00,3780.00)%
    \gplgaddtomacro\gplbacktext{%
      \csname LTb\endcsname%
      \put(600,640){\makebox(0,0)[r]{\strut{} 0}}%
      \put(600,1123){\makebox(0,0)[r]{\strut{} 5}}%
      \put(600,1606){\makebox(0,0)[r]{\strut{} 10}}%
      \put(600,2090){\makebox(0,0)[r]{\strut{} 15}}%
      \put(600,2573){\makebox(0,0)[r]{\strut{} 20}}%
      \put(600,3056){\makebox(0,0)[r]{\strut{} 25}}%
      \put(600,3539){\makebox(0,0)[r]{\strut{} 30}}%
      \put(928,440){\makebox(0,0){\strut{} 2}}%
      \put(1345,440){\makebox(0,0){\strut{} 3}}%
      \put(1761,440){\makebox(0,0){\strut{} 4}}%
      \put(2177,440){\makebox(0,0){\strut{} 5}}%
      \put(2594,440){\makebox(0,0){\strut{} 6}}%
      \put(3010,440){\makebox(0,0){\strut{} 7}}%
      \put(3426,440){\makebox(0,0){\strut{} 8}}%
      \put(3843,440){\makebox(0,0){\strut{} 9}}%
      \put(4259,440){\makebox(0,0){\strut{} 10}}%
      \put(140,2089){\rotatebox{-270}{\makebox(0,0){\strut{}$\sigma_\perp$ [$\mu$m]}}}%
      \put(2489,140){\makebox(0,0){\strut{}$n_{pixel}$}}%
      \put(897,3249){\makebox(0,0)[l]{\strut{}Pixels}}%
    }%
    \gplgaddtomacro\gplfronttext{%
      \csname LTb\endcsname%
      \put(3299,3376){\makebox(0,0)[l]{\strut{}Weighted}}%
      \csname LTb\endcsname%
      \put(3299,3176){\makebox(0,0)[l]{\strut{}First-last}}%
      \csname LTb\endcsname%
      \put(3299,2976){\makebox(0,0)[l]{\strut{}Fitter}}%
    }%
    \gplbacktext
    \put(0,0){\includegraphics{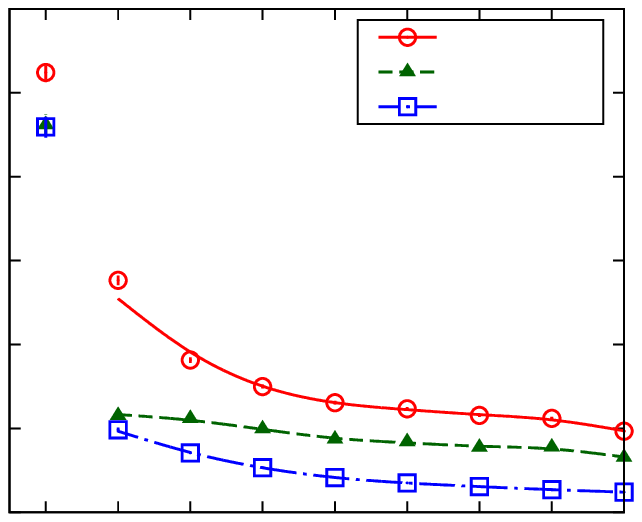}}%
    \gplfronttext
  \end{picture}%
\endgroup

%% file: residual_fit_para.tex
\begingroup
  \makeatletter
  \providecommand\color[2][]{%
    \GenericError{(gnuplot) \space\space\space\@spaces}{%
      Package color not loaded in conjunction with
      terminal option `colourtext'%
    }{See the gnuplot documentation for explanation.%
    }{Either use 'blacktext' in gnuplot or load the package
      color.sty in LaTeX.}%
    \renewcommand\color[2][]{}%
  }%
  \providecommand\includegraphics[2][]{%
    \GenericError{(gnuplot) \space\space\space\@spaces}{%
      Package graphicx or graphics not loaded%
    }{See the gnuplot documentation for explanation.%
    }{The gnuplot epslatex terminal needs graphicx.sty or graphics.sty.}%
    \renewcommand\includegraphics[2][]{}%
  }%
  \providecommand\rotatebox[2]{#2}%
  \@ifundefined{ifGPcolor}{%
    \newif\ifGPcolor
    \GPcolortrue
  }{}%
  \@ifundefined{ifGPblacktext}{%
    \newif\ifGPblacktext
    \GPblacktexttrue
  }{}%
  \let\gplgaddtomacro\g@addto@macro
  \gdef\gplbacktext{}%
  \gdef\gplfronttext{}%
  \makeatother
  \ifGPblacktext
    \def\colorrgb#1{}%
    \def\colorgray#1{}%
  \else
    \ifGPcolor
      \def\colorrgb#1{\color[rgb]{#1}}%
      \def\colorgray#1{\color[gray]{#1}}%
      \expandafter\def\csname LTw\endcsname{\color{white}}%
      \expandafter\def\csname LTb\endcsname{\color{black}}%
      \expandafter\def\csname LTa\endcsname{\color{black}}%
      \expandafter\def\csname LT0\endcsname{\color[rgb]{1,0,0}}%
      \expandafter\def\csname LT1\endcsname{\color[rgb]{0,1,0}}%
      \expandafter\def\csname LT2\endcsname{\color[rgb]{0,0,1}}%
      \expandafter\def\csname LT3\endcsname{\color[rgb]{1,0,1}}%
      \expandafter\def\csname LT4\endcsname{\color[rgb]{0,1,1}}%
      \expandafter\def\csname LT5\endcsname{\color[rgb]{1,1,0}}%
      \expandafter\def\csname LT6\endcsname{\color[rgb]{0,0,0}}%
      \expandafter\def\csname LT7\endcsname{\color[rgb]{1,0.3,0}}%
      \expandafter\def\csname LT8\endcsname{\color[rgb]{0.5,0.5,0.5}}%
    \else
      \def\colorrgb#1{\color{black}}%
      \def\colorgray#1{\color[gray]{#1}}%
      \expandafter\def\csname LTw\endcsname{\color{white}}%
      \expandafter\def\csname LTb\endcsname{\color{black}}%
      \expandafter\def\csname LTa\endcsname{\color{black}}%
      \expandafter\def\csname LT0\endcsname{\color{black}}%
      \expandafter\def\csname LT1\endcsname{\color{black}}%
      \expandafter\def\csname LT2\endcsname{\color{black}}%
      \expandafter\def\csname LT3\endcsname{\color{black}}%
      \expandafter\def\csname LT4\endcsname{\color{black}}%
      \expandafter\def\csname LT5\endcsname{\color{black}}%
      \expandafter\def\csname LT6\endcsname{\color{black}}%
      \expandafter\def\csname LT7\endcsname{\color{black}}%
      \expandafter\def\csname LT8\endcsname{\color{black}}%
    \fi
  \fi
  \setlength{\unitlength}{0.0500bp}%
  \begin{picture}(4500.00,3780.00)%
    \gplgaddtomacro\gplbacktext{%
      \csname LTb\endcsname%
      \put(600,640){\makebox(0,0)[r]{\strut{} 0}}%
      \put(600,1123){\makebox(0,0)[r]{\strut{} 5}}%
      \put(600,1606){\makebox(0,0)[r]{\strut{} 10}}%
      \put(600,2090){\makebox(0,0)[r]{\strut{} 15}}%
      \put(600,2573){\makebox(0,0)[r]{\strut{} 20}}%
      \put(600,3056){\makebox(0,0)[r]{\strut{} 25}}%
      \put(600,3539){\makebox(0,0)[r]{\strut{} 30}}%
      \put(928,440){\makebox(0,0){\strut{} 2}}%
      \put(1345,440){\makebox(0,0){\strut{} 3}}%
      \put(1761,440){\makebox(0,0){\strut{} 4}}%
      \put(2177,440){\makebox(0,0){\strut{} 5}}%
      \put(2594,440){\makebox(0,0){\strut{} 6}}%
      \put(3010,440){\makebox(0,0){\strut{} 7}}%
      \put(3426,440){\makebox(0,0){\strut{} 8}}%
      \put(3843,440){\makebox(0,0){\strut{} 9}}%
      \put(4259,440){\makebox(0,0){\strut{} 10}}%
      \put(140,2089){\rotatebox{-270}{\makebox(0,0){\strut{}$\sigma_\parallel$ [$\mu$m]}}}%
      \put(2489,140){\makebox(0,0){\strut{}$n_{pixel}$}}%
      \put(897,3249){\makebox(0,0)[l]{\strut{}Pixels}}%
    }%
    \gplgaddtomacro\gplfronttext{%
      \csname LTb\endcsname%
      \put(3299,1203){\makebox(0,0)[l]{\strut{}Weighted}}%
      \csname LTb\endcsname%
      \put(3299,1003){\makebox(0,0)[l]{\strut{}First-last}}%
      \csname LTb\endcsname%
      \put(3299,803){\makebox(0,0)[l]{\strut{}Fitter}}%
    }%
    \gplbacktext
    \put(0,0){\includegraphics{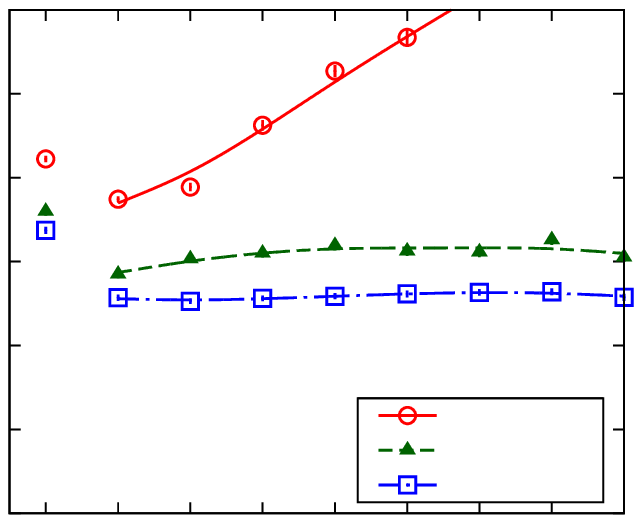}}%
    \gplfronttext
  \end{picture}%
\endgroup

%% file: residual_x_strips.tex
\begingroup
  \makeatletter
  \providecommand\color[2][]{%
    \GenericError{(gnuplot) \space\space\space\@spaces}{%
      Package color not loaded in conjunction with
      terminal option `colourtext'%
    }{See the gnuplot documentation for explanation.%
    }{Either use 'blacktext' in gnuplot or load the package
      color.sty in LaTeX.}%
    \renewcommand\color[2][]{}%
  }%
  \providecommand\includegraphics[2][]{%
    \GenericError{(gnuplot) \space\space\space\@spaces}{%
      Package graphicx or graphics not loaded%
    }{See the gnuplot documentation for explanation.%
    }{The gnuplot epslatex terminal needs graphicx.sty or graphics.sty.}%
    \renewcommand\includegraphics[2][]{}%
  }%
  \providecommand\rotatebox[2]{#2}%
  \@ifundefined{ifGPcolor}{%
    \newif\ifGPcolor
    \GPcolortrue
  }{}%
  \@ifundefined{ifGPblacktext}{%
    \newif\ifGPblacktext
    \GPblacktexttrue
  }{}%
  \let\gplgaddtomacro\g@addto@macro
  \gdef\gplbacktext{}%
  \gdef\gplfronttext{}%
  \makeatother
  \ifGPblacktext
    \def\colorrgb#1{}%
    \def\colorgray#1{}%
  \else
    \ifGPcolor
      \def\colorrgb#1{\color[rgb]{#1}}%
      \def\colorgray#1{\color[gray]{#1}}%
      \expandafter\def\csname LTw\endcsname{\color{white}}%
      \expandafter\def\csname LTb\endcsname{\color{black}}%
      \expandafter\def\csname LTa\endcsname{\color{black}}%
      \expandafter\def\csname LT0\endcsname{\color[rgb]{1,0,0}}%
      \expandafter\def\csname LT1\endcsname{\color[rgb]{0,1,0}}%
      \expandafter\def\csname LT2\endcsname{\color[rgb]{0,0,1}}%
      \expandafter\def\csname LT3\endcsname{\color[rgb]{1,0,1}}%
      \expandafter\def\csname LT4\endcsname{\color[rgb]{0,1,1}}%
      \expandafter\def\csname LT5\endcsname{\color[rgb]{1,1,0}}%
      \expandafter\def\csname LT6\endcsname{\color[rgb]{0,0,0}}%
      \expandafter\def\csname LT7\endcsname{\color[rgb]{1,0.3,0}}%
      \expandafter\def\csname LT8\endcsname{\color[rgb]{0.5,0.5,0.5}}%
    \else
      \def\colorrgb#1{\color{black}}%
      \def\colorgray#1{\color[gray]{#1}}%
      \expandafter\def\csname LTw\endcsname{\color{white}}%
      \expandafter\def\csname LTb\endcsname{\color{black}}%
      \expandafter\def\csname LTa\endcsname{\color{black}}%
      \expandafter\def\csname LT0\endcsname{\color{black}}%
      \expandafter\def\csname LT1\endcsname{\color{black}}%
      \expandafter\def\csname LT2\endcsname{\color{black}}%
      \expandafter\def\csname LT3\endcsname{\color{black}}%
      \expandafter\def\csname LT4\endcsname{\color{black}}%
      \expandafter\def\csname LT5\endcsname{\color{black}}%
      \expandafter\def\csname LT6\endcsname{\color{black}}%
      \expandafter\def\csname LT7\endcsname{\color{black}}%
      \expandafter\def\csname LT8\endcsname{\color{black}}%
    \fi
  \fi
  \setlength{\unitlength}{0.0500bp}%
  \begin{picture}(4500.00,3780.00)%
    \gplgaddtomacro\gplbacktext{%
      \csname LTb\endcsname%
      \put(600,640){\makebox(0,0)[r]{\strut{} 0}}%
      \put(600,1002){\makebox(0,0)[r]{\strut{} 500}}%
      \put(600,1365){\makebox(0,0)[r]{\strut{} 1000}}%
      \put(600,1727){\makebox(0,0)[r]{\strut{} 1500}}%
      \put(600,2090){\makebox(0,0)[r]{\strut{} 2000}}%
      \put(600,2452){\makebox(0,0)[r]{\strut{} 2500}}%
      \put(600,2814){\makebox(0,0)[r]{\strut{} 3000}}%
      \put(600,3177){\makebox(0,0)[r]{\strut{} 3500}}%
      \put(600,3539){\makebox(0,0)[r]{\strut{} 4000}}%
      \put(720,440){\makebox(0,0){\strut{}-100}}%
      \put(1074,440){\makebox(0,0){\strut{}-80}}%
      \put(1428,440){\makebox(0,0){\strut{}-60}}%
      \put(1782,440){\makebox(0,0){\strut{}-40}}%
      \put(2136,440){\makebox(0,0){\strut{}-20}}%
      \put(2490,440){\makebox(0,0){\strut{} 0}}%
      \put(2843,440){\makebox(0,0){\strut{} 20}}%
      \put(3197,440){\makebox(0,0){\strut{} 40}}%
      \put(3551,440){\makebox(0,0){\strut{} 60}}%
      \put(3905,440){\makebox(0,0){\strut{} 80}}%
      \put(4259,440){\makebox(0,0){\strut{} 100}}%
      \put(2489,140){\makebox(0,0){\strut{}$\delta x$ [$\mu$m]}}%
      \put(897,3249){\makebox(0,0)[l]{\strut{}Strips}}%
    }%
    \gplgaddtomacro\gplfronttext{%
      \csname LTb\endcsname%
      \put(3299,3376){\makebox(0,0)[l]{\strut{}Weighted}}%
      \csname LTb\endcsname%
      \put(3299,3176){\makebox(0,0)[l]{\strut{}First-last}}%
      \csname LTb\endcsname%
      \put(3299,2976){\makebox(0,0)[l]{\strut{}Fitter}}%
    }%
    \gplbacktext
    \put(0,0){\includegraphics{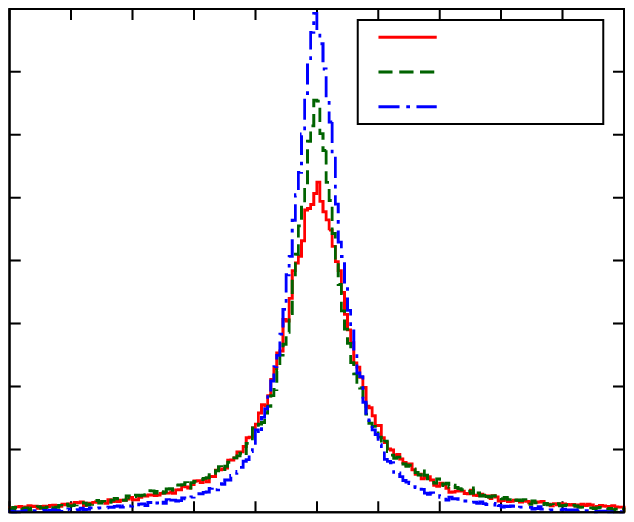}}%
    \gplfronttext
  \end{picture}%
\endgroup

%% file: residual_fit_strips.tex
\begingroup
  \makeatletter
  \providecommand\color[2][]{%
    \GenericError{(gnuplot) \space\space\space\@spaces}{%
      Package color not loaded in conjunction with
      terminal option `colourtext'%
    }{See the gnuplot documentation for explanation.%
    }{Either use 'blacktext' in gnuplot or load the package
      color.sty in LaTeX.}%
    \renewcommand\color[2][]{}%
  }%
  \providecommand\includegraphics[2][]{%
    \GenericError{(gnuplot) \space\space\space\@spaces}{%
      Package graphicx or graphics not loaded%
    }{See the gnuplot documentation for explanation.%
    }{The gnuplot epslatex terminal needs graphicx.sty or graphics.sty.}%
    \renewcommand\includegraphics[2][]{}%
  }%
  \providecommand\rotatebox[2]{#2}%
  \@ifundefined{ifGPcolor}{%
    \newif\ifGPcolor
    \GPcolortrue
  }{}%
  \@ifundefined{ifGPblacktext}{%
    \newif\ifGPblacktext
    \GPblacktexttrue
  }{}%
  \let\gplgaddtomacro\g@addto@macro
  \gdef\gplbacktext{}%
  \gdef\gplfronttext{}%
  \makeatother
  \ifGPblacktext
    \def\colorrgb#1{}%
    \def\colorgray#1{}%
  \else
    \ifGPcolor
      \def\colorrgb#1{\color[rgb]{#1}}%
      \def\colorgray#1{\color[gray]{#1}}%
      \expandafter\def\csname LTw\endcsname{\color{white}}%
      \expandafter\def\csname LTb\endcsname{\color{black}}%
      \expandafter\def\csname LTa\endcsname{\color{black}}%
      \expandafter\def\csname LT0\endcsname{\color[rgb]{1,0,0}}%
      \expandafter\def\csname LT1\endcsname{\color[rgb]{0,1,0}}%
      \expandafter\def\csname LT2\endcsname{\color[rgb]{0,0,1}}%
      \expandafter\def\csname LT3\endcsname{\color[rgb]{1,0,1}}%
      \expandafter\def\csname LT4\endcsname{\color[rgb]{0,1,1}}%
      \expandafter\def\csname LT5\endcsname{\color[rgb]{1,1,0}}%
      \expandafter\def\csname LT6\endcsname{\color[rgb]{0,0,0}}%
      \expandafter\def\csname LT7\endcsname{\color[rgb]{1,0.3,0}}%
      \expandafter\def\csname LT8\endcsname{\color[rgb]{0.5,0.5,0.5}}%
    \else
      \def\colorrgb#1{\color{black}}%
      \def\colorgray#1{\color[gray]{#1}}%
      \expandafter\def\csname LTw\endcsname{\color{white}}%
      \expandafter\def\csname LTb\endcsname{\color{black}}%
      \expandafter\def\csname LTa\endcsname{\color{black}}%
      \expandafter\def\csname LT0\endcsname{\color{black}}%
      \expandafter\def\csname LT1\endcsname{\color{black}}%
      \expandafter\def\csname LT2\endcsname{\color{black}}%
      \expandafter\def\csname LT3\endcsname{\color{black}}%
      \expandafter\def\csname LT4\endcsname{\color{black}}%
      \expandafter\def\csname LT5\endcsname{\color{black}}%
      \expandafter\def\csname LT6\endcsname{\color{black}}%
      \expandafter\def\csname LT7\endcsname{\color{black}}%
      \expandafter\def\csname LT8\endcsname{\color{black}}%
    \fi
  \fi
  \setlength{\unitlength}{0.0500bp}%
  \begin{picture}(4500.00,3780.00)%
    \gplgaddtomacro\gplbacktext{%
      \csname LTb\endcsname%
      \put(600,640){\makebox(0,0)[r]{\strut{} 0}}%
      \put(600,1123){\makebox(0,0)[r]{\strut{} 5}}%
      \put(600,1606){\makebox(0,0)[r]{\strut{} 10}}%
      \put(600,2090){\makebox(0,0)[r]{\strut{} 15}}%
      \put(600,2573){\makebox(0,0)[r]{\strut{} 20}}%
      \put(600,3056){\makebox(0,0)[r]{\strut{} 25}}%
      \put(600,3539){\makebox(0,0)[r]{\strut{} 30}}%
      \put(928,440){\makebox(0,0){\strut{} 2}}%
      \put(1345,440){\makebox(0,0){\strut{} 3}}%
      \put(1761,440){\makebox(0,0){\strut{} 4}}%
      \put(2177,440){\makebox(0,0){\strut{} 5}}%
      \put(2594,440){\makebox(0,0){\strut{} 6}}%
      \put(3010,440){\makebox(0,0){\strut{} 7}}%
      \put(3426,440){\makebox(0,0){\strut{} 8}}%
      \put(3843,440){\makebox(0,0){\strut{} 9}}%
      \put(4259,440){\makebox(0,0){\strut{} 10}}%
      \put(140,2089){\rotatebox{-270}{\makebox(0,0){\strut{}$\sigma_\parallel$ [$\mu$m]}}}%
      \put(2489,140){\makebox(0,0){\strut{}$n_{strip}$}}%
      \put(897,3249){\makebox(0,0)[l]{\strut{}Strips}}%
    }%
    \gplgaddtomacro\gplfronttext{%
      \csname LTb\endcsname%
      \put(3299,3376){\makebox(0,0)[l]{\strut{}Weighted}}%
      \csname LTb\endcsname%
      \put(3299,3176){\makebox(0,0)[l]{\strut{}First-last}}%
      \csname LTb\endcsname%
      \put(3299,2976){\makebox(0,0)[l]{\strut{}Fitter}}%
    }%
    \gplbacktext
    \put(0,0){\includegraphics{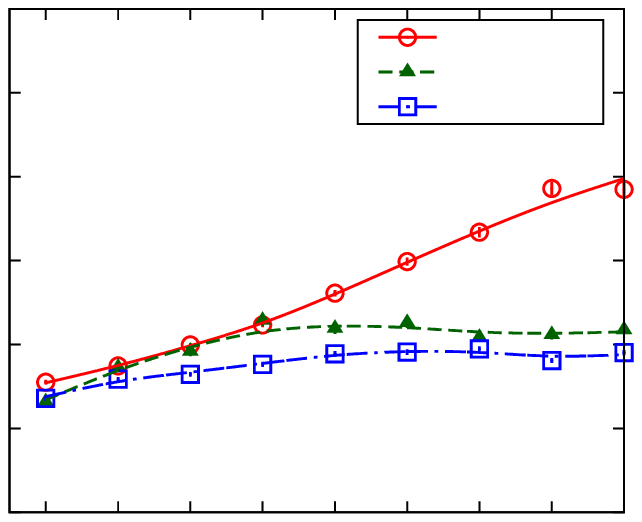}}%
    \gplfronttext
  \end{picture}%
\endgroup

%% file: resolution_x_pixels.tex
\begingroup
  \makeatletter
  \providecommand\color[2][]{%
    \GenericError{(gnuplot) \space\space\space\@spaces}{%
      Package color not loaded in conjunction with
      terminal option `colourtext'%
    }{See the gnuplot documentation for explanation.%
    }{Either use 'blacktext' in gnuplot or load the package
      color.sty in LaTeX.}%
    \renewcommand\color[2][]{}%
  }%
  \providecommand\includegraphics[2][]{%
    \GenericError{(gnuplot) \space\space\space\@spaces}{%
      Package graphicx or graphics not loaded%
    }{See the gnuplot documentation for explanation.%
    }{The gnuplot epslatex terminal needs graphicx.sty or graphics.sty.}%
    \renewcommand\includegraphics[2][]{}%
  }%
  \providecommand\rotatebox[2]{#2}%
  \@ifundefined{ifGPcolor}{%
    \newif\ifGPcolor
    \GPcolortrue
  }{}%
  \@ifundefined{ifGPblacktext}{%
    \newif\ifGPblacktext
    \GPblacktexttrue
  }{}%
  \let\gplgaddtomacro\g@addto@macro
  \gdef\gplbacktext{}%
  \gdef\gplfronttext{}%
  \makeatother
  \ifGPblacktext
    \def\colorrgb#1{}%
    \def\colorgray#1{}%
  \else
    \ifGPcolor
      \def\colorrgb#1{\color[rgb]{#1}}%
      \def\colorgray#1{\color[gray]{#1}}%
      \expandafter\def\csname LTw\endcsname{\color{white}}%
      \expandafter\def\csname LTb\endcsname{\color{black}}%
      \expandafter\def\csname LTa\endcsname{\color{black}}%
      \expandafter\def\csname LT0\endcsname{\color[rgb]{1,0,0}}%
      \expandafter\def\csname LT1\endcsname{\color[rgb]{0,1,0}}%
      \expandafter\def\csname LT2\endcsname{\color[rgb]{0,0,1}}%
      \expandafter\def\csname LT3\endcsname{\color[rgb]{1,0,1}}%
      \expandafter\def\csname LT4\endcsname{\color[rgb]{0,1,1}}%
      \expandafter\def\csname LT5\endcsname{\color[rgb]{1,1,0}}%
      \expandafter\def\csname LT6\endcsname{\color[rgb]{0,0,0}}%
      \expandafter\def\csname LT7\endcsname{\color[rgb]{1,0.3,0}}%
      \expandafter\def\csname LT8\endcsname{\color[rgb]{0.5,0.5,0.5}}%
    \else
      \def\colorrgb#1{\color{black}}%
      \def\colorgray#1{\color[gray]{#1}}%
      \expandafter\def\csname LTw\endcsname{\color{white}}%
      \expandafter\def\csname LTb\endcsname{\color{black}}%
      \expandafter\def\csname LTa\endcsname{\color{black}}%
      \expandafter\def\csname LT0\endcsname{\color{black}}%
      \expandafter\def\csname LT1\endcsname{\color{black}}%
      \expandafter\def\csname LT2\endcsname{\color{black}}%
      \expandafter\def\csname LT3\endcsname{\color{black}}%
      \expandafter\def\csname LT4\endcsname{\color{black}}%
      \expandafter\def\csname LT5\endcsname{\color{black}}%
      \expandafter\def\csname LT6\endcsname{\color{black}}%
      \expandafter\def\csname LT7\endcsname{\color{black}}%
      \expandafter\def\csname LT8\endcsname{\color{black}}%
    \fi
  \fi
  \setlength{\unitlength}{0.0500bp}%
  \begin{picture}(2970.00,3780.00)%
    \gplgaddtomacro\gplbacktext{%
      \csname LTb\endcsname%
      \put(480,640){\makebox(0,0)[r]{\strut{} 0}}%
      \put(480,981){\makebox(0,0)[r]{\strut{} 1000}}%
      \put(480,1322){\makebox(0,0)[r]{\strut{} 2000}}%
      \put(480,1663){\makebox(0,0)[r]{\strut{} 3000}}%
      \put(480,2004){\makebox(0,0)[r]{\strut{} 4000}}%
      \put(480,2345){\makebox(0,0)[r]{\strut{} 5000}}%
      \put(480,2686){\makebox(0,0)[r]{\strut{} 6000}}%
      \put(480,3027){\makebox(0,0)[r]{\strut{} 7000}}%
      \put(480,3368){\makebox(0,0)[r]{\strut{} 8000}}%
      \put(594,440){\makebox(0,0){\strut{}-100}}%
      \put(1188,440){\makebox(0,0){\strut{}-50}}%
      \put(1782,440){\makebox(0,0){\strut{} 0}}%
      \put(2375,440){\makebox(0,0){\strut{} 50}}%
      \put(2969,440){\makebox(0,0){\strut{} 100}}%
      \put(1784,140){\makebox(0,0){\strut{}$\delta x$ [$\mu$m]}}%
      \put(718,3249){\makebox(0,0)[l]{\strut{}Pixels}}%
    }%
    \gplgaddtomacro\gplfronttext{%
      \csname LTb\endcsname%
      \put(2009,3376){\makebox(0,0)[l]{\strut{}Fitter}}%
      \csname LTb\endcsname%
      \put(2009,3176){\makebox(0,0)[l]{\strut{}Predicted}}%
    }%
    \gplbacktext
    \put(0,0){\includegraphics{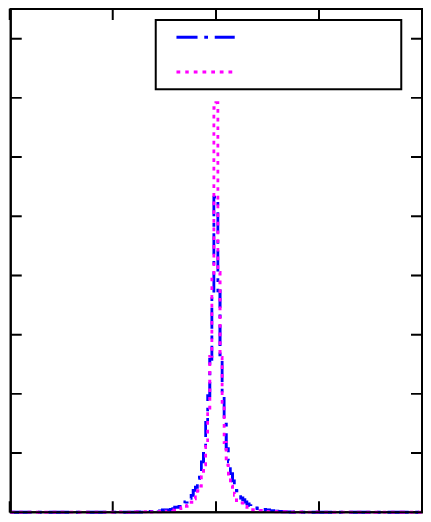}}%
    \gplfronttext
  \end{picture}%
\endgroup

%% file: resolution_y_pixels.tex
\begingroup
  \makeatletter
  \providecommand\color[2][]{%
    \GenericError{(gnuplot) \space\space\space\@spaces}{%
      Package color not loaded in conjunction with
      terminal option `colourtext'%
    }{See the gnuplot documentation for explanation.%
    }{Either use 'blacktext' in gnuplot or load the package
      color.sty in LaTeX.}%
    \renewcommand\color[2][]{}%
  }%
  \providecommand\includegraphics[2][]{%
    \GenericError{(gnuplot) \space\space\space\@spaces}{%
      Package graphicx or graphics not loaded%
    }{See the gnuplot documentation for explanation.%
    }{The gnuplot epslatex terminal needs graphicx.sty or graphics.sty.}%
    \renewcommand\includegraphics[2][]{}%
  }%
  \providecommand\rotatebox[2]{#2}%
  \@ifundefined{ifGPcolor}{%
    \newif\ifGPcolor
    \GPcolortrue
  }{}%
  \@ifundefined{ifGPblacktext}{%
    \newif\ifGPblacktext
    \GPblacktexttrue
  }{}%
  \let\gplgaddtomacro\g@addto@macro
  \gdef\gplbacktext{}%
  \gdef\gplfronttext{}%
  \makeatother
  \ifGPblacktext
    \def\colorrgb#1{}%
    \def\colorgray#1{}%
  \else
    \ifGPcolor
      \def\colorrgb#1{\color[rgb]{#1}}%
      \def\colorgray#1{\color[gray]{#1}}%
      \expandafter\def\csname LTw\endcsname{\color{white}}%
      \expandafter\def\csname LTb\endcsname{\color{black}}%
      \expandafter\def\csname LTa\endcsname{\color{black}}%
      \expandafter\def\csname LT0\endcsname{\color[rgb]{1,0,0}}%
      \expandafter\def\csname LT1\endcsname{\color[rgb]{0,1,0}}%
      \expandafter\def\csname LT2\endcsname{\color[rgb]{0,0,1}}%
      \expandafter\def\csname LT3\endcsname{\color[rgb]{1,0,1}}%
      \expandafter\def\csname LT4\endcsname{\color[rgb]{0,1,1}}%
      \expandafter\def\csname LT5\endcsname{\color[rgb]{1,1,0}}%
      \expandafter\def\csname LT6\endcsname{\color[rgb]{0,0,0}}%
      \expandafter\def\csname LT7\endcsname{\color[rgb]{1,0.3,0}}%
      \expandafter\def\csname LT8\endcsname{\color[rgb]{0.5,0.5,0.5}}%
    \else
      \def\colorrgb#1{\color{black}}%
      \def\colorgray#1{\color[gray]{#1}}%
      \expandafter\def\csname LTw\endcsname{\color{white}}%
      \expandafter\def\csname LTb\endcsname{\color{black}}%
      \expandafter\def\csname LTa\endcsname{\color{black}}%
      \expandafter\def\csname LT0\endcsname{\color{black}}%
      \expandafter\def\csname LT1\endcsname{\color{black}}%
      \expandafter\def\csname LT2\endcsname{\color{black}}%
      \expandafter\def\csname LT3\endcsname{\color{black}}%
      \expandafter\def\csname LT4\endcsname{\color{black}}%
      \expandafter\def\csname LT5\endcsname{\color{black}}%
      \expandafter\def\csname LT6\endcsname{\color{black}}%
      \expandafter\def\csname LT7\endcsname{\color{black}}%
      \expandafter\def\csname LT8\endcsname{\color{black}}%
    \fi
  \fi
  \setlength{\unitlength}{0.0500bp}%
  \begin{picture}(2970.00,3780.00)%
    \gplgaddtomacro\gplbacktext{%
      \csname LTb\endcsname%
      \put(480,640){\makebox(0,0)[r]{\strut{} 0}}%
      \put(480,1220){\makebox(0,0)[r]{\strut{} 500}}%
      \put(480,1800){\makebox(0,0)[r]{\strut{} 1000}}%
      \put(480,2379){\makebox(0,0)[r]{\strut{} 1500}}%
      \put(480,2959){\makebox(0,0)[r]{\strut{} 2000}}%
      \put(480,3539){\makebox(0,0)[r]{\strut{} 2500}}%
      \put(594,440){\makebox(0,0){\strut{}-100}}%
      \put(1188,440){\makebox(0,0){\strut{}-50}}%
      \put(1782,440){\makebox(0,0){\strut{} 0}}%
      \put(2375,440){\makebox(0,0){\strut{} 50}}%
      \put(2969,440){\makebox(0,0){\strut{} 100}}%
      \put(1784,140){\makebox(0,0){\strut{}$\delta y$ [$\mu$m]}}%
      \put(718,3249){\makebox(0,0)[l]{\strut{}Pixels}}%
    }%
    \gplgaddtomacro\gplfronttext{%
      \csname LTb\endcsname%
      \put(2009,3376){\makebox(0,0)[l]{\strut{}Fitter}}%
      \csname LTb\endcsname%
      \put(2009,3176){\makebox(0,0)[l]{\strut{}Predicted}}%
    }%
    \gplbacktext
    \put(0,0){\includegraphics{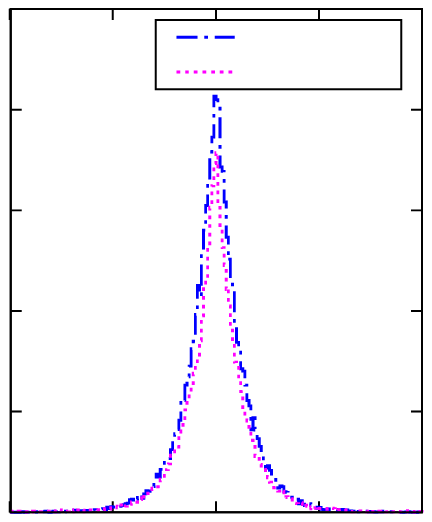}}%
    \gplfronttext
  \end{picture}%
\endgroup

%% file: resolution_x_strips.tex
\begingroup
  \makeatletter
  \providecommand\color[2][]{%
    \GenericError{(gnuplot) \space\space\space\@spaces}{%
      Package color not loaded in conjunction with
      terminal option `colourtext'%
    }{See the gnuplot documentation for explanation.%
    }{Either use 'blacktext' in gnuplot or load the package
      color.sty in LaTeX.}%
    \renewcommand\color[2][]{}%
  }%
  \providecommand\includegraphics[2][]{%
    \GenericError{(gnuplot) \space\space\space\@spaces}{%
      Package graphicx or graphics not loaded%
    }{See the gnuplot documentation for explanation.%
    }{The gnuplot epslatex terminal needs graphicx.sty or graphics.sty.}%
    \renewcommand\includegraphics[2][]{}%
  }%
  \providecommand\rotatebox[2]{#2}%
  \@ifundefined{ifGPcolor}{%
    \newif\ifGPcolor
    \GPcolortrue
  }{}%
  \@ifundefined{ifGPblacktext}{%
    \newif\ifGPblacktext
    \GPblacktexttrue
  }{}%
  \let\gplgaddtomacro\g@addto@macro
  \gdef\gplbacktext{}%
  \gdef\gplfronttext{}%
  \makeatother
  \ifGPblacktext
    \def\colorrgb#1{}%
    \def\colorgray#1{}%
  \else
    \ifGPcolor
      \def\colorrgb#1{\color[rgb]{#1}}%
      \def\colorgray#1{\color[gray]{#1}}%
      \expandafter\def\csname LTw\endcsname{\color{white}}%
      \expandafter\def\csname LTb\endcsname{\color{black}}%
      \expandafter\def\csname LTa\endcsname{\color{black}}%
      \expandafter\def\csname LT0\endcsname{\color[rgb]{1,0,0}}%
      \expandafter\def\csname LT1\endcsname{\color[rgb]{0,1,0}}%
      \expandafter\def\csname LT2\endcsname{\color[rgb]{0,0,1}}%
      \expandafter\def\csname LT3\endcsname{\color[rgb]{1,0,1}}%
      \expandafter\def\csname LT4\endcsname{\color[rgb]{0,1,1}}%
      \expandafter\def\csname LT5\endcsname{\color[rgb]{1,1,0}}%
      \expandafter\def\csname LT6\endcsname{\color[rgb]{0,0,0}}%
      \expandafter\def\csname LT7\endcsname{\color[rgb]{1,0.3,0}}%
      \expandafter\def\csname LT8\endcsname{\color[rgb]{0.5,0.5,0.5}}%
    \else
      \def\colorrgb#1{\color{black}}%
      \def\colorgray#1{\color[gray]{#1}}%
      \expandafter\def\csname LTw\endcsname{\color{white}}%
      \expandafter\def\csname LTb\endcsname{\color{black}}%
      \expandafter\def\csname LTa\endcsname{\color{black}}%
      \expandafter\def\csname LT0\endcsname{\color{black}}%
      \expandafter\def\csname LT1\endcsname{\color{black}}%
      \expandafter\def\csname LT2\endcsname{\color{black}}%
      \expandafter\def\csname LT3\endcsname{\color{black}}%
      \expandafter\def\csname LT4\endcsname{\color{black}}%
      \expandafter\def\csname LT5\endcsname{\color{black}}%
      \expandafter\def\csname LT6\endcsname{\color{black}}%
      \expandafter\def\csname LT7\endcsname{\color{black}}%
      \expandafter\def\csname LT8\endcsname{\color{black}}%
    \fi
  \fi
  \setlength{\unitlength}{0.0500bp}%
  \begin{picture}(2970.00,3780.00)%
    \gplgaddtomacro\gplbacktext{%
      \csname LTb\endcsname%
      \put(480,640){\makebox(0,0)[r]{\strut{} 0}}%
      \put(480,1220){\makebox(0,0)[r]{\strut{} 1000}}%
      \put(480,1800){\makebox(0,0)[r]{\strut{} 2000}}%
      \put(480,2379){\makebox(0,0)[r]{\strut{} 3000}}%
      \put(480,2959){\makebox(0,0)[r]{\strut{} 4000}}%
      \put(480,3539){\makebox(0,0)[r]{\strut{} 5000}}%
      \put(594,440){\makebox(0,0){\strut{}-100}}%
      \put(1188,440){\makebox(0,0){\strut{}-50}}%
      \put(1782,440){\makebox(0,0){\strut{} 0}}%
      \put(2375,440){\makebox(0,0){\strut{} 50}}%
      \put(2969,440){\makebox(0,0){\strut{} 100}}%
      \put(1784,140){\makebox(0,0){\strut{}$\delta x$ [$\mu$m]}}%
      \put(718,3249){\makebox(0,0)[l]{\strut{}Strips}}%
    }%
    \gplgaddtomacro\gplfronttext{%
      \csname LTb\endcsname%
      \put(2009,3376){\makebox(0,0)[l]{\strut{}Fitter}}%
      \csname LTb\endcsname%
      \put(2009,3176){\makebox(0,0)[l]{\strut{}Predicted}}%
    }%
    \gplbacktext
    \put(0,0){\includegraphics{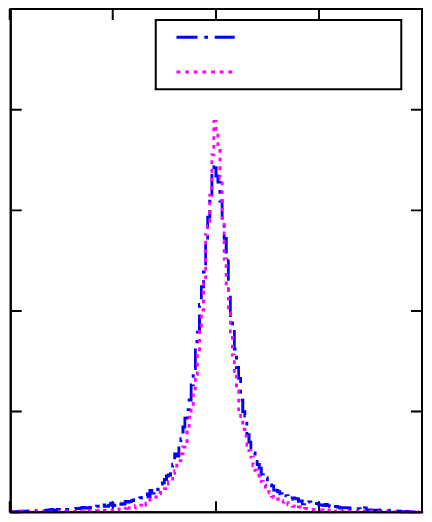}}%
    \gplfronttext
  \end{picture}%
\endgroup

%% file: corrected_3.tex
\begingroup
  \makeatletter
  \providecommand\color[2][]{%
    \GenericError{(gnuplot) \space\space\space\@spaces}{%
      Package color not loaded in conjunction with
      terminal option `colourtext'%
    }{See the gnuplot documentation for explanation.%
    }{Either use 'blacktext' in gnuplot or load the package
      color.sty in LaTeX.}%
    \renewcommand\color[2][]{}%
  }%
  \providecommand\includegraphics[2][]{%
    \GenericError{(gnuplot) \space\space\space\@spaces}{%
      Package graphicx or graphics not loaded%
    }{See the gnuplot documentation for explanation.%
    }{The gnuplot epslatex terminal needs graphicx.sty or graphics.sty.}%
    \renewcommand\includegraphics[2][]{}%
  }%
  \providecommand\rotatebox[2]{#2}%
  \@ifundefined{ifGPcolor}{%
    \newif\ifGPcolor
    \GPcolortrue
  }{}%
  \@ifundefined{ifGPblacktext}{%
    \newif\ifGPblacktext
    \GPblacktexttrue
  }{}%
  \let\gplgaddtomacro\g@addto@macro
  \gdef\gplbacktext{}%
  \gdef\gplfronttext{}%
  \makeatother
  \ifGPblacktext
    \def\colorrgb#1{}%
    \def\colorgray#1{}%
  \else
    \ifGPcolor
      \def\colorrgb#1{\color[rgb]{#1}}%
      \def\colorgray#1{\color[gray]{#1}}%
      \expandafter\def\csname LTw\endcsname{\color{white}}%
      \expandafter\def\csname LTb\endcsname{\color{black}}%
      \expandafter\def\csname LTa\endcsname{\color{black}}%
      \expandafter\def\csname LT0\endcsname{\color[rgb]{1,0,0}}%
      \expandafter\def\csname LT1\endcsname{\color[rgb]{0,1,0}}%
      \expandafter\def\csname LT2\endcsname{\color[rgb]{0,0,1}}%
      \expandafter\def\csname LT3\endcsname{\color[rgb]{1,0,1}}%
      \expandafter\def\csname LT4\endcsname{\color[rgb]{0,1,1}}%
      \expandafter\def\csname LT5\endcsname{\color[rgb]{1,1,0}}%
      \expandafter\def\csname LT6\endcsname{\color[rgb]{0,0,0}}%
      \expandafter\def\csname LT7\endcsname{\color[rgb]{1,0.3,0}}%
      \expandafter\def\csname LT8\endcsname{\color[rgb]{0.5,0.5,0.5}}%
    \else
      \def\colorrgb#1{\color{black}}%
      \def\colorgray#1{\color[gray]{#1}}%
      \expandafter\def\csname LTw\endcsname{\color{white}}%
      \expandafter\def\csname LTb\endcsname{\color{black}}%
      \expandafter\def\csname LTa\endcsname{\color{black}}%
      \expandafter\def\csname LT0\endcsname{\color{black}}%
      \expandafter\def\csname LT1\endcsname{\color{black}}%
      \expandafter\def\csname LT2\endcsname{\color{black}}%
      \expandafter\def\csname LT3\endcsname{\color{black}}%
      \expandafter\def\csname LT4\endcsname{\color{black}}%
      \expandafter\def\csname LT5\endcsname{\color{black}}%
      \expandafter\def\csname LT6\endcsname{\color{black}}%
      \expandafter\def\csname LT7\endcsname{\color{black}}%
      \expandafter\def\csname LT8\endcsname{\color{black}}%
    \fi
  \fi
  \setlength{\unitlength}{0.0500bp}%
  \begin{picture}(4500.00,3780.00)%
    \gplgaddtomacro\gplbacktext{%
      \csname LTb\endcsname%
      \put(600,640){\makebox(0,0)[r]{\strut{} 0}}%
      \put(600,1123){\makebox(0,0)[r]{\strut{} 2}}%
      \put(600,1606){\makebox(0,0)[r]{\strut{} 4}}%
      \put(600,2090){\makebox(0,0)[r]{\strut{} 6}}%
      \put(600,2573){\makebox(0,0)[r]{\strut{} 8}}%
      \put(600,3056){\makebox(0,0)[r]{\strut{} 10}}%
      \put(600,3539){\makebox(0,0)[r]{\strut{} 12}}%
      \put(720,440){\makebox(0,0){\strut{} 0}}%
      \put(1310,440){\makebox(0,0){\strut{} 20}}%
      \put(1900,440){\makebox(0,0){\strut{} 40}}%
      \put(2490,440){\makebox(0,0){\strut{} 60}}%
      \put(3079,440){\makebox(0,0){\strut{} 80}}%
      \put(3669,440){\makebox(0,0){\strut{} 100}}%
      \put(4259,440){\makebox(0,0){\strut{} 120}}%
      \put(140,2089){\rotatebox{-270}{\makebox(0,0){\strut{}$\langle y \rangle$ [keV]}}}%
      \put(2489,140){\makebox(0,0){\strut{}l [$\mu$m]}}%
      \put(897,3249){\makebox(0,0)[l]{\strut{}$\beta\gamma = 3.16$}}%
    }%
    \gplgaddtomacro\gplfronttext{%
      \csname LTb\endcsname%
      \put(3179,1203){\makebox(0,0)[l]{\strut{}$t =$  6~keV}}%
      \csname LTb\endcsname%
      \put(3179,1003){\makebox(0,0)[l]{\strut{}$t =$  8~keV}}%
      \csname LTb\endcsname%
      \put(3179,803){\makebox(0,0)[l]{\strut{}$t =$ 10~keV}}%
    }%
    \gplbacktext
    \put(0,0){\includegraphics{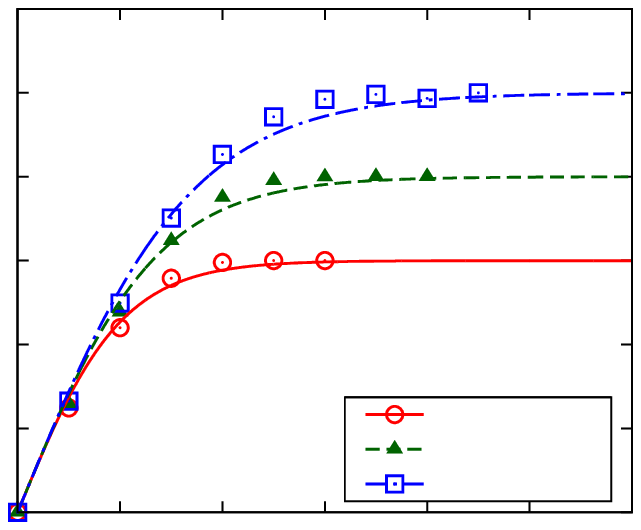}}%
    \gplfronttext
  \end{picture}%
\endgroup

%% file: corrected_high_0.tex
\begingroup
  \makeatletter
  \providecommand\color[2][]{%
    \GenericError{(gnuplot) \space\space\space\@spaces}{%
      Package color not loaded in conjunction with
      terminal option `colourtext'%
    }{See the gnuplot documentation for explanation.%
    }{Either use 'blacktext' in gnuplot or load the package
      color.sty in LaTeX.}%
    \renewcommand\color[2][]{}%
  }%
  \providecommand\includegraphics[2][]{%
    \GenericError{(gnuplot) \space\space\space\@spaces}{%
      Package graphicx or graphics not loaded%
    }{See the gnuplot documentation for explanation.%
    }{The gnuplot epslatex terminal needs graphicx.sty or graphics.sty.}%
    \renewcommand\includegraphics[2][]{}%
  }%
  \providecommand\rotatebox[2]{#2}%
  \@ifundefined{ifGPcolor}{%
    \newif\ifGPcolor
    \GPcolortrue
  }{}%
  \@ifundefined{ifGPblacktext}{%
    \newif\ifGPblacktext
    \GPblacktexttrue
  }{}%
  \let\gplgaddtomacro\g@addto@macro
  \gdef\gplbacktext{}%
  \gdef\gplfronttext{}%
  \makeatother
  \ifGPblacktext
    \def\colorrgb#1{}%
    \def\colorgray#1{}%
  \else
    \ifGPcolor
      \def\colorrgb#1{\color[rgb]{#1}}%
      \def\colorgray#1{\color[gray]{#1}}%
      \expandafter\def\csname LTw\endcsname{\color{white}}%
      \expandafter\def\csname LTb\endcsname{\color{black}}%
      \expandafter\def\csname LTa\endcsname{\color{black}}%
      \expandafter\def\csname LT0\endcsname{\color[rgb]{1,0,0}}%
      \expandafter\def\csname LT1\endcsname{\color[rgb]{0,1,0}}%
      \expandafter\def\csname LT2\endcsname{\color[rgb]{0,0,1}}%
      \expandafter\def\csname LT3\endcsname{\color[rgb]{1,0,1}}%
      \expandafter\def\csname LT4\endcsname{\color[rgb]{0,1,1}}%
      \expandafter\def\csname LT5\endcsname{\color[rgb]{1,1,0}}%
      \expandafter\def\csname LT6\endcsname{\color[rgb]{0,0,0}}%
      \expandafter\def\csname LT7\endcsname{\color[rgb]{1,0.3,0}}%
      \expandafter\def\csname LT8\endcsname{\color[rgb]{0.5,0.5,0.5}}%
    \else
      \def\colorrgb#1{\color{black}}%
      \def\colorgray#1{\color[gray]{#1}}%
      \expandafter\def\csname LTw\endcsname{\color{white}}%
      \expandafter\def\csname LTb\endcsname{\color{black}}%
      \expandafter\def\csname LTa\endcsname{\color{black}}%
      \expandafter\def\csname LT0\endcsname{\color{black}}%
      \expandafter\def\csname LT1\endcsname{\color{black}}%
      \expandafter\def\csname LT2\endcsname{\color{black}}%
      \expandafter\def\csname LT3\endcsname{\color{black}}%
      \expandafter\def\csname LT4\endcsname{\color{black}}%
      \expandafter\def\csname LT5\endcsname{\color{black}}%
      \expandafter\def\csname LT6\endcsname{\color{black}}%
      \expandafter\def\csname LT7\endcsname{\color{black}}%
      \expandafter\def\csname LT8\endcsname{\color{black}}%
    \fi
  \fi
  \setlength{\unitlength}{0.0500bp}%
  \begin{picture}(4500.00,3780.00)%
    \gplgaddtomacro\gplbacktext{%
      \csname LTb\endcsname%
      \put(600,640){\makebox(0,0)[r]{\strut{} 0}}%
      \put(600,1220){\makebox(0,0)[r]{\strut{} 100}}%
      \put(600,1800){\makebox(0,0)[r]{\strut{} 200}}%
      \put(600,2379){\makebox(0,0)[r]{\strut{} 300}}%
      \put(600,2959){\makebox(0,0)[r]{\strut{} 400}}%
      \put(600,3539){\makebox(0,0)[r]{\strut{} 500}}%
      \put(720,440){\makebox(0,0){\strut{} 0}}%
      \put(1457,440){\makebox(0,0){\strut{} 50}}%
      \put(2195,440){\makebox(0,0){\strut{} 100}}%
      \put(2932,440){\makebox(0,0){\strut{} 150}}%
      \put(3669,440){\makebox(0,0){\strut{} 200}}%
      \put(20,2089){\rotatebox{-270}{\makebox(0,0){\strut{}$\langle y \rangle$ [keV]}}}%
      \put(2489,140){\makebox(0,0){\strut{}l [$\mu$m]}}%
      \put(897,3249){\makebox(0,0)[l]{\strut{}$\beta\gamma = 0.32$}}%
    }%
    \gplgaddtomacro\gplfronttext{%
      \csname LTb\endcsname%
      \put(3059,1203){\makebox(0,0)[l]{\strut{}$t =$ 100~keV}}%
      \csname LTb\endcsname%
      \put(3059,1003){\makebox(0,0)[l]{\strut{}$t =$ 150~keV}}%
      \csname LTb\endcsname%
      \put(3059,803){\makebox(0,0)[l]{\strut{}$t =$ 200~keV}}%
    }%
    \gplbacktext
    \put(0,0){\includegraphics{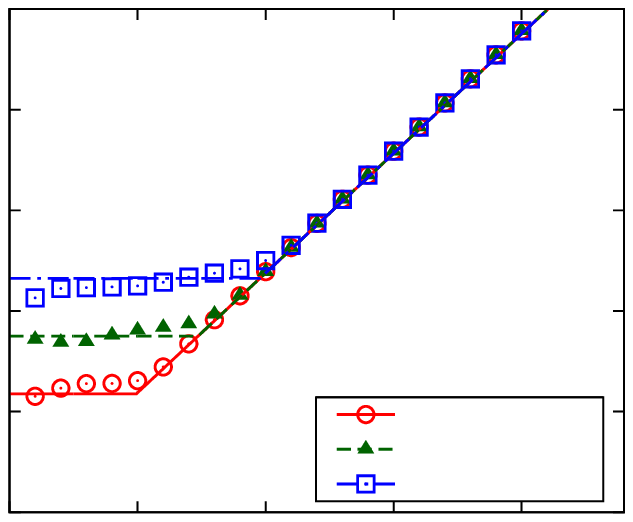}}%
    \gplfronttext
  \end{picture}%
\endgroup

%% file: residual_charge_pixels.tex
\begingroup
  \makeatletter
  \providecommand\color[2][]{%
    \GenericError{(gnuplot) \space\space\space\@spaces}{%
      Package color not loaded in conjunction with
      terminal option `colourtext'%
    }{See the gnuplot documentation for explanation.%
    }{Either use 'blacktext' in gnuplot or load the package
      color.sty in LaTeX.}%
    \renewcommand\color[2][]{}%
  }%
  \providecommand\includegraphics[2][]{%
    \GenericError{(gnuplot) \space\space\space\@spaces}{%
      Package graphicx or graphics not loaded%
    }{See the gnuplot documentation for explanation.%
    }{The gnuplot epslatex terminal needs graphicx.sty or graphics.sty.}%
    \renewcommand\includegraphics[2][]{}%
  }%
  \providecommand\rotatebox[2]{#2}%
  \@ifundefined{ifGPcolor}{%
    \newif\ifGPcolor
    \GPcolortrue
  }{}%
  \@ifundefined{ifGPblacktext}{%
    \newif\ifGPblacktext
    \GPblacktexttrue
  }{}%
  \let\gplgaddtomacro\g@addto@macro
  \gdef\gplbacktext{}%
  \gdef\gplfronttext{}%
  \makeatother
  \ifGPblacktext
    \def\colorrgb#1{}%
    \def\colorgray#1{}%
  \else
    \ifGPcolor
      \def\colorrgb#1{\color[rgb]{#1}}%
      \def\colorgray#1{\color[gray]{#1}}%
      \expandafter\def\csname LTw\endcsname{\color{white}}%
      \expandafter\def\csname LTb\endcsname{\color{black}}%
      \expandafter\def\csname LTa\endcsname{\color{black}}%
      \expandafter\def\csname LT0\endcsname{\color[rgb]{1,0,0}}%
      \expandafter\def\csname LT1\endcsname{\color[rgb]{0,1,0}}%
      \expandafter\def\csname LT2\endcsname{\color[rgb]{0,0,1}}%
      \expandafter\def\csname LT3\endcsname{\color[rgb]{1,0,1}}%
      \expandafter\def\csname LT4\endcsname{\color[rgb]{0,1,1}}%
      \expandafter\def\csname LT5\endcsname{\color[rgb]{1,1,0}}%
      \expandafter\def\csname LT6\endcsname{\color[rgb]{0,0,0}}%
      \expandafter\def\csname LT7\endcsname{\color[rgb]{1,0.3,0}}%
      \expandafter\def\csname LT8\endcsname{\color[rgb]{0.5,0.5,0.5}}%
    \else
      \def\colorrgb#1{\color{black}}%
      \def\colorgray#1{\color[gray]{#1}}%
      \expandafter\def\csname LTw\endcsname{\color{white}}%
      \expandafter\def\csname LTb\endcsname{\color{black}}%
      \expandafter\def\csname LTa\endcsname{\color{black}}%
      \expandafter\def\csname LT0\endcsname{\color{black}}%
      \expandafter\def\csname LT1\endcsname{\color{black}}%
      \expandafter\def\csname LT2\endcsname{\color{black}}%
      \expandafter\def\csname LT3\endcsname{\color{black}}%
      \expandafter\def\csname LT4\endcsname{\color{black}}%
      \expandafter\def\csname LT5\endcsname{\color{black}}%
      \expandafter\def\csname LT6\endcsname{\color{black}}%
      \expandafter\def\csname LT7\endcsname{\color{black}}%
      \expandafter\def\csname LT8\endcsname{\color{black}}%
    \fi
  \fi
  \setlength{\unitlength}{0.0500bp}%
  \begin{picture}(4500.00,3780.00)%
    \gplgaddtomacro\gplbacktext{%
      \csname LTb\endcsname%
      \put(600,640){\makebox(0,0)[r]{\strut{} 0}}%
      \put(600,1054){\makebox(0,0)[r]{\strut{} 200}}%
      \put(600,1468){\makebox(0,0)[r]{\strut{} 400}}%
      \put(600,1882){\makebox(0,0)[r]{\strut{} 600}}%
      \put(600,2297){\makebox(0,0)[r]{\strut{} 800}}%
      \put(600,2711){\makebox(0,0)[r]{\strut{} 1000}}%
      \put(600,3125){\makebox(0,0)[r]{\strut{} 1200}}%
      \put(600,3539){\makebox(0,0)[r]{\strut{} 1400}}%
      \put(720,440){\makebox(0,0){\strut{}-20}}%
      \put(1162,440){\makebox(0,0){\strut{}-15}}%
      \put(1605,440){\makebox(0,0){\strut{}-10}}%
      \put(2047,440){\makebox(0,0){\strut{}-5}}%
      \put(2489,440){\makebox(0,0){\strut{} 0}}%
      \put(2932,440){\makebox(0,0){\strut{} 5}}%
      \put(3374,440){\makebox(0,0){\strut{} 10}}%
      \put(3817,440){\makebox(0,0){\strut{} 15}}%
      \put(4259,440){\makebox(0,0){\strut{} 20}}%
      \put(2489,140){\makebox(0,0){\strut{}$\delta\Delta$ [keV]}}%
      \put(897,3249){\makebox(0,0)[l]{\strut{}Pixels}}%
    }%
    \gplgaddtomacro\gplfronttext{%
      \csname LTb\endcsname%
      \put(3659,3376){\makebox(0,0)[l]{\strut{}Sum}}%
      \csname LTb\endcsname%
      \put(3659,3176){\makebox(0,0)[l]{\strut{}Fitter}}%
    }%
    \gplbacktext
    \put(0,0){\includegraphics{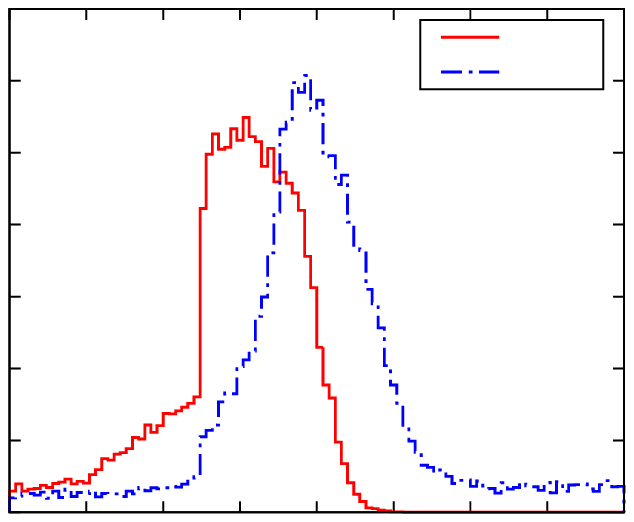}}%
    \gplfronttext
  \end{picture}%
\endgroup

%% file: residual_charge_strips.tex
\begingroup
  \makeatletter
  \providecommand\color[2][]{%
    \GenericError{(gnuplot) \space\space\space\@spaces}{%
      Package color not loaded in conjunction with
      terminal option `colourtext'%
    }{See the gnuplot documentation for explanation.%
    }{Either use 'blacktext' in gnuplot or load the package
      color.sty in LaTeX.}%
    \renewcommand\color[2][]{}%
  }%
  \providecommand\includegraphics[2][]{%
    \GenericError{(gnuplot) \space\space\space\@spaces}{%
      Package graphicx or graphics not loaded%
    }{See the gnuplot documentation for explanation.%
    }{The gnuplot epslatex terminal needs graphicx.sty or graphics.sty.}%
    \renewcommand\includegraphics[2][]{}%
  }%
  \providecommand\rotatebox[2]{#2}%
  \@ifundefined{ifGPcolor}{%
    \newif\ifGPcolor
    \GPcolortrue
  }{}%
  \@ifundefined{ifGPblacktext}{%
    \newif\ifGPblacktext
    \GPblacktexttrue
  }{}%
  \let\gplgaddtomacro\g@addto@macro
  \gdef\gplbacktext{}%
  \gdef\gplfronttext{}%
  \makeatother
  \ifGPblacktext
    \def\colorrgb#1{}%
    \def\colorgray#1{}%
  \else
    \ifGPcolor
      \def\colorrgb#1{\color[rgb]{#1}}%
      \def\colorgray#1{\color[gray]{#1}}%
      \expandafter\def\csname LTw\endcsname{\color{white}}%
      \expandafter\def\csname LTb\endcsname{\color{black}}%
      \expandafter\def\csname LTa\endcsname{\color{black}}%
      \expandafter\def\csname LT0\endcsname{\color[rgb]{1,0,0}}%
      \expandafter\def\csname LT1\endcsname{\color[rgb]{0,1,0}}%
      \expandafter\def\csname LT2\endcsname{\color[rgb]{0,0,1}}%
      \expandafter\def\csname LT3\endcsname{\color[rgb]{1,0,1}}%
      \expandafter\def\csname LT4\endcsname{\color[rgb]{0,1,1}}%
      \expandafter\def\csname LT5\endcsname{\color[rgb]{1,1,0}}%
      \expandafter\def\csname LT6\endcsname{\color[rgb]{0,0,0}}%
      \expandafter\def\csname LT7\endcsname{\color[rgb]{1,0.3,0}}%
      \expandafter\def\csname LT8\endcsname{\color[rgb]{0.5,0.5,0.5}}%
    \else
      \def\colorrgb#1{\color{black}}%
      \def\colorgray#1{\color[gray]{#1}}%
      \expandafter\def\csname LTw\endcsname{\color{white}}%
      \expandafter\def\csname LTb\endcsname{\color{black}}%
      \expandafter\def\csname LTa\endcsname{\color{black}}%
      \expandafter\def\csname LT0\endcsname{\color{black}}%
      \expandafter\def\csname LT1\endcsname{\color{black}}%
      \expandafter\def\csname LT2\endcsname{\color{black}}%
      \expandafter\def\csname LT3\endcsname{\color{black}}%
      \expandafter\def\csname LT4\endcsname{\color{black}}%
      \expandafter\def\csname LT5\endcsname{\color{black}}%
      \expandafter\def\csname LT6\endcsname{\color{black}}%
      \expandafter\def\csname LT7\endcsname{\color{black}}%
      \expandafter\def\csname LT8\endcsname{\color{black}}%
    \fi
  \fi
  \setlength{\unitlength}{0.0500bp}%
  \begin{picture}(4500.00,3780.00)%
    \gplgaddtomacro\gplbacktext{%
      \csname LTb\endcsname%
      \put(600,640){\makebox(0,0)[r]{\strut{} 0}}%
      \put(600,1002){\makebox(0,0)[r]{\strut{} 500}}%
      \put(600,1365){\makebox(0,0)[r]{\strut{} 1000}}%
      \put(600,1727){\makebox(0,0)[r]{\strut{} 1500}}%
      \put(600,2090){\makebox(0,0)[r]{\strut{} 2000}}%
      \put(600,2452){\makebox(0,0)[r]{\strut{} 2500}}%
      \put(600,2814){\makebox(0,0)[r]{\strut{} 3000}}%
      \put(600,3177){\makebox(0,0)[r]{\strut{} 3500}}%
      \put(600,3539){\makebox(0,0)[r]{\strut{} 4000}}%
      \put(720,440){\makebox(0,0){\strut{}-20}}%
      \put(1162,440){\makebox(0,0){\strut{}-15}}%
      \put(1605,440){\makebox(0,0){\strut{}-10}}%
      \put(2047,440){\makebox(0,0){\strut{}-5}}%
      \put(2489,440){\makebox(0,0){\strut{} 0}}%
      \put(2932,440){\makebox(0,0){\strut{} 5}}%
      \put(3374,440){\makebox(0,0){\strut{} 10}}%
      \put(3817,440){\makebox(0,0){\strut{} 15}}%
      \put(4259,440){\makebox(0,0){\strut{} 20}}%
      \put(2489,140){\makebox(0,0){\strut{}$\delta\Delta$ [keV]}}%
      \put(897,3249){\makebox(0,0)[l]{\strut{}Strips}}%
    }%
    \gplgaddtomacro\gplfronttext{%
      \csname LTb\endcsname%
      \put(3659,3376){\makebox(0,0)[l]{\strut{}Sum}}%
      \csname LTb\endcsname%
      \put(3659,3176){\makebox(0,0)[l]{\strut{}Fitter}}%
    }%
    \gplbacktext
    \put(0,0){\includegraphics{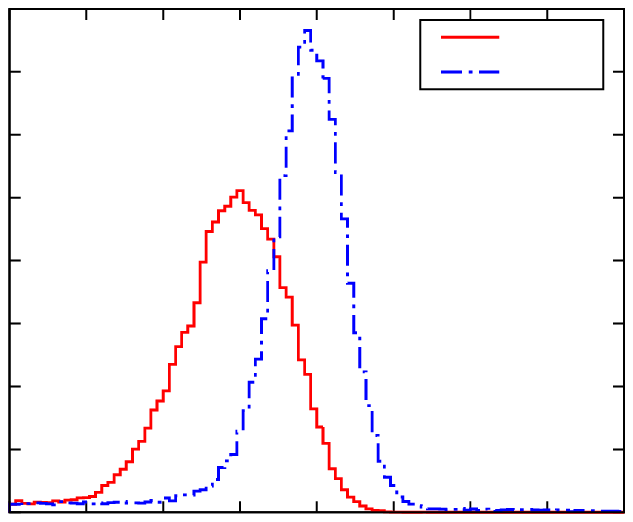}}%
    \gplfronttext
  \end{picture}%
\endgroup

%% file: fisher.tex
\begingroup
  \makeatletter
  \providecommand\color[2][]{%
    \GenericError{(gnuplot) \space\space\space\@spaces}{%
      Package color not loaded in conjunction with
      terminal option `colourtext'%
    }{See the gnuplot documentation for explanation.%
    }{Either use 'blacktext' in gnuplot or load the package
      color.sty in LaTeX.}%
    \renewcommand\color[2][]{}%
  }%
  \providecommand\includegraphics[2][]{%
    \GenericError{(gnuplot) \space\space\space\@spaces}{%
      Package graphicx or graphics not loaded%
    }{See the gnuplot documentation for explanation.%
    }{The gnuplot epslatex terminal needs graphicx.sty or graphics.sty.}%
    \renewcommand\includegraphics[2][]{}%
  }%
  \providecommand\rotatebox[2]{#2}%
  \@ifundefined{ifGPcolor}{%
    \newif\ifGPcolor
    \GPcolortrue
  }{}%
  \@ifundefined{ifGPblacktext}{%
    \newif\ifGPblacktext
    \GPblacktexttrue
  }{}%
  \let\gplgaddtomacro\g@addto@macro
  \gdef\gplbacktext{}%
  \gdef\gplfronttext{}%
  \makeatother
  \ifGPblacktext
    \def\colorrgb#1{}%
    \def\colorgray#1{}%
  \else
    \ifGPcolor
      \def\colorrgb#1{\color[rgb]{#1}}%
      \def\colorgray#1{\color[gray]{#1}}%
      \expandafter\def\csname LTw\endcsname{\color{white}}%
      \expandafter\def\csname LTb\endcsname{\color{black}}%
      \expandafter\def\csname LTa\endcsname{\color{black}}%
      \expandafter\def\csname LT0\endcsname{\color[rgb]{1,0,0}}%
      \expandafter\def\csname LT1\endcsname{\color[rgb]{0,1,0}}%
      \expandafter\def\csname LT2\endcsname{\color[rgb]{0,0,1}}%
      \expandafter\def\csname LT3\endcsname{\color[rgb]{1,0,1}}%
      \expandafter\def\csname LT4\endcsname{\color[rgb]{0,1,1}}%
      \expandafter\def\csname LT5\endcsname{\color[rgb]{1,1,0}}%
      \expandafter\def\csname LT6\endcsname{\color[rgb]{0,0,0}}%
      \expandafter\def\csname LT7\endcsname{\color[rgb]{1,0.3,0}}%
      \expandafter\def\csname LT8\endcsname{\color[rgb]{0.5,0.5,0.5}}%
    \else
      \def\colorrgb#1{\color{black}}%
      \def\colorgray#1{\color[gray]{#1}}%
      \expandafter\def\csname LTw\endcsname{\color{white}}%
      \expandafter\def\csname LTb\endcsname{\color{black}}%
      \expandafter\def\csname LTa\endcsname{\color{black}}%
      \expandafter\def\csname LT0\endcsname{\color{black}}%
      \expandafter\def\csname LT1\endcsname{\color{black}}%
      \expandafter\def\csname LT2\endcsname{\color{black}}%
      \expandafter\def\csname LT3\endcsname{\color{black}}%
      \expandafter\def\csname LT4\endcsname{\color{black}}%
      \expandafter\def\csname LT5\endcsname{\color{black}}%
      \expandafter\def\csname LT6\endcsname{\color{black}}%
      \expandafter\def\csname LT7\endcsname{\color{black}}%
      \expandafter\def\csname LT8\endcsname{\color{black}}%
    \fi
  \fi
  \setlength{\unitlength}{0.0500bp}%
  \begin{picture}(4500.00,3780.00)%
    \gplgaddtomacro\gplbacktext{%
      \csname LTb\endcsname%
      \put(600,1147){\makebox(0,0)[r]{\strut{}$10^{-4}$}}%
      \put(600,1871){\makebox(0,0)[r]{\strut{}$10^{-3}$}}%
      \put(600,2596){\makebox(0,0)[r]{\strut{}$10^{-2}$}}%
      \put(600,3321){\makebox(0,0)[r]{\strut{}$10^{-1}$}}%
      \put(906,440){\makebox(0,0){\strut{} 100}}%
      \put(1279,440){\makebox(0,0){\strut{} 200}}%
      \put(1651,440){\makebox(0,0){\strut{} 300}}%
      \put(2024,440){\makebox(0,0){\strut{} 400}}%
      \put(2396,440){\makebox(0,0){\strut{} 500}}%
      \put(2769,440){\makebox(0,0){\strut{} 600}}%
      \put(3141,440){\makebox(0,0){\strut{} 700}}%
      \put(3514,440){\makebox(0,0){\strut{} 800}}%
      \put(3886,440){\makebox(0,0){\strut{} 900}}%
      \put(4259,440){\makebox(0,0){\strut{} 1000}}%
      \put(20,2089){\rotatebox{-270}{\makebox(0,0){\strut{}$\left\langle \sigma_\Delta^{-2} \right\rangle$ [$\mathrm{keV^{-2}}$]}}}%
      \put(2489,140){\makebox(0,0){\strut{}$l$ [$\mu$m]}}%
      \put(897,1075){\makebox(0,0)[l]{\strut{}$n_\varepsilon = -1.58\pm 0.02$}}%
      \put(897,857){\makebox(0,0)[l]{\strut{}$n_l = -1.75\pm 0.02$}}%
    }%
    \gplgaddtomacro\gplfronttext{%
      \csname LTb\endcsname%
      \put(3299,3376){\makebox(0,0)[l]{\strut{}$\beta\gamma = 0.56$}}%
      \csname LTb\endcsname%
      \put(3299,3176){\makebox(0,0)[l]{\strut{}$\beta\gamma = 1.00$}}%
      \csname LTb\endcsname%
      \put(3299,2976){\makebox(0,0)[l]{\strut{}$\beta\gamma = 3.16$}}%
      \csname LTb\endcsname%
      \put(3299,2776){\makebox(0,0)[l]{\strut{}$\beta\gamma = 10.0$}}%
    }%
    \gplbacktext
    \put(0,0){\includegraphics{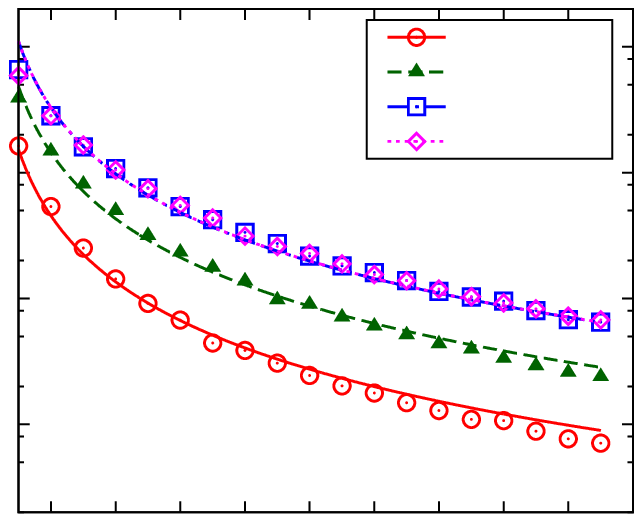}}%
    \gplfronttext
  \end{picture}%
\endgroup

%% file: estimates_0.tex
\begingroup
  \makeatletter
  \providecommand\color[2][]{%
    \GenericError{(gnuplot) \space\space\space\@spaces}{%
      Package color not loaded in conjunction with
      terminal option `colourtext'%
    }{See the gnuplot documentation for explanation.%
    }{Either use 'blacktext' in gnuplot or load the package
      color.sty in LaTeX.}%
    \renewcommand\color[2][]{}%
  }%
  \providecommand\includegraphics[2][]{%
    \GenericError{(gnuplot) \space\space\space\@spaces}{%
      Package graphicx or graphics not loaded%
    }{See the gnuplot documentation for explanation.%
    }{The gnuplot epslatex terminal needs graphicx.sty or graphics.sty.}%
    \renewcommand\includegraphics[2][]{}%
  }%
  \providecommand\rotatebox[2]{#2}%
  \@ifundefined{ifGPcolor}{%
    \newif\ifGPcolor
    \GPcolortrue
  }{}%
  \@ifundefined{ifGPblacktext}{%
    \newif\ifGPblacktext
    \GPblacktexttrue
  }{}%
  \let\gplgaddtomacro\g@addto@macro
  \gdef\gplbacktext{}%
  \gdef\gplfronttext{}%
  \makeatother
  \ifGPblacktext
    \def\colorrgb#1{}%
    \def\colorgray#1{}%
  \else
    \ifGPcolor
      \def\colorrgb#1{\color[rgb]{#1}}%
      \def\colorgray#1{\color[gray]{#1}}%
      \expandafter\def\csname LTw\endcsname{\color{white}}%
      \expandafter\def\csname LTb\endcsname{\color{black}}%
      \expandafter\def\csname LTa\endcsname{\color{black}}%
      \expandafter\def\csname LT0\endcsname{\color[rgb]{1,0,0}}%
      \expandafter\def\csname LT1\endcsname{\color[rgb]{0,1,0}}%
      \expandafter\def\csname LT2\endcsname{\color[rgb]{0,0,1}}%
      \expandafter\def\csname LT3\endcsname{\color[rgb]{1,0,1}}%
      \expandafter\def\csname LT4\endcsname{\color[rgb]{0,1,1}}%
      \expandafter\def\csname LT5\endcsname{\color[rgb]{1,1,0}}%
      \expandafter\def\csname LT6\endcsname{\color[rgb]{0,0,0}}%
      \expandafter\def\csname LT7\endcsname{\color[rgb]{1,0.3,0}}%
      \expandafter\def\csname LT8\endcsname{\color[rgb]{0.5,0.5,0.5}}%
    \else
      \def\colorrgb#1{\color{black}}%
      \def\colorgray#1{\color[gray]{#1}}%
      \expandafter\def\csname LTw\endcsname{\color{white}}%
      \expandafter\def\csname LTb\endcsname{\color{black}}%
      \expandafter\def\csname LTa\endcsname{\color{black}}%
      \expandafter\def\csname LT0\endcsname{\color{black}}%
      \expandafter\def\csname LT1\endcsname{\color{black}}%
      \expandafter\def\csname LT2\endcsname{\color{black}}%
      \expandafter\def\csname LT3\endcsname{\color{black}}%
      \expandafter\def\csname LT4\endcsname{\color{black}}%
      \expandafter\def\csname LT5\endcsname{\color{black}}%
      \expandafter\def\csname LT6\endcsname{\color{black}}%
      \expandafter\def\csname LT7\endcsname{\color{black}}%
      \expandafter\def\csname LT8\endcsname{\color{black}}%
    \fi
  \fi
  \setlength{\unitlength}{0.0500bp}%
  \begin{picture}(4500.00,3780.00)%
    \gplgaddtomacro\gplbacktext{%
      \csname LTb\endcsname%
      \put(600,640){\makebox(0,0)[r]{\strut{} 0}}%
      \put(600,1123){\makebox(0,0)[r]{\strut{} 1000}}%
      \put(600,1606){\makebox(0,0)[r]{\strut{} 2000}}%
      \put(600,2090){\makebox(0,0)[r]{\strut{} 3000}}%
      \put(600,2573){\makebox(0,0)[r]{\strut{} 4000}}%
      \put(600,3056){\makebox(0,0)[r]{\strut{} 5000}}%
      \put(600,3539){\makebox(0,0)[r]{\strut{} 6000}}%
      \put(720,440){\makebox(0,0){\strut{} 0.8}}%
      \put(1428,440){\makebox(0,0){\strut{} 1}}%
      \put(2136,440){\makebox(0,0){\strut{} 1.2}}%
      \put(2843,440){\makebox(0,0){\strut{} 1.4}}%
      \put(3551,440){\makebox(0,0){\strut{} 1.6}}%
      \put(4259,440){\makebox(0,0){\strut{} 1.8}}%
      \put(-100,2089){\rotatebox{-270}{\makebox(0,0){\strut{}Reconstructed relative gain}}}%
      \put(2489,140){\makebox(0,0){\strut{}$\log$($\varepsilon$ [MeV/cm])}}%
      \put(4082,2090){\makebox(0,0)[r]{\strut{}300~$\mu$m}}%
    }%
    \gplgaddtomacro\gplfronttext{%
      \csname LTb\endcsname%
      \put(2939,3376){\makebox(0,0)[l]{\strut{}Fitter}}%
      \csname LTb\endcsname%
      \put(2939,3176){\makebox(0,0)[l]{\strut{}Truncated}}%
      \csname LTb\endcsname%
      \put(2939,2976){\makebox(0,0)[l]{\strut{}Power (-2)}}%
      \csname LTb\endcsname%
      \put(2939,2776){\makebox(0,0)[l]{\strut{}Harmonic (-1)}}%
      \csname LTb\endcsname%
      \put(2939,2576){\makebox(0,0)[l]{\strut{}Arithmetic (1)}}%
    }%
    \gplbacktext
    \put(0,0){\includegraphics{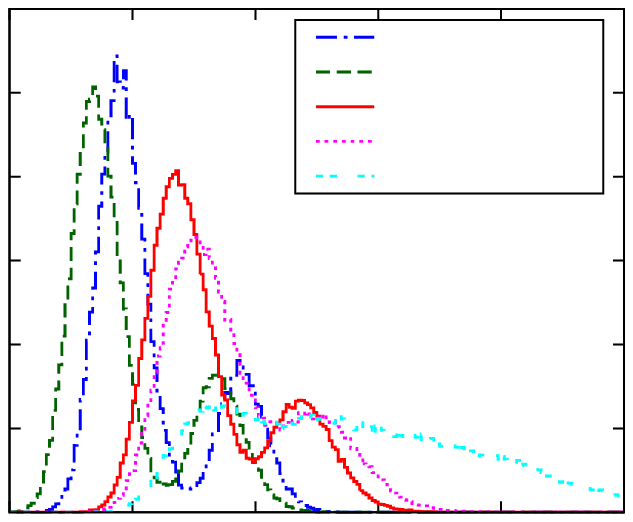}}%
    \gplfronttext
  \end{picture}%
\endgroup

%% file: estimates_1.tex
\begingroup
  \makeatletter
  \providecommand\color[2][]{%
    \GenericError{(gnuplot) \space\space\space\@spaces}{%
      Package color not loaded in conjunction with
      terminal option `colourtext'%
    }{See the gnuplot documentation for explanation.%
    }{Either use 'blacktext' in gnuplot or load the package
      color.sty in LaTeX.}%
    \renewcommand\color[2][]{}%
  }%
  \providecommand\includegraphics[2][]{%
    \GenericError{(gnuplot) \space\space\space\@spaces}{%
      Package graphicx or graphics not loaded%
    }{See the gnuplot documentation for explanation.%
    }{The gnuplot epslatex terminal needs graphicx.sty or graphics.sty.}%
    \renewcommand\includegraphics[2][]{}%
  }%
  \providecommand\rotatebox[2]{#2}%
  \@ifundefined{ifGPcolor}{%
    \newif\ifGPcolor
    \GPcolortrue
  }{}%
  \@ifundefined{ifGPblacktext}{%
    \newif\ifGPblacktext
    \GPblacktexttrue
  }{}%
  \let\gplgaddtomacro\g@addto@macro
  \gdef\gplbacktext{}%
  \gdef\gplfronttext{}%
  \makeatother
  \ifGPblacktext
    \def\colorrgb#1{}%
    \def\colorgray#1{}%
  \else
    \ifGPcolor
      \def\colorrgb#1{\color[rgb]{#1}}%
      \def\colorgray#1{\color[gray]{#1}}%
      \expandafter\def\csname LTw\endcsname{\color{white}}%
      \expandafter\def\csname LTb\endcsname{\color{black}}%
      \expandafter\def\csname LTa\endcsname{\color{black}}%
      \expandafter\def\csname LT0\endcsname{\color[rgb]{1,0,0}}%
      \expandafter\def\csname LT1\endcsname{\color[rgb]{0,1,0}}%
      \expandafter\def\csname LT2\endcsname{\color[rgb]{0,0,1}}%
      \expandafter\def\csname LT3\endcsname{\color[rgb]{1,0,1}}%
      \expandafter\def\csname LT4\endcsname{\color[rgb]{0,1,1}}%
      \expandafter\def\csname LT5\endcsname{\color[rgb]{1,1,0}}%
      \expandafter\def\csname LT6\endcsname{\color[rgb]{0,0,0}}%
      \expandafter\def\csname LT7\endcsname{\color[rgb]{1,0.3,0}}%
      \expandafter\def\csname LT8\endcsname{\color[rgb]{0.5,0.5,0.5}}%
    \else
      \def\colorrgb#1{\color{black}}%
      \def\colorgray#1{\color[gray]{#1}}%
      \expandafter\def\csname LTw\endcsname{\color{white}}%
      \expandafter\def\csname LTb\endcsname{\color{black}}%
      \expandafter\def\csname LTa\endcsname{\color{black}}%
      \expandafter\def\csname LT0\endcsname{\color{black}}%
      \expandafter\def\csname LT1\endcsname{\color{black}}%
      \expandafter\def\csname LT2\endcsname{\color{black}}%
      \expandafter\def\csname LT3\endcsname{\color{black}}%
      \expandafter\def\csname LT4\endcsname{\color{black}}%
      \expandafter\def\csname LT5\endcsname{\color{black}}%
      \expandafter\def\csname LT6\endcsname{\color{black}}%
      \expandafter\def\csname LT7\endcsname{\color{black}}%
      \expandafter\def\csname LT8\endcsname{\color{black}}%
    \fi
  \fi
  \setlength{\unitlength}{0.0500bp}%
  \begin{picture}(4500.00,3780.00)%
    \gplgaddtomacro\gplbacktext{%
      \csname LTb\endcsname%
      \put(600,640){\makebox(0,0)[r]{\strut{} 0}}%
      \put(600,1123){\makebox(0,0)[r]{\strut{} 1000}}%
      \put(600,1606){\makebox(0,0)[r]{\strut{} 2000}}%
      \put(600,2090){\makebox(0,0)[r]{\strut{} 3000}}%
      \put(600,2573){\makebox(0,0)[r]{\strut{} 4000}}%
      \put(600,3056){\makebox(0,0)[r]{\strut{} 5000}}%
      \put(600,3539){\makebox(0,0)[r]{\strut{} 6000}}%
      \put(720,440){\makebox(0,0){\strut{} 0.8}}%
      \put(1428,440){\makebox(0,0){\strut{} 1}}%
      \put(2136,440){\makebox(0,0){\strut{} 1.2}}%
      \put(2843,440){\makebox(0,0){\strut{} 1.4}}%
      \put(3551,440){\makebox(0,0){\strut{} 1.6}}%
      \put(4259,440){\makebox(0,0){\strut{} 1.8}}%
      \put(-100,2089){\rotatebox{-270}{\makebox(0,0){\strut{}Reconstructed relative gain}}}%
      \put(2489,140){\makebox(0,0){\strut{}$\log$($\varepsilon$ [MeV/cm])}}%
      \put(4082,2090){\makebox(0,0)[r]{\strut{}300-600~$\mu$m}}%
    }%
    \gplgaddtomacro\gplfronttext{%
      \csname LTb\endcsname%
      \put(2939,3376){\makebox(0,0)[l]{\strut{}Fitter}}%
      \csname LTb\endcsname%
      \put(2939,3176){\makebox(0,0)[l]{\strut{}Truncated}}%
      \csname LTb\endcsname%
      \put(2939,2976){\makebox(0,0)[l]{\strut{}Power (-2)}}%
      \csname LTb\endcsname%
      \put(2939,2776){\makebox(0,0)[l]{\strut{}Harmonic (-1)}}%
      \csname LTb\endcsname%
      \put(2939,2576){\makebox(0,0)[l]{\strut{}Arithmetic (1)}}%
    }%
    \gplbacktext
    \put(0,0){\includegraphics{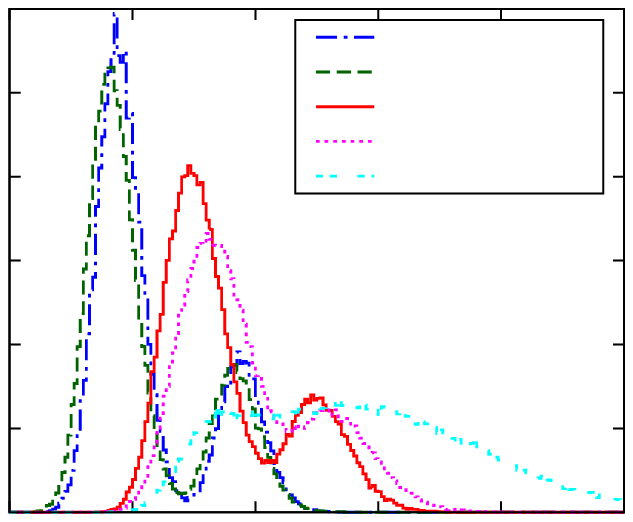}}%
    \gplfronttext
  \end{picture}%
\endgroup

%% file: estimates_2.tex
\begingroup
  \makeatletter
  \providecommand\color[2][]{%
    \GenericError{(gnuplot) \space\space\space\@spaces}{%
      Package color not loaded in conjunction with
      terminal option `colourtext'%
    }{See the gnuplot documentation for explanation.%
    }{Either use 'blacktext' in gnuplot or load the package
      color.sty in LaTeX.}%
    \renewcommand\color[2][]{}%
  }%
  \providecommand\includegraphics[2][]{%
    \GenericError{(gnuplot) \space\space\space\@spaces}{%
      Package graphicx or graphics not loaded%
    }{See the gnuplot documentation for explanation.%
    }{The gnuplot epslatex terminal needs graphicx.sty or graphics.sty.}%
    \renewcommand\includegraphics[2][]{}%
  }%
  \providecommand\rotatebox[2]{#2}%
  \@ifundefined{ifGPcolor}{%
    \newif\ifGPcolor
    \GPcolortrue
  }{}%
  \@ifundefined{ifGPblacktext}{%
    \newif\ifGPblacktext
    \GPblacktexttrue
  }{}%
  \let\gplgaddtomacro\g@addto@macro
  \gdef\gplbacktext{}%
  \gdef\gplfronttext{}%
  \makeatother
  \ifGPblacktext
    \def\colorrgb#1{}%
    \def\colorgray#1{}%
  \else
    \ifGPcolor
      \def\colorrgb#1{\color[rgb]{#1}}%
      \def\colorgray#1{\color[gray]{#1}}%
      \expandafter\def\csname LTw\endcsname{\color{white}}%
      \expandafter\def\csname LTb\endcsname{\color{black}}%
      \expandafter\def\csname LTa\endcsname{\color{black}}%
      \expandafter\def\csname LT0\endcsname{\color[rgb]{1,0,0}}%
      \expandafter\def\csname LT1\endcsname{\color[rgb]{0,1,0}}%
      \expandafter\def\csname LT2\endcsname{\color[rgb]{0,0,1}}%
      \expandafter\def\csname LT3\endcsname{\color[rgb]{1,0,1}}%
      \expandafter\def\csname LT4\endcsname{\color[rgb]{0,1,1}}%
      \expandafter\def\csname LT5\endcsname{\color[rgb]{1,1,0}}%
      \expandafter\def\csname LT6\endcsname{\color[rgb]{0,0,0}}%
      \expandafter\def\csname LT7\endcsname{\color[rgb]{1,0.3,0}}%
      \expandafter\def\csname LT8\endcsname{\color[rgb]{0.5,0.5,0.5}}%
    \else
      \def\colorrgb#1{\color{black}}%
      \def\colorgray#1{\color[gray]{#1}}%
      \expandafter\def\csname LTw\endcsname{\color{white}}%
      \expandafter\def\csname LTb\endcsname{\color{black}}%
      \expandafter\def\csname LTa\endcsname{\color{black}}%
      \expandafter\def\csname LT0\endcsname{\color{black}}%
      \expandafter\def\csname LT1\endcsname{\color{black}}%
      \expandafter\def\csname LT2\endcsname{\color{black}}%
      \expandafter\def\csname LT3\endcsname{\color{black}}%
      \expandafter\def\csname LT4\endcsname{\color{black}}%
      \expandafter\def\csname LT5\endcsname{\color{black}}%
      \expandafter\def\csname LT6\endcsname{\color{black}}%
      \expandafter\def\csname LT7\endcsname{\color{black}}%
      \expandafter\def\csname LT8\endcsname{\color{black}}%
    \fi
  \fi
  \setlength{\unitlength}{0.0500bp}%
  \begin{picture}(4500.00,3780.00)%
    \gplgaddtomacro\gplbacktext{%
      \csname LTb\endcsname%
      \put(600,640){\makebox(0,0)[r]{\strut{} 0}}%
      \put(600,1054){\makebox(0,0)[r]{\strut{} 1000}}%
      \put(600,1468){\makebox(0,0)[r]{\strut{} 2000}}%
      \put(600,1882){\makebox(0,0)[r]{\strut{} 3000}}%
      \put(600,2297){\makebox(0,0)[r]{\strut{} 4000}}%
      \put(600,2711){\makebox(0,0)[r]{\strut{} 5000}}%
      \put(600,3125){\makebox(0,0)[r]{\strut{} 6000}}%
      \put(600,3539){\makebox(0,0)[r]{\strut{} 7000}}%
      \put(720,440){\makebox(0,0){\strut{} 0.8}}%
      \put(1428,440){\makebox(0,0){\strut{} 1}}%
      \put(2136,440){\makebox(0,0){\strut{} 1.2}}%
      \put(2843,440){\makebox(0,0){\strut{} 1.4}}%
      \put(3551,440){\makebox(0,0){\strut{} 1.6}}%
      \put(4259,440){\makebox(0,0){\strut{} 1.8}}%
      \put(-100,2089){\rotatebox{-270}{\makebox(0,0){\strut{}Reconstructed relative gain}}}%
      \put(2489,140){\makebox(0,0){\strut{}$\log$($\varepsilon$ [MeV/cm])}}%
      \put(4082,2090){\makebox(0,0)[r]{\strut{}300-900~$\mu$m}}%
    }%
    \gplgaddtomacro\gplfronttext{%
      \csname LTb\endcsname%
      \put(2939,3376){\makebox(0,0)[l]{\strut{}Fitter}}%
      \csname LTb\endcsname%
      \put(2939,3176){\makebox(0,0)[l]{\strut{}Truncated}}%
      \csname LTb\endcsname%
      \put(2939,2976){\makebox(0,0)[l]{\strut{}Power (-2)}}%
      \csname LTb\endcsname%
      \put(2939,2776){\makebox(0,0)[l]{\strut{}Harmonic (-1)}}%
      \csname LTb\endcsname%
      \put(2939,2576){\makebox(0,0)[l]{\strut{}Arithmetic (1)}}%
    }%
    \gplbacktext
    \put(0,0){\includegraphics{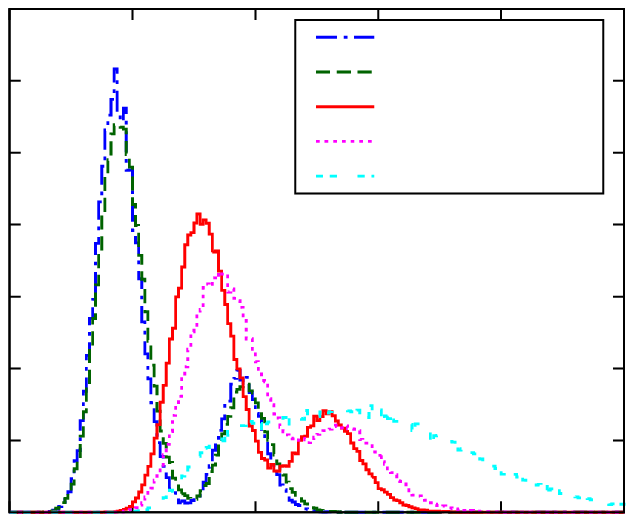}}%
    \gplfronttext
  \end{picture}%
\endgroup

%% file: estimates_3.tex
\begingroup
  \makeatletter
  \providecommand\color[2][]{%
    \GenericError{(gnuplot) \space\space\space\@spaces}{%
      Package color not loaded in conjunction with
      terminal option `colourtext'%
    }{See the gnuplot documentation for explanation.%
    }{Either use 'blacktext' in gnuplot or load the package
      color.sty in LaTeX.}%
    \renewcommand\color[2][]{}%
  }%
  \providecommand\includegraphics[2][]{%
    \GenericError{(gnuplot) \space\space\space\@spaces}{%
      Package graphicx or graphics not loaded%
    }{See the gnuplot documentation for explanation.%
    }{The gnuplot epslatex terminal needs graphicx.sty or graphics.sty.}%
    \renewcommand\includegraphics[2][]{}%
  }%
  \providecommand\rotatebox[2]{#2}%
  \@ifundefined{ifGPcolor}{%
    \newif\ifGPcolor
    \GPcolortrue
  }{}%
  \@ifundefined{ifGPblacktext}{%
    \newif\ifGPblacktext
    \GPblacktexttrue
  }{}%
  \let\gplgaddtomacro\g@addto@macro
  \gdef\gplbacktext{}%
  \gdef\gplfronttext{}%
  \makeatother
  \ifGPblacktext
    \def\colorrgb#1{}%
    \def\colorgray#1{}%
  \else
    \ifGPcolor
      \def\colorrgb#1{\color[rgb]{#1}}%
      \def\colorgray#1{\color[gray]{#1}}%
      \expandafter\def\csname LTw\endcsname{\color{white}}%
      \expandafter\def\csname LTb\endcsname{\color{black}}%
      \expandafter\def\csname LTa\endcsname{\color{black}}%
      \expandafter\def\csname LT0\endcsname{\color[rgb]{1,0,0}}%
      \expandafter\def\csname LT1\endcsname{\color[rgb]{0,1,0}}%
      \expandafter\def\csname LT2\endcsname{\color[rgb]{0,0,1}}%
      \expandafter\def\csname LT3\endcsname{\color[rgb]{1,0,1}}%
      \expandafter\def\csname LT4\endcsname{\color[rgb]{0,1,1}}%
      \expandafter\def\csname LT5\endcsname{\color[rgb]{1,1,0}}%
      \expandafter\def\csname LT6\endcsname{\color[rgb]{0,0,0}}%
      \expandafter\def\csname LT7\endcsname{\color[rgb]{1,0.3,0}}%
      \expandafter\def\csname LT8\endcsname{\color[rgb]{0.5,0.5,0.5}}%
    \else
      \def\colorrgb#1{\color{black}}%
      \def\colorgray#1{\color[gray]{#1}}%
      \expandafter\def\csname LTw\endcsname{\color{white}}%
      \expandafter\def\csname LTb\endcsname{\color{black}}%
      \expandafter\def\csname LTa\endcsname{\color{black}}%
      \expandafter\def\csname LT0\endcsname{\color{black}}%
      \expandafter\def\csname LT1\endcsname{\color{black}}%
      \expandafter\def\csname LT2\endcsname{\color{black}}%
      \expandafter\def\csname LT3\endcsname{\color{black}}%
      \expandafter\def\csname LT4\endcsname{\color{black}}%
      \expandafter\def\csname LT5\endcsname{\color{black}}%
      \expandafter\def\csname LT6\endcsname{\color{black}}%
      \expandafter\def\csname LT7\endcsname{\color{black}}%
      \expandafter\def\csname LT8\endcsname{\color{black}}%
    \fi
  \fi
  \setlength{\unitlength}{0.0500bp}%
  \begin{picture}(4500.00,3780.00)%
    \gplgaddtomacro\gplbacktext{%
      \csname LTb\endcsname%
      \put(600,640){\makebox(0,0)[r]{\strut{} 0}}%
      \put(600,1054){\makebox(0,0)[r]{\strut{} 1000}}%
      \put(600,1468){\makebox(0,0)[r]{\strut{} 2000}}%
      \put(600,1882){\makebox(0,0)[r]{\strut{} 3000}}%
      \put(600,2297){\makebox(0,0)[r]{\strut{} 4000}}%
      \put(600,2711){\makebox(0,0)[r]{\strut{} 5000}}%
      \put(600,3125){\makebox(0,0)[r]{\strut{} 6000}}%
      \put(600,3539){\makebox(0,0)[r]{\strut{} 7000}}%
      \put(720,440){\makebox(0,0){\strut{} 0.8}}%
      \put(1428,440){\makebox(0,0){\strut{} 1}}%
      \put(2136,440){\makebox(0,0){\strut{} 1.2}}%
      \put(2843,440){\makebox(0,0){\strut{} 1.4}}%
      \put(3551,440){\makebox(0,0){\strut{} 1.6}}%
      \put(4259,440){\makebox(0,0){\strut{} 1.8}}%
      \put(-100,2089){\rotatebox{-270}{\makebox(0,0){\strut{}Reconstructed relative gain}}}%
      \put(2489,140){\makebox(0,0){\strut{}$\log$($\varepsilon$ [MeV/cm])}}%
      \put(4082,2090){\makebox(0,0)[r]{\strut{}300-1200~$\mu$m}}%
    }%
    \gplgaddtomacro\gplfronttext{%
      \csname LTb\endcsname%
      \put(2939,3376){\makebox(0,0)[l]{\strut{}Fitter}}%
      \csname LTb\endcsname%
      \put(2939,3176){\makebox(0,0)[l]{\strut{}Truncated}}%
      \csname LTb\endcsname%
      \put(2939,2976){\makebox(0,0)[l]{\strut{}Power (-2)}}%
      \csname LTb\endcsname%
      \put(2939,2776){\makebox(0,0)[l]{\strut{}Harmonic (-1)}}%
      \csname LTb\endcsname%
      \put(2939,2576){\makebox(0,0)[l]{\strut{}Arithmetic (1)}}%
    }%
    \gplbacktext
    \put(0,0){\includegraphics{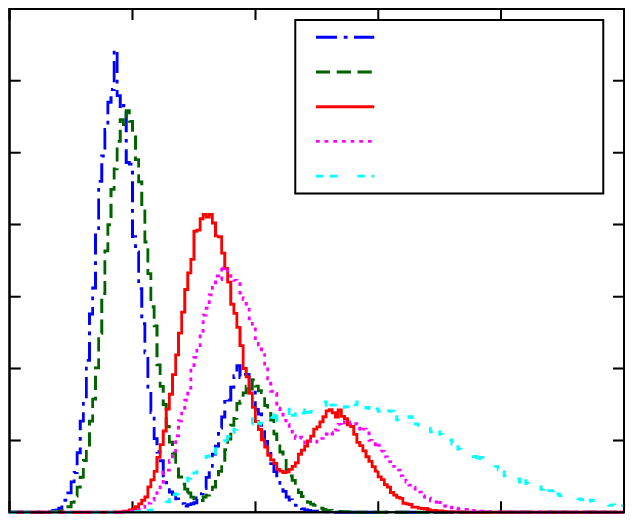}}%
    \gplfronttext
  \end{picture}%
\endgroup

%% file: separation.tex
\begingroup
  \makeatletter
  \providecommand\color[2][]{%
    \GenericError{(gnuplot) \space\space\space\@spaces}{%
      Package color not loaded in conjunction with
      terminal option `colourtext'%
    }{See the gnuplot documentation for explanation.%
    }{Either use 'blacktext' in gnuplot or load the package
      color.sty in LaTeX.}%
    \renewcommand\color[2][]{}%
  }%
  \providecommand\includegraphics[2][]{%
    \GenericError{(gnuplot) \space\space\space\@spaces}{%
      Package graphicx or graphics not loaded%
    }{See the gnuplot documentation for explanation.%
    }{The gnuplot epslatex terminal needs graphicx.sty or graphics.sty.}%
    \renewcommand\includegraphics[2][]{}%
  }%
  \providecommand\rotatebox[2]{#2}%
  \@ifundefined{ifGPcolor}{%
    \newif\ifGPcolor
    \GPcolortrue
  }{}%
  \@ifundefined{ifGPblacktext}{%
    \newif\ifGPblacktext
    \GPblacktexttrue
  }{}%
  \let\gplgaddtomacro\g@addto@macro
  \gdef\gplbacktext{}%
  \gdef\gplfronttext{}%
  \makeatother
  \ifGPblacktext
    \def\colorrgb#1{}%
    \def\colorgray#1{}%
  \else
    \ifGPcolor
      \def\colorrgb#1{\color[rgb]{#1}}%
      \def\colorgray#1{\color[gray]{#1}}%
      \expandafter\def\csname LTw\endcsname{\color{white}}%
      \expandafter\def\csname LTb\endcsname{\color{black}}%
      \expandafter\def\csname LTa\endcsname{\color{black}}%
      \expandafter\def\csname LT0\endcsname{\color[rgb]{1,0,0}}%
      \expandafter\def\csname LT1\endcsname{\color[rgb]{0,1,0}}%
      \expandafter\def\csname LT2\endcsname{\color[rgb]{0,0,1}}%
      \expandafter\def\csname LT3\endcsname{\color[rgb]{1,0,1}}%
      \expandafter\def\csname LT4\endcsname{\color[rgb]{0,1,1}}%
      \expandafter\def\csname LT5\endcsname{\color[rgb]{1,1,0}}%
      \expandafter\def\csname LT6\endcsname{\color[rgb]{0,0,0}}%
      \expandafter\def\csname LT7\endcsname{\color[rgb]{1,0.3,0}}%
      \expandafter\def\csname LT8\endcsname{\color[rgb]{0.5,0.5,0.5}}%
    \else
      \def\colorrgb#1{\color{black}}%
      \def\colorgray#1{\color[gray]{#1}}%
      \expandafter\def\csname LTw\endcsname{\color{white}}%
      \expandafter\def\csname LTb\endcsname{\color{black}}%
      \expandafter\def\csname LTa\endcsname{\color{black}}%
      \expandafter\def\csname LT0\endcsname{\color{black}}%
      \expandafter\def\csname LT1\endcsname{\color{black}}%
      \expandafter\def\csname LT2\endcsname{\color{black}}%
      \expandafter\def\csname LT3\endcsname{\color{black}}%
      \expandafter\def\csname LT4\endcsname{\color{black}}%
      \expandafter\def\csname LT5\endcsname{\color{black}}%
      \expandafter\def\csname LT6\endcsname{\color{black}}%
      \expandafter\def\csname LT7\endcsname{\color{black}}%
      \expandafter\def\csname LT8\endcsname{\color{black}}%
    \fi
  \fi
  \setlength{\unitlength}{0.0500bp}%
  \begin{picture}(4500.00,3780.00)%
    \gplgaddtomacro\gplbacktext{%
      \csname LTb\endcsname%
      \put(600,709){\makebox(0,0)[r]{\strut{} 3}}%
      \put(600,1275){\makebox(0,0)[r]{\strut{} 4}}%
      \put(600,1841){\makebox(0,0)[r]{\strut{} 5}}%
      \put(600,2407){\makebox(0,0)[r]{\strut{} 6}}%
      \put(600,2973){\makebox(0,0)[r]{\strut{} 7}}%
      \put(600,3539){\makebox(0,0)[r]{\strut{} 8}}%
      \put(1162,289){\rotatebox{45}{\makebox(0,0){\strut{}300~$\mu$m}}}%
      \put(2047,289){\rotatebox{45}{\makebox(0,0){\strut{}300--600~$\mu$m}}}%
      \put(2932,289){\rotatebox{45}{\makebox(0,0){\strut{}300--900~$\mu$m}}}%
      \put(3817,289){\rotatebox{45}{\makebox(0,0){\strut{}300--1200~$\mu$m}}}%
      \put(260,2124){\rotatebox{-270}{\makebox(0,0){\strut{}Pion-kaon separation}}}%
    }%
    \gplgaddtomacro\gplfronttext{%
      \csname LTb\endcsname%
      \put(2939,3376){\makebox(0,0)[l]{\strut{}Fitter}}%
      \csname LTb\endcsname%
      \put(2939,3176){\makebox(0,0)[l]{\strut{}Truncated}}%
      \csname LTb\endcsname%
      \put(2939,2976){\makebox(0,0)[l]{\strut{}Power (-2)}}%
      \csname LTb\endcsname%
      \put(2939,2776){\makebox(0,0)[l]{\strut{}Harmonic (-1)}}%
    }%
    \gplbacktext
    \put(0,0){\includegraphics{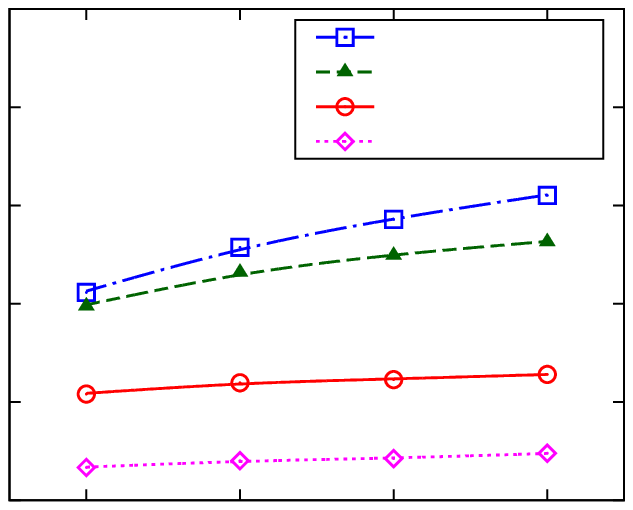}}%
    \gplfronttext
  \end{picture}%
\endgroup

%% file: dependence.tex
\begingroup
  \makeatletter
  \providecommand\color[2][]{%
    \GenericError{(gnuplot) \space\space\space\@spaces}{%
      Package color not loaded in conjunction with
      terminal option `colourtext'%
    }{See the gnuplot documentation for explanation.%
    }{Either use 'blacktext' in gnuplot or load the package
      color.sty in LaTeX.}%
    \renewcommand\color[2][]{}%
  }%
  \providecommand\includegraphics[2][]{%
    \GenericError{(gnuplot) \space\space\space\@spaces}{%
      Package graphicx or graphics not loaded%
    }{See the gnuplot documentation for explanation.%
    }{The gnuplot epslatex terminal needs graphicx.sty or graphics.sty.}%
    \renewcommand\includegraphics[2][]{}%
  }%
  \providecommand\rotatebox[2]{#2}%
  \@ifundefined{ifGPcolor}{%
    \newif\ifGPcolor
    \GPcolortrue
  }{}%
  \@ifundefined{ifGPblacktext}{%
    \newif\ifGPblacktext
    \GPblacktexttrue
  }{}%
  \let\gplgaddtomacro\g@addto@macro
  \gdef\gplbacktext{}%
  \gdef\gplfronttext{}%
  \makeatother
  \ifGPblacktext
    \def\colorrgb#1{}%
    \def\colorgray#1{}%
  \else
    \ifGPcolor
      \def\colorrgb#1{\color[rgb]{#1}}%
      \def\colorgray#1{\color[gray]{#1}}%
      \expandafter\def\csname LTw\endcsname{\color{white}}%
      \expandafter\def\csname LTb\endcsname{\color{black}}%
      \expandafter\def\csname LTa\endcsname{\color{black}}%
      \expandafter\def\csname LT0\endcsname{\color[rgb]{1,0,0}}%
      \expandafter\def\csname LT1\endcsname{\color[rgb]{0,1,0}}%
      \expandafter\def\csname LT2\endcsname{\color[rgb]{0,0,1}}%
      \expandafter\def\csname LT3\endcsname{\color[rgb]{1,0,1}}%
      \expandafter\def\csname LT4\endcsname{\color[rgb]{0,1,1}}%
      \expandafter\def\csname LT5\endcsname{\color[rgb]{1,1,0}}%
      \expandafter\def\csname LT6\endcsname{\color[rgb]{0,0,0}}%
      \expandafter\def\csname LT7\endcsname{\color[rgb]{1,0.3,0}}%
      \expandafter\def\csname LT8\endcsname{\color[rgb]{0.5,0.5,0.5}}%
    \else
      \def\colorrgb#1{\color{black}}%
      \def\colorgray#1{\color[gray]{#1}}%
      \expandafter\def\csname LTw\endcsname{\color{white}}%
      \expandafter\def\csname LTb\endcsname{\color{black}}%
      \expandafter\def\csname LTa\endcsname{\color{black}}%
      \expandafter\def\csname LT0\endcsname{\color{black}}%
      \expandafter\def\csname LT1\endcsname{\color{black}}%
      \expandafter\def\csname LT2\endcsname{\color{black}}%
      \expandafter\def\csname LT3\endcsname{\color{black}}%
      \expandafter\def\csname LT4\endcsname{\color{black}}%
      \expandafter\def\csname LT5\endcsname{\color{black}}%
      \expandafter\def\csname LT6\endcsname{\color{black}}%
      \expandafter\def\csname LT7\endcsname{\color{black}}%
      \expandafter\def\csname LT8\endcsname{\color{black}}%
    \fi
  \fi
  \setlength{\unitlength}{0.0500bp}%
  \begin{picture}(4500.00,3780.00)%
    \gplgaddtomacro\gplbacktext{%
      \csname LTb\endcsname%
      \put(600,709){\makebox(0,0)[r]{\strut{} 0.9}}%
      \put(600,1416){\makebox(0,0)[r]{\strut{} 1}}%
      \put(600,2124){\makebox(0,0)[r]{\strut{} 1.1}}%
      \put(600,2832){\makebox(0,0)[r]{\strut{} 1.2}}%
      \put(600,3539){\makebox(0,0)[r]{\strut{} 1.3}}%
      \put(1162,289){\rotatebox{45}{\makebox(0,0){\strut{}300~$\mu$m}}}%
      \put(2047,289){\rotatebox{45}{\makebox(0,0){\strut{}300--600~$\mu$m}}}%
      \put(2932,289){\rotatebox{45}{\makebox(0,0){\strut{}300--900~$\mu$m}}}%
      \put(3817,289){\rotatebox{45}{\makebox(0,0){\strut{}300--1200~$\mu$m}}}%
      \put(140,2124){\rotatebox{-270}{\makebox(0,0){\strut{}$\log$($\varepsilon$ [MeV/cm])}}}%
    }%
    \gplgaddtomacro\gplfronttext{%
      \csname LTb\endcsname%
      \put(2939,3376){\makebox(0,0)[l]{\strut{}Fitter}}%
      \csname LTb\endcsname%
      \put(2939,3176){\makebox(0,0)[l]{\strut{}Truncated}}%
      \csname LTb\endcsname%
      \put(2939,2976){\makebox(0,0)[l]{\strut{}Power (-2)}}%
      \csname LTb\endcsname%
      \put(2939,2776){\makebox(0,0)[l]{\strut{}Harmonic (-1)}}%
    }%
    \gplbacktext
    \put(0,0){\includegraphics{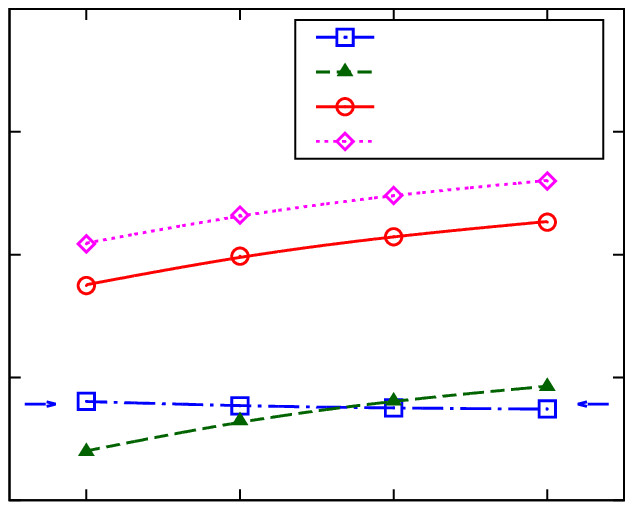}}%
    \gplfronttext
  \end{picture}%
\endgroup

%% file: gains_1.tex
\begingroup
  \makeatletter
  \providecommand\color[2][]{%
    \GenericError{(gnuplot) \space\space\space\@spaces}{%
      Package color not loaded in conjunction with
      terminal option `colourtext'%
    }{See the gnuplot documentation for explanation.%
    }{Either use 'blacktext' in gnuplot or load the package
      color.sty in LaTeX.}%
    \renewcommand\color[2][]{}%
  }%
  \providecommand\includegraphics[2][]{%
    \GenericError{(gnuplot) \space\space\space\@spaces}{%
      Package graphicx or graphics not loaded%
    }{See the gnuplot documentation for explanation.%
    }{The gnuplot epslatex terminal needs graphicx.sty or graphics.sty.}%
    \renewcommand\includegraphics[2][]{}%
  }%
  \providecommand\rotatebox[2]{#2}%
  \@ifundefined{ifGPcolor}{%
    \newif\ifGPcolor
    \GPcolortrue
  }{}%
  \@ifundefined{ifGPblacktext}{%
    \newif\ifGPblacktext
    \GPblacktexttrue
  }{}%
  \let\gplgaddtomacro\g@addto@macro
  \gdef\gplbacktext{}%
  \gdef\gplfronttext{}%
  \makeatother
  \ifGPblacktext
    \def\colorrgb#1{}%
    \def\colorgray#1{}%
  \else
    \ifGPcolor
      \def\colorrgb#1{\color[rgb]{#1}}%
      \def\colorgray#1{\color[gray]{#1}}%
      \expandafter\def\csname LTw\endcsname{\color{white}}%
      \expandafter\def\csname LTb\endcsname{\color{black}}%
      \expandafter\def\csname LTa\endcsname{\color{black}}%
      \expandafter\def\csname LT0\endcsname{\color[rgb]{1,0,0}}%
      \expandafter\def\csname LT1\endcsname{\color[rgb]{0,1,0}}%
      \expandafter\def\csname LT2\endcsname{\color[rgb]{0,0,1}}%
      \expandafter\def\csname LT3\endcsname{\color[rgb]{1,0,1}}%
      \expandafter\def\csname LT4\endcsname{\color[rgb]{0,1,1}}%
      \expandafter\def\csname LT5\endcsname{\color[rgb]{1,1,0}}%
      \expandafter\def\csname LT6\endcsname{\color[rgb]{0,0,0}}%
      \expandafter\def\csname LT7\endcsname{\color[rgb]{1,0.3,0}}%
      \expandafter\def\csname LT8\endcsname{\color[rgb]{0.5,0.5,0.5}}%
    \else
      \def\colorrgb#1{\color{black}}%
      \def\colorgray#1{\color[gray]{#1}}%
      \expandafter\def\csname LTw\endcsname{\color{white}}%
      \expandafter\def\csname LTb\endcsname{\color{black}}%
      \expandafter\def\csname LTa\endcsname{\color{black}}%
      \expandafter\def\csname LT0\endcsname{\color{black}}%
      \expandafter\def\csname LT1\endcsname{\color{black}}%
      \expandafter\def\csname LT2\endcsname{\color{black}}%
      \expandafter\def\csname LT3\endcsname{\color{black}}%
      \expandafter\def\csname LT4\endcsname{\color{black}}%
      \expandafter\def\csname LT5\endcsname{\color{black}}%
      \expandafter\def\csname LT6\endcsname{\color{black}}%
      \expandafter\def\csname LT7\endcsname{\color{black}}%
      \expandafter\def\csname LT8\endcsname{\color{black}}%
    \fi
  \fi
  \setlength{\unitlength}{0.0500bp}%
  \begin{picture}(4500.00,3780.00)%
    \gplgaddtomacro\gplbacktext{%
      \csname LTb\endcsname%
      \put(920,930){\makebox(0,0)[r]{\strut{} 0.6}}%
      \put(920,1510){\makebox(0,0)[r]{\strut{} 0.8}}%
      \put(920,2090){\makebox(0,0)[r]{\strut{} 1}}%
      \put(920,2669){\makebox(0,0)[r]{\strut{} 1.2}}%
      \put(920,3249){\makebox(0,0)[r]{\strut{} 1.4}}%
      \put(1330,440){\makebox(0,0){\strut{} 0.6}}%
      \put(1910,440){\makebox(0,0){\strut{} 0.8}}%
      \put(2490,440){\makebox(0,0){\strut{} 1}}%
      \put(3069,440){\makebox(0,0){\strut{} 1.2}}%
      \put(3649,440){\makebox(0,0){\strut{} 1.4}}%
      \put(340,2089){\rotatebox{-270}{\makebox(0,0){\strut{}Reconstructed relative gain}}}%
      \put(2489,140){\makebox(0,0){\strut{}Real relative gain}}%
      \put(3794,2090){\makebox(0,0)[r]{\strut{}300-600~$\mu$m}}%
    }%
    \gplgaddtomacro\gplfronttext{%
    }%
    \gplbacktext
    \put(0,0){\includegraphics{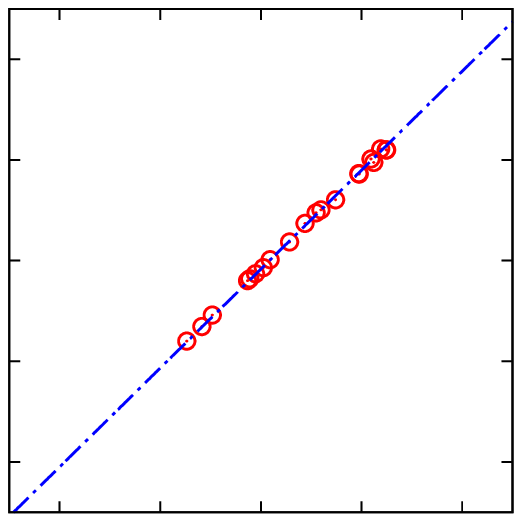}}%
    \gplfronttext
  \end{picture}%
\endgroup